\documentclass[acmlarge]{acmart}

\AtBeginDocument{%
  }

\setcopyright{none}

\usepackage[utf8]{inputenc}
\usepackage[english]{babel}
\usepackage{booktabs}
\usepackage[absolute,overlay]{textpos}
\usepackage{etoolbox}
\usepackage[figurename=Fig.,font={small},labelfont={up}]{caption}
\usepackage[export]{adjustbox}
\usepackage[linesnumbered,ruled,vlined]{algorithm2e}
\usepackage{color, colortbl}
\usepackage{longtable}
\usepackage[dvipsnames]{xcolor}
\usepackage{soul}
\usepackage{comment}
\usepackage{url}
\usepackage{balance}
\usepackage{graphicx}
\usepackage{subfigure}
\usepackage{multirow}
\usepackage{textcomp}
\usepackage{lipsum}
\usepackage{tikz}
\usepackage{hyperref}
\usepackage[capitalise]{cleveref}
\usepackage{float}
\usepackage{xspace}
\usepackage{enumitem}
\usepackage{tabularx}
\usepackage{lscape}
\usepackage{tcolorbox}
\usepackage{makecell}
\usepackage{longtable}
\usepackage{booktabs}
\usepackage{tabularx}
\usepackage{array}
\usepackage[table]{xcolor}
\usepackage{xurl}

\usepackage{seqsplit}
\usepackage{xltabular}
\usepackage{svg}

\tcbuselibrary{listingsutf8}

\usepackage[normalem]{ulem}
\DeclareRobustCommand{\resdel}[1]{{\color{gray}\sout{#1}}}

\usepackage{diagbox}

\newbool{showComments}
\booltrue{showComments}

\ifbool{showComments}{%

}

\begin{document}
\title{Smart Glasses for Ubiquitous Computing: A Layered Survey and the Road Ahead}

\author{Linlu Gao}
\authornote{Co-first authors with equal contribution.}
\email{lg26i@fsu.edu}
\affiliation{%
  \institution{Florida State University}
  \country{United States}
}

\author{Yueyuan Sui}
\authornotemark[1]
\email{yueyuansui2024@u.northwestern.edu}
\affiliation{%
  \institution{Northwestern University}
  \country{United States}
}

\author{Junxi Xia}
\authornotemark[1]
\affiliation{%
  \institution{Northwestern University}
  \country{United States}
}

\author{Xin Liu}
\authornote{Co-corresponding authors.}
\email{xliu15@fsu.edu}
\affiliation{%
  \institution{Florida State University}
  \country{United States}
}

\author{Stephen Xia}
\authornotemark[2]
\email{stephen.xia@northwestern.edu}
\affiliation{%
  \institution{Northwestern University}
  \country{United States}
}

\author{Yang Liu}
\authornotemark[2]
\email{yl25r@fsu.edu}
\affiliation{%
  \institution{Florida State University}
  \country{United States}
}

\begin{abstract}
Smart glasses are evolving from display-centric augmented reality devices into general-purpose wearable computing platforms that bring sensing, computation, interaction, and intelligence into everyday environments.
Recent advances in multimodal sensing, egocentric perception, foundation models, wearable artificial intelligence (AI), and commercial smart-glasses platforms are expanding their role in ubiquitous computing.
This momentum is reflected not only in commercial growth, but also in a sharp recent increase in academic research.
Our literature search emphasized ubiquitous computing venues and focused on work published from 2020 to October 2026, supplemented by selected earlier papers. Of the reviewed papers, 63.2\% were published in 2024 or later.
However, existing surveys do not capture this rapid expansion.
Broad reviews of smart glasses are now outdated, while recent surveys focus mainly on healthcare settings such as distributed care, surgery, and digital health.

To address this gap, we conduct a systematic survey of smart-glasses research for ubiquitous computing.
Our search covers work published from 2020 to October 2026, with selected earlier work included for historical context.
After screening and title-level deduplication, we identify 163 smart-glasses papers and separately compile related resources, including 14 representative hardware platforms, their associated software tools and pretrained models, and 10 datasets.
We organize this literature through four complementary layers. \emph{Research Resources} covers hardware platforms, software tools and models, and datasets that support development and evaluation. \emph{Sensing \& Perception} examines how glasses measure and interpret information about users and their surroundings. \emph{Interaction \& Assistance} addresses how users express intentions and receive feedback and assistance. \emph{Applications} situates these contributions in tasks and use contexts. \textcolor{black}{Energy, latency, wearability, reliability, privacy, security, and social acceptability} are treated as deployment constraints across all four layers.
This framework provides a unified view of current research and a vision for how advances across the four layers could come together to support smart glasses as a ubiquitous computing platform. \textcolor{black}{Building on this synthesis, we outline research directions for extending capabilities across tasks and over time, supporting assistance under human control, and achieving reliable, efficient, and sustained everyday use.}
\end{abstract}

\maketitle
\section{Introduction}

Smart glasses are wearable computing devices worn on or around the eyes that combine sensing, computation, display, and interaction in a head-worn form factor.
Since the introduction of early commercial devices such as Google Glass, smart glasses have often been associated with augmented reality (AR), heads-up information display, and micro-interactions~\cite{lee2018interaction}.
However, recent research has broadened this view substantially.
Modern smart glasses are increasingly studied not only as display devices, but also as sensing and interaction platforms that can perceive the user, the surrounding environment, and social context from an egocentric perspective~\cite{engel2023project,lv2024ariaeverydayactivitiesdataset}.
Their position on the face provides access to visual, auditory, ocular, head-motion, and physiological signals, while also enabling hands-free feedback through visual, audio, and haptic channels.
These properties make smart glasses a compelling platform for ubiquitous computing.
This growing research interest is mirrored by commercial momentum: Counterpoint Research reported that global shipments of smart glasses without displays increased by 210\% from 2023 to 2024. In 2025, shipments in the first and second halves were 110\% and 139\% higher, respectively, than in the same periods of 2024~\cite{counterpoint2025h1smartglasses,counterpoint2026h2smartglasses}. Looking ahead, International Data Corporation (IDC) forecasts that annual shipments of glasses without displays will approximately double between 2026 and 2030. Over the same period, shipments of optical see-through glasses, which overlay digital information while allowing users to see the physical world, are expected to reach approximately four times their 2026 level by 2030.

This shift is driven by several converging trends.
First, advances in miniature cameras, inertial sensors, microphones, eye trackers, physiological sensors, and low-power processors have made it possible to capture rich multimodal signals from wearable eyewear~\cite{lv2024ariaeverydayactivitiesdataset}.
Second, recent progress in foundation models, multimodal learning, and egocentric perception has expanded the ability of smart glasses to interpret first-person visual, auditory, motion, and physiological data in context~\cite{liu2023llava,engel2023project}.
Third, commercial smart-glasses platforms are becoming more capable, lightweight, and integrated with everyday mobile ecosystems, lowering the barrier for research prototypes and real-world deployments.
Together, these developments have expanded the design space of smart glasses beyond visual augmentation to include continuous sensing, context-aware interaction, multimodal AI assistance, accessibility and health support, and interaction with connected devices and robots.

In our literature search, which emphasizes ubiquitous computing venues, we identified 163 unique smart-glasses papers, focusing on publications from 2020 to October 2026 while retaining selected earlier work for historical context.
Of these papers, 103 (63.2\%) were published in 2024 or later, indicating a sharp recent surge of research activity around smart glasses for ubiquitous computing.
This rapid expansion has produced a broad and increasingly difficult-to-navigate body of work, spanning diverse technical foundations, system designs, application contexts, and deployment questions.
This creates a timely need to synthesize the field, clarify how major research threads relate to one another, and identify the challenges that must be addressed for smart glasses to become practical ubiquitous computing platforms.
We therefore conduct a systematic survey of smart-glasses research for ubiquitous computing.

Although prior surveys provide useful foundations, they do not capture this recent expansion of smart-glasses research.
Broad surveys are now outdated: one early review of wearable smart-glasses features, applications, progress, and challenges was published in 2018~\cite{nallapaneni2018wearable}; another interaction-focused survey was also published in 2018~\cite{lee2018interaction}; and a later systematic review analyzed 57 papers published between January 2014 and October 2020~\cite{kim2021applications}.
In contrast, more recent surveys have mainly focused on domain-specific healthcare settings, such as distributed care and telemedicine, surgery, or AI-powered digital health~\cite{zhang2023smart,gollapalli2024smart,wang2025ai}.
Consequently, existing surveys do not provide an up-to-date synthesis of the rapid expansion of smart-glasses research or a unified view of its implications for ubiquitous computing.

To build this survey, we combined keyword-based retrieval, targeted venue checking, and backward reference tracing, focusing on work published from 2020 to October 2026 while retaining selected earlier work for historical context.
After screening and title-level deduplication, we identified 163 unique smart-glasses papers and separately compiled related resources, including 14 representative hardware platforms, their associated software tools and pretrained models, and 10 datasets.
We organize this literature through four complementary layers that describe the resources available, the capabilities developed, and the tasks and settings in which they are used. The \textbf{Research Resources} layer covers hardware platforms, software tools and models, and datasets \textcolor{black}{and associated benchmarks} that support development and evaluation. The \textbf{Sensing \& Perception} layer examines how glasses measure or infer the state of users and their surroundings. The \textbf{Interaction \& Assistance} layer addresses how users express intentions and how systems present information and provide help. The \textbf{Applications} layer situates these contributions in everyday assistance, learning, communication, accessibility, health, professional work, \textcolor{black}{interaction with connected devices and robots, and robot learning}. A single study may contribute to several layers. Across all four, \textbf{deployment constraints}, including energy, latency, reliability, wearability, privacy, security, and \textcolor{black}{social acceptability}, shape the feasibility and experience of everyday use.

The framework provides a unified view of current research and a vision for how advances across the four layers could come together to support smart glasses as a ubiquitous computing platform. Individual studies often combine resources and capabilities to address a specific task. Viewing these studies together helps us examine what has been established, how contributions across the layers relate to one another, and what could be shared, combined, or extended beyond the original tasks and settings. Building on this perspective, \textcolor{black}{we identify seven research directions spanning generalizable egocentric AI, understanding users and context over time, proactive assistance under human control, open research infrastructure, energy-efficient operation, adaptation in everyday use, and sustained application value.}

\noindent\textbf{In summary, this survey makes the following contributions:}

\begin{itemize}[leftmargin=*,itemsep=3pt,topsep=4pt]
    \item We provide a \textbf{systematic and up-to-date review} of smart-glasses research for ubiquitous computing, covering work from 2020 to October 2026 and incorporating selected earlier studies for historical context.
    \item We construct a \textbf{literature corpus} of 163 unique smart-glasses papers and compile related resources, including 14 representative hardware platforms, their associated software tools and pretrained models, and 10 datasets, providing an empirical basis for analyzing the recent growth of the field.
    \item We introduce a \textbf{taxonomy of four complementary layers}: Research Resources, Sensing \& Perception, Interaction \& Assistance, and Applications. With deployment constraints considered across all four layers, the taxonomy provides a unified picture of current research and a vision for smart glasses as a ubiquitous computing platform.
    \item \textcolor{black}{We synthesize \textbf{research progress, recurring limitations, and future directions} across the framework, using evidence from the reviewed studies to identify development needs and evaluation priorities for extending capabilities across tasks and settings and establishing sustained application value.}
\end{itemize}

\section{Methodology}
\label{sec:methodology}
\label{sec:methodology-revised}
We conducted a structured literature review of smart glasses for ubiquitous computing, combining keyword-based retrieval, venue checking, backward reference tracing, screening, and deduplication. Research resources were identified and assessed separately.

\subsection{Paper Retrieval}
\label{sec:methodology-revised-retrieval}
Our search focused on peer-reviewed conference and journal publications from 2020 to October 2026, supplemented by selected relevant earlier studies and surveys dating back to 2018. We retrieved literature through digital libraries and publisher websites, including the ACM Digital Library, IEEE Xplore, Springer, ScienceDirect, MDPI, and the nature.com website. Unpublished preprints, including arXiv-only papers, were excluded.

\noindent{\small$\blacksquare$}~\textbf{Venue and subject coverage.}
Table~\ref{tab:methodology-coverage} summarizes core coverage in ubiquitous computing, HCI, accessibility, and AR, together with supplementary searches in related technical and interdisciplinary fields. Venues guided coverage; eligibility depended on the criteria in Section~\ref{selection}, regardless of publication venue.

\begin{table}[tbp]
\centering
\caption{Venue and subject coverage of the literature search. The groups describe the organization of the search; all retrieved papers were assessed using the same selection criteria.}
\label{tab:methodology-coverage}
\small
\renewcommand{\arraystretch}{1.08}
\begin{tabular}{@{}>{\raggedright\arraybackslash}p{\linewidth}@{}}
\toprule
\rowcolor{gray!10}
\textbf{Core coverage} \\
\midrule
\textbf{Ubiquitous, wearable, and mobile computing:} ACM IMWUT/UbiComp; ISWC; PerCom; MobiSys; MobiCom; SenSys; MobiHoc; IEEE TMC \\[2pt]
\textbf{Human--computer interaction:} CHI; UIST; DIS; IUI; MobileHCI; MUM; ACM TOCHI; IJHCS; CSCW (PACMHCI) \\[2pt]
\textbf{Accessibility and augmentation:} ASSETS; PETRA; Augmented Human (AHs); ICCHP \\[2pt]
\textbf{Gaze interaction, AR, and visualization:} ETRA; ISMAR; IEEE TVCG \\
\midrule
\rowcolor{gray!10}
\textbf{Supplementary coverage} \\
\midrule
\textbf{Speech and audio processing:} INTERSPEECH; ICASSP; CHiME \\[2pt]
\textbf{Egocentric vision and AI:} CVPR; ECCV; ICCV; NeurIPS; ACM MM \\[2pt]
\textbf{Privacy and security:} IEEE Security \& Privacy; NSPW; SOUPS; USENIX Security; ACM CCS; NDSS \\[2pt]
\textbf{Agents, robots, and connected systems:} The Web Conference; ICRA; IROS; IEEE TASE; HRI Companion; CHASE; CoRL \\
\midrule
\rowcolor{gray!10}
\textbf{Broader technical and interdisciplinary research} \\
\midrule
IEEE Access; Sensors; Applied Sciences; Scientific Reports; Eye; PLOS ONE; Multimedia Tools and Applications; Augmented Human Research; AI and Ethics; Technological Forecasting and Social Change; Humanities and Social Sciences Communications; JMIR Formative Research \\
\bottomrule
\end{tabular}
\end{table}

\noindent{\small$\blacksquare$}~\textbf{Keyword combinations.}
We combined the device terms below with specific keywords, refining the set as relevant terms emerged from retrieved papers and prior surveys. We searched titles, abstracts, keywords, and metadata where supported by each digital library.

\noindent\textbf{Device terms:} ``smart glasses'', ``smart eyewear'', ``smart eyeglasses'', ``AR glasses'', ``augmented-reality glasses'', ``mixed-reality glasses'', ``camera glasses'', ``wearable glasses'', ``eyewear'', ``head-worn device'', and ``head-mounted device''.

\noindent\textbf{Additional search terms:}
\begin{enumerate}[leftmargin=*,itemsep=2pt,topsep=3pt]
    \item \textbf{General technical terms:} ``benchmark'', ``commercial device'', ``dataset'', ``hardware platform'', ``perception'', ``recognition'', ``sensing'', ``tracking''.
    \item \textbf{Vision and context:} ``attention'', ``context awareness'', ``eye tracking'', ``gaze'', ``object recognition'', ``scene understanding''.
    \item \textbf{Speech and communication:} ``acoustic sensing'', ``audio'', ``communication assistance'', ``speech''.
    \item \textbf{Actions and interfaces:} ``feedback'', ``gesture'', ``input'', ``interaction'', ``robot interaction'', ``user interface''.
    \item \textbf{Accessibility and everyday tasks:} ``accessibility'', ``assistive technology'', ``education'', ``navigation'', ``personal assistance'', ``visual impairment'', ``workplace''.
    \item \textbf{Activity, health, and care:} ``activity recognition'', ``clinical'', ``healthcare'', ``intervention'', ``physiological sensing'', ``rehabilitation'', ``well-being''.
    \item \textbf{Privacy and user experience:} ``authentication'', ``bystander privacy'', ``consent'', ``privacy'', ``security'', ``social acceptance'', ``usability''.
\end{enumerate}

\noindent{\small$\blacksquare$}~\textbf{Supplementary retrieval.}
Venue checking and backward reference tracing complemented keyword retrieval. We examined references in selected papers and prior surveys to identify additional relevant work. Candidates discovered through these routes were considered under the same selection criteria as database results and incorporated into the screening process described below.

\noindent{\small$\blacksquare$}~\textbf{Resource identification.}
\label{sec:methodology-resources}
Alongside the paper search, we identified research resources through relevant publications and official manufacturer, developer, and project sources. For hardware, we consulted product pages and technical specifications for commercial eyewear and related head-worn research platforms, checking their sensors, computation, connectivity, power arrangements, and research access. For software, we examined developer documentation and project repositories for sensor-access interfaces, recording and streaming tools, software development kits (SDKs), application-development frameworks, and released pretrained models. Examples include Research Mode and hl2ss for sensor access and streaming, and MRTK and OpenXR for application development. For datasets, we examined dataset papers and release pages describing recordings collected with glasses or related head-worn devices, including multimodal collections that combine these recordings with external views or other sensors. We recorded the capture setup, sensing modalities, annotations, collection scale, supported tasks, and conditions of data access. Section~\ref{sec: Resources} presents the resulting representative inventory and its supporting sources.\par

\subsection{Selection Criteria}
\label{selection}
Eligibility considered both the device and the research contribution. The target platform is frame-worn smart glasses that combine sensing with computation or connectivity while maintaining access to the physical world. It includes glasses with visual displays as well as audio-only and camera-based configurations.

\noindent{\small$\blacksquare$}~\textbf{Target platform.}
Table~\ref{tab:methodology-device-criteria} defines five device characteristics: form factor (F), sensing on frame (S), computation or connectivity (C), wearability in public (W), and world visibility (V). Platforms meeting all five satisfied device eligibility; others were assessed using the research-relevance criteria below. Examples of target platforms include Ray-Ban Meta, Google Glass, Project Aria, and lightweight AR glasses.

\begin{table}[t]
\centering
\caption{Device characteristics used to identify the target smart-glasses platform. Related research platforms were also retained on the basis of the research relevance explained in the text.}
\label{tab:methodology-device-criteria}
\small
\setlength{\tabcolsep}{5pt}
\renewcommand{\arraystretch}{1.16}
\begin{tabular}{@{}>{\raggedright\arraybackslash}p{0.27\linewidth}>{\raggedright\arraybackslash}p{0.69\linewidth}@{}}
\toprule
\rowcolor{gray!15}
\textbf{Characteristic} & \textbf{Target-platform description} \\
\midrule
\rowcolor{gray!5}
\textbf{Form factor (F)} & An eyeglass or sunglasses frame designed to be worn like conventional glasses, with an emphasis on a compact and lightweight form factor. \\
\textbf{Sensing on frame (S)} & Cameras, microphones, IMUs, eye-tracking sensors, radar, or other sensing components integrated into the frame or lenses and collecting signals through the glasses. \\
\rowcolor{gray!5}
\textbf{Compute/connectivity (C)} & Onboard computation, data storage, or connectivity to a smartphone, PC, or cloud/edge infrastructure, supporting sensing, perception, or interaction through live or subsequent processing. \\
\textbf{Wearability in public (W)} & A design intended for everyday or repeated use in public environments, with attention to physical and social wearability. \\
\rowcolor{gray!5}
\textbf{World visibility (V)} & Continued access to the surrounding physical environment, through optical see-through displays or non-occluding configurations such as audio-only and displayless glasses. \\
\bottomrule
\end{tabular}
\end{table}

\noindent{\small$\blacksquare$}~\textbf{Relevant research platforms and prototypes.}
For studies using platforms that did not meet all five device criteria in Table~\ref{tab:methodology-device-criteria}, we defined three research-relevance criteria. A study was eligible for inclusion if its contribution met at least one of the following:
\begin{enumerate}[leftmargin=*,itemsep=2pt,topsep=3pt]
    \item \textbf{Sensing methods:} studies techniques and/or introduced findings for measuring or interpreting signals available, or expected to become available, on smart glasses.
    \item \textbf{Interaction techniques:} studies methods and/or findings that involve inputs, feedback, interfaces, or assistance applicable to a glasses form-factor.
    \item \textbf{Tasks and use conditions:} provides insights into tasks, user needs, or deployment conditions relevant to smart glasses.
\end{enumerate}

This includes relevant work using AR research platforms such as HoloLens and Magic Leap, whose sensing and spatial interaction capabilities support investigations of smart-glasses use. It also accommodates experimental configurations that use external components or have yet to achieve the target platform's lightweight, everyday form, as well as selected studies on smartphones or other wearable devices whose methods or findings address the research-relevance criteria above.

The consolidated list contained 166 unique publications after deduplication and separation of resource-only entries. Applying the criteria above, we excluded three publications and retained 163 papers, including all 15 studies identified as borderline during device assessment. These borderline studies were retained because their contributions addressed the research-relevance criteria above, despite differences between their device configurations and the target smart-glasses platform.

\noindent{\small$\blacksquare$}~\textbf{Exclusion boundaries.}
Work on immersive VR, standalone smartphones, other wearables, general AR, or solely optical display design, rendering, and graphics was excluded when it failed the research-relevance criteria above. Merely mentioning smart glasses as a future interface was insufficient. Related prototypes were assessed for substantive relevance rather than conformity to existing consumer form factors.

\noindent{\small$\blacksquare$}~\textbf{Publication and resource types.}
The paper corpus comprised peer-reviewed conference and journal publications, including papers introducing relevant research resources. Unpublished preprints, including arXiv-only papers, were excluded. When a work was available both as a preprint and as a peer-reviewed publication, the published version was considered for inclusion under the same platform and research-relevance criteria. Prior surveys and selected earlier studies provided historical context and supported reference tracing.

Resources were selected for their practical use in developing or studying smart-glasses systems: relevant hardware platforms, software for application development or signal processing, and datasets collected with glasses or related devices or directly supporting relevant tasks. Datasets could combine eyewear recordings with external views or sensors. Resource entities were counted separately from papers. Product documentation, repositories, project pages, and resource reports supported their descriptions without adding papers to the corpus; a peer-reviewed resource paper could also enter the corpus if it met the paper-selection criteria.

\subsection{Screening, Deduplication, and Corpus}
\label{sec:methodology-revised-screening}
We compiled candidate papers into a spreadsheet and screened them using their titles, abstracts, publication venues and years, application areas, sensing modalities, and main contributions. Device characteristics and research relevance were assessed using the criteria in Section~\ref{selection}. For ambiguous cases, we inspected the full text; disagreements were resolved through discussion. Candidates identified through reference tracing underwent the same assessment.

Title-level deduplication consolidated papers retrieved through multiple searches or categories. For each included paper, we recorded venue, year, application area, sensing modality, device type, inclusion classification, and main contribution for comparison and thematic organization. Papers could address several topics while counting once in the corpus.

The resulting corpus contained 163 unique papers, including three earlier papers retained for historical context; 103 (63.2\%) were published in 2024 or later. We separately compiled 14 representative hardware platforms, their associated software tools and pretrained models, and 10 datasets, reviewed in Section~\ref{sec: Resources}.

\section{A Layered Taxonomy of Smart-Glasses Research}
\label{sec:taxonomy-revised}
\label{sec:taxonomy}
\graphicspath{{./}}
\providecommand{\Description}[1]{}
Smart-glasses research spans the resources used to build and study systems, the capabilities those systems provide, and the tasks they support. We organize this work into four complementary layers: \emph{Research Resources}, \emph{Sensing \& Perception}, \emph{Interaction \& Assistance}, and \emph{Applications}. \emph{Cross-cutting Deployment Constraints} capture the conditions that shape how these resources and capabilities can be used in everyday settings.

The taxonomy provides a common view of progress across these research areas and a basis for considering smart glasses as a ubiquitous-computing platform. Individual studies often bring together several elements of the framework around a particular task. Viewing these studies together helps us consider how their advances can support a broader range of uses, what can be shared or adapted across systems, and what further development and validation everyday use requires. The framework therefore serves both to organize existing research and to guide discussion of the field's future development.

\subsection{Four Complementary Layers}
\label{sec:taxonomy-revised-layers}

\begin{figure}[t]
  \centering
  \includegraphics[width=\linewidth]{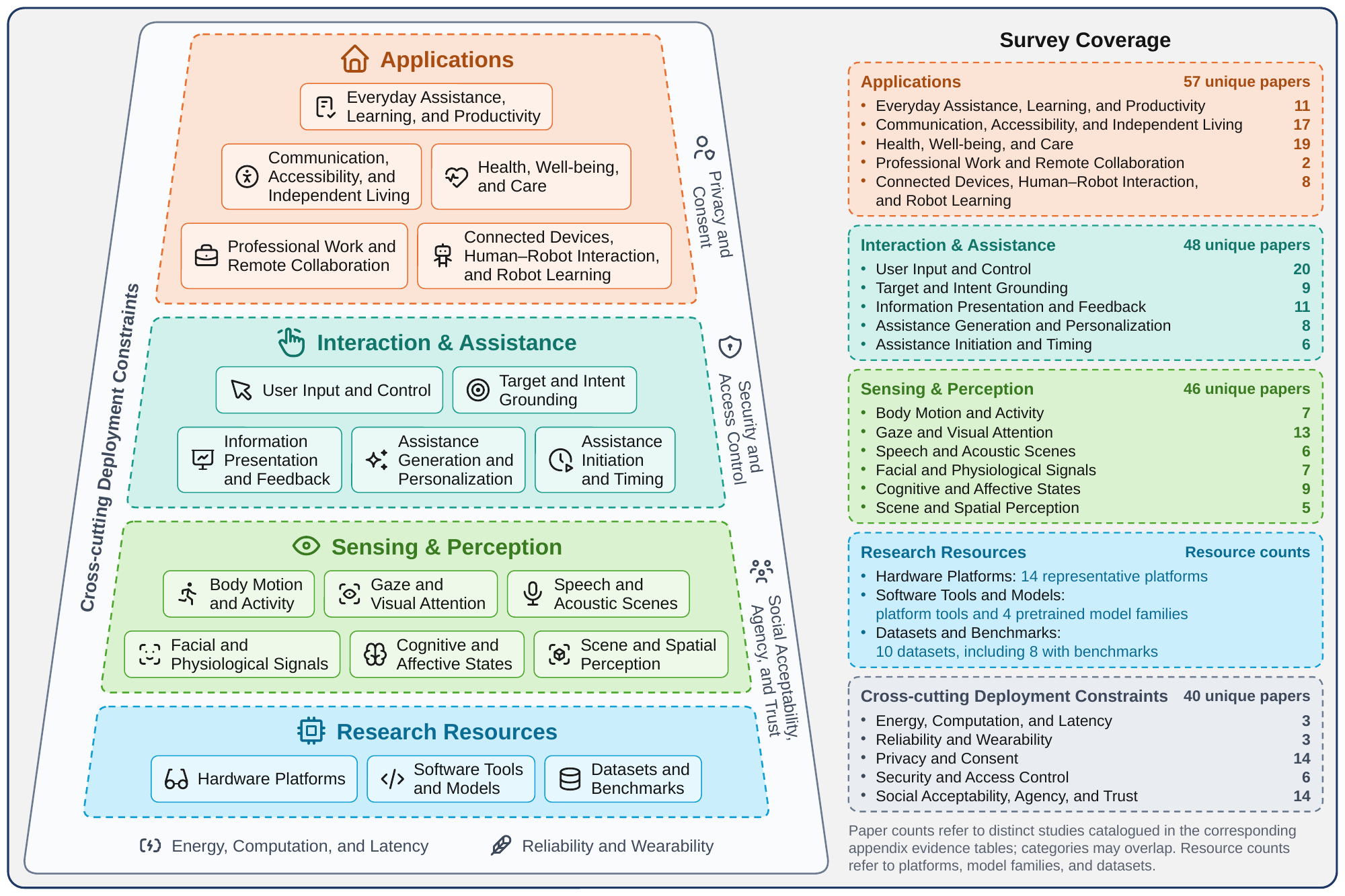}
  \Description{A layered framework places Research Resources at the foundation, followed by Sensing and Perception, Interaction and Assistance, and Applications. Deployment constraints span the framework. Each layer identifies its main research areas, and individual studies may contribute to multiple layers.}
  \caption{A layered taxonomy of smart-glasses research. Research resources form a shared foundation for sensing and perception, interaction and assistance, and applications. Deployment constraints shape decisions across all four layers. A study may contribute to multiple layers, with application needs and deployment experience informing further development throughout the framework.}
  \label{fig:taxonomy-revised-framework}
  \label{fig:taxonomy}
\end{figure}

Figure~\ref{fig:taxonomy-revised-framework} presents the framework. Each layer addresses a different aspect of smart-glasses research and provides an organizing perspective for the following sections.

\noindent{\small$\blacksquare$}~\textbf{Research Resources:} \textit{what enables development and evaluation?}
This foundational layer includes hardware platforms, software tools and models, and datasets and associated benchmarks. These resources support development, training, operation, and evaluation across the \emph{Sensing \& Perception}, \emph{Interaction \& Assistance}, and \emph{Applications} layers.

\noindent{\small$\blacksquare$}~\textbf{Sensing \& Perception:} \textit{what can glasses measure and understand?}
This layer examines how glasses measure or infer the state of the wearer and the surrounding world. Within its research fields, sensing modalities and operating mechanisms distinguish how capabilities are achieved. Their outputs provide user and context information for interaction, assistance, and applications.

\noindent{\small$\blacksquare$}~\textbf{Interaction \& Assistance:} \textit{how are capabilities used to communicate and help?}
This layer examines how users express intentions and how systems use available information to support interaction and assistance. Its fields address input and control, target and intent grounding, information presentation and feedback, assistance generation and personalization, and assistance initiation and timing.

\noindent{\small$\blacksquare$}~\textbf{Applications:} \textit{for whom, in which situations, and toward which outcomes?}
This layer situates resources and capabilities within tasks and use contexts. It examines the role glasses play in an activity and the evidence about their use, complementing the functions and mechanisms examined in the \emph{Sensing \& Perception} and \emph{Interaction \& Assistance} layers. Applications provide settings to assess usefulness and identify requirements for further development.

\noindent{\small$\blacksquare$}~\textbf{Cross-cutting Deployment Constraints.}
Energy, computation, communication, and latency; robustness, reliability, and wearability; privacy and consent; security and access control; and social acceptability, agency, and trust shape decisions across all four layers. These conditions influence which resources are available, which capabilities are practical, and how systems can support users in different settings. Research that examines or addresses these conditions is an integral part of the survey.

The layers describe complementary research roles, and a single paper may contribute to several of them. Resources can directly support work across the framework, while application needs and deployment experience can guide the development of resources, capabilities, and interaction designs. We use these relationships to retain a view of the systems and tasks in which individual advances become useful.\par

\subsection{Using the Framework to Understand Progress and Future Directions}
\label{sec:taxonomy-revised-perspective}
We use the framework to examine the literature through three questions that guide the synthesis across the survey.

\noindent\textbf{What has been established?}
We identify the resources available to researchers, the capabilities that smart glasses can support, the interaction and assistance approaches developed, and the applications studied. Considering these areas together shows where progress has been made and the tasks and settings in which its value has been examined.

\noindent\textbf{How do resources and capabilities come together in smart-glasses systems?}
 We examine how research resources support sensing, perception, interaction, and assistance, and how these capabilities are combined in applications. This includes recurring mechanisms, combinations of capabilities, documented reuse of methods and resources, and requirements arising from particular tasks. The framework helps relate findings across research areas and understand the conditions that make particular combinations useful.

\noindent\textbf{What is needed to advance smart glasses as a ubiquitous-computing platform?}
We consider how existing advances could support more reusable and adaptable systems for everyday use. This involves identifying capabilities that could be combined or extended, the validation needed beyond their current tasks and settings, and deployment constraints that affect their use. These questions guide the research directions developed in the later sections.

\section{Resources for Smart-Glasses Research}
\label{sec: Resources}

\begin{figure}[tp]
  \centering
  \includegraphics[width=\linewidth]{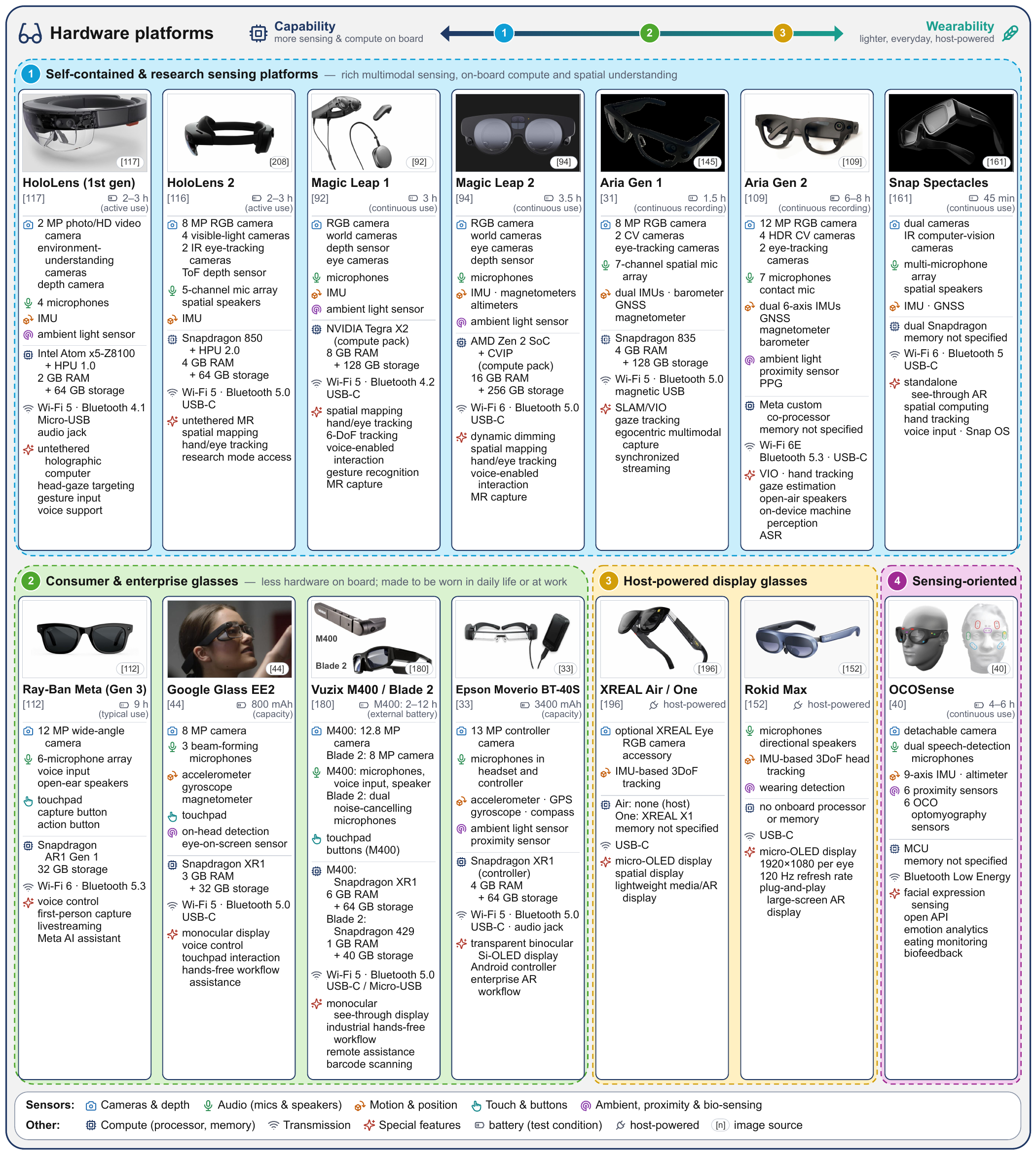}
  \Description{Fourteen smart-glasses hardware platforms shown as cards in four groups. Each card gives a photo of one platform and lists its sensors, compute, transmission, battery, and special features.}
  \caption{Representative hardware platforms for smart-glasses research, grouped by their intended use and hardware configuration. The figure summarizes sensors, computation, connectivity, battery information, and distinctive functions. The groups illustrate different design priorities, including onboard sensing and processing, everyday use, and reliance on external devices. Battery figures reflect the reported operating conditions for each platform.}
  \label{fig:hardware-platforms}
  \nocite{zari2023magic}
\end{figure}

This section examines how hardware platforms, software tools and models, and publicly available datasets and benchmarks enable research on smart glasses. Together, they define what researchers can sense, implement, and evaluate under different constraints on computation, wearability, access, and deployment.

\subsection{Hardware Platforms}

 Hardware platforms distribute sensing, computation, displays, and power between the frame and external devices. Figure~\ref{fig:hardware-platforms} shows configurations ranging from integrated AR and research sensing platforms to lightweight displays that rely on a connected host. These choices shape both the capabilities available for prototyping and the conditions under which they can be studied.

 Figure \mbox{\ref{fig:hardware-platforms}} compares 14 representative platforms by sensors, compute, transmission, battery, and special features.

\noindent{\small$\blacksquare$}~\textbf{Sensors.}  Almost every platform carries a camera, an inertial measurement unit (IMU), and microphones that support scene, speech, and motion sensing. Fusing camera observations with inertial signals enables visual-inertial odometry (VIO), linking the wearer's movement to the surrounding space. Eye-tracking and depth cameras (on HoloLens 2, Magic Leap, and Aria) extend these measurements to visual attention and scene geometry. Research platforms also add physiological signals: Aria Gen 2 uses photoplethysmography (PPG) for heart rate, while OCOSense measures facial movement through optomyography. This sensor combination determines which aspects of the wearer and surroundings can be observed together.

\noindent{\small$\blacksquare$}~\textbf{Compute.}  Seven of the 14 platforms run on a Qualcomm Snapdragon system on chip (SoC), which integrates CPU, GPU, and AI processing in a system on chip (SoC)~\cite{qualcommxr1}. Platforms performing continuous spatial tracking often add a co-processor, such as the HoloLens holographic processing unit, to handle perception while leaving the main processor available for applications. Computation can also move off the frame: Magic Leap uses a hip-worn compute pack, while XREAL Air and Rokid Max depend on a host. These arrangements reduce processing on the head but introduce external hardware dependencies. OCOSense instead uses a microcontroller for sampling and streaming.

 The compute class of the onboard processor sets how much AI a platform runs on board. A microcontroller handles sensor sampling and simple processing; a SoC with a neural processing unit (NPU), such as the AR1 Gen 1 in Ray-Ban Meta, can execute compact image and audio models~\cite{qualcommar1,qualcommar1brief}. Aria Gen 2 performs VIO, eye tracking, hand tracking, and speech recognition on device~\cite{ariaGen2docs}. Larger multimodal assistants rely on a paired phone or cloud service~\cite{metaaicloud}.

\noindent{\small$\blacksquare$}~\textbf{Transmission.}  Bluetooth and USB are standard on almost every platform, and every platform with a built-in camera supports at least Wi-Fi 5. Wi-Fi requires more energy but supports high-rate video and offloaded computation~\cite{ieee80211ac,ieee80211ax}, whereas Bluetooth Low Energy (BLE) suits control and low-rate sensor streams~\cite{bluetooth5}. XREAL and Rokid use a single USB-C cable that carries video and power from a host device.

\noindent{\small$\blacksquare$}~\textbf{Battery.}  The battery must fit inside a small eyeglass frame, and how long it can sustain the system depends on the workload and compute location. Figure~\ref{fig:hardware-platforms} reports battery figures with their operating conditions, which vary wildly, from the Snap Spectacles that need to render see-through AR, resulting in a battery life of just 45 minutes, to Ray-Ban Meta that has no display, resulting in a typical battery life of 9 hours. XREAL and Rokid carry no battery and draw power from a host device. Platforms that need more energy have the option to include external low-power processors, compute packs, or batteries (e.g., Aria Gen 2, Vuzix M400, Epson Moverio).

\noindent{\small$\blacksquare$}~\textbf{Special features.}  Each platform’s distinctive capabilities reflect its intended applications. HoloLens, Magic Leap, and Snap Spectacles emphasize spatial interaction through binocular see-through displays and hand tracking, whereas Google Glass and Vuzix support hands-free workflows through monocular displays and voice and touchpad input. XREAL and Rokid emphasize wearable displays that serve as large private screens, while Ray-Ban Meta emphasizes first-person capture and interaction with a built-in AI assistant.

 Taken together, these hardware characteristics and intended use cases provide a basis for organizing the platforms in Figure \mbox{\ref{fig:hardware-platforms}} into four broad groups, highlighting the capabilities and constraints each brings to smart-glasses research:

\begin{enumerate}[leftmargin=*,itemsep=2pt,topsep=3pt]
\item \textbf{Self-contained and research sensing platforms.}  HoloLens, Magic Leap, Aria, and Spectacles provide rich sensing for multimodal perception and, where supported, spatial interaction~\cite{hololens1,hololens2hardware,magicleap1,magicleap2,engel2023project,ariaGen2docs,snap_spectacles2024}. Their sensing access makes them useful research platforms, while their increased weight, power, and cost can limit how closely studies represent everyday eyewear.

\item \textbf{Consumer and enterprise glasses.}  Ray-Ban Meta, Google Glass EE2, Vuzix, and Epson Moverio are closer to consumer eyewear because these glassessupport studies of daily and workplace use~\cite{metaaiglasses,googleglassee2,vuzix_m400_blade,epson_moverio_bt40s}. They expose practical constraints on social acceptability, battery life, API access, and hands-free workflows, often with fewer research-facing sensors.

\item \textbf{Host-powered display glasses.}  XREAL Air / One and Rokid Max suit studies of wearable display ergonomics, visual feedback, and split computing~\cite{xreal_glasses,rokid_max}. Their limited onboard sensing and reliance on a host constrain fully integrated sensing experiments. Their light head-worn form should be assessed together with the host and connection required for the application.

\item \textbf{Sensing-oriented glasses.}  OCOSense uses optomyography and proximity sensing for facial expressions~\cite{10.1145/3544549.3583918}, illustrating how specialized sensing can reduce reliance on always-on cameras.
\end{enumerate}

\subsection{Software Tools and Models}
\label{subsec:software-tools-models}

 Table~\ref{tab:software-tools-models} summarizes data-access tools, application-development tools, and pretrained models that researchers reuse to enable perception and reasoning for the platforms in Figure~\ref{fig:hardware-platforms}.

\begin{table*}[t]
\centering
\caption{Software tools and models for each hardware platform in Figure~\ref{fig:hardware-platforms}.}
\label{tab:software-tools-models}
\small
\setlength{\tabcolsep}{3pt}
\renewcommand{\arraystretch}{1.1}
\begin{tabularx}{\textwidth}{
>{\raggedright\arraybackslash\hspace{0pt}}p{3.0cm}
>{\raggedright\arraybackslash\hspace{0pt}}X
>{\raggedright\arraybackslash\hspace{0pt}}p{3.8cm}
>{\raggedright\arraybackslash\hspace{0pt}}p{4.1cm}
}
\toprule
& \multicolumn{2}{c}{\textbf{Software tools}} & \\
\cmidrule(lr){2-3}
\textbf{Platform} & \textbf{Data access} & \textbf{Development} & \textbf{Models} \\
\midrule
\rowcolor{gray!10} HoloLens (1st gen) & Research Mode & Unity, MRTK & EgoVLP, Whisper \\
\rowcolor{gray!10} HoloLens 2 & Research Mode, hl2ss & Unity, MRTK, OpenXR & EgoVLP, Whisper \\
\rowcolor{gray!10} Magic Leap 1 & MLCamera (RGB photo, video) & Unity, Unreal, Lumin SDK & EgoVLP, Whisper \\
\rowcolor{gray!10} Magic Leap 2 & Pixel Sensor API & Unity, Unreal, OpenXR, C~API & EgoVLP, Whisper \\
\rowcolor{gray!10} Aria Gen 1 & Client SDK, Aria Tools, MPS & No developer access & EgoVLP, Whisper, EgoBlur, EVL \\
\rowcolor{gray!10} Aria Gen 2 & Client SDK, Aria Tools, MPS & No developer access & EgoVLP, Whisper, EgoBlur \\
\rowcolor{gray!10} Snap Spectacles & Camera Module & Lens Studio & EgoVLP, Whisper \\
Ray-Ban Meta (Gen 3) & Device Access Toolkit & Android SDK & EgoVLP, Whisper \\
Google Glass EE2 & Android APIs & Android Studio & EgoVLP, Whisper \\
Vuzix M400 / Blade 2 & Android APIs, Vuzix SDKs & Android Studio & EgoVLP, Whisper \\
Epson Moverio BT-40S & Moverio Basic Function SDK & Android Studio & Whisper \\
\rowcolor{gray!10} XREAL Air / One & Camera with XREAL Eye (One) & Unity, XREAL SDK & EgoVLP \\
\rowcolor{gray!10} Rokid Max & Android APIs (Station) & Unity, UXR SDK & Whisper \\
OCOSense & Supervision App, LSL & No public SDK & \textcolor{black}{No raw video or audio} \\
\bottomrule
\end{tabularx}
\par\smallskip
\begin{minipage}{\textwidth}\small
Rows follow the four groups of Figure~\ref{fig:hardware-platforms}, and shading marks groups 1 and 3.
\end{minipage}
\end{table*}

\noindent{\small$\blacksquare$}~\textbf{Data access.}  On self-contained and research platforms, access to sensor streams remains platform-specific, even when tools are open source. HoloLens Research Mode exposes tracking and depth streams, with inertial sensors added on HoloLens 2~\cite{hl1researchmode,ungureanu2020hololens,hl2researchmodedocs}; hl2ss streams these alongside front-camera, microphone, eye-tracking, and hand-tracking data to a computer~\cite{dibene2022hl2ss}. Magic Leap 2 also exposes world, depth, and eye cameras through its Pixel Sensor API~\cite{ml2pixelsensor}, while Spectacles supplies camera frames through the Camera Module~\cite{snapcameramodule}. Aria provides a complementary pipeline: the Client SDK streams live data, Project Aria Tools reads recordings, and Machine Perception Services (MPS) process recordings in the cloud to return six-degree-of-freedom trajectories, point clouds, gaze, and hand landmarks~\cite{ariaclientsdk,ariatools,ariamps}.

 Consumer and enterprise platforms expose more limited streams, mainly cameras, microphones, and motion sensors. Ray-Ban Meta routes access through a paired phone~\cite{metawearablestoolkit,metawearablesfaq}, while Android camera and sensor APIs serve enterprise platforms, supplemented by vendor-specific functions~\cite{glassee2dev,vuzixsdk,epsonmoveriosdk}.

 Host-powered platforms expose limited sensing. XREAL provides camera frames only when the optional XREAL Eye is attached to a supported One-series device~\cite{xrealsdk}. Rokid Max has no camera, but applications on its companion hardware can read motion sensors~\cite{rokidstation}.

 The sensing-oriented OCOSense records and streams its raw signals through the Supervision App and Lab Streaming Layer (LSL)~\cite{ocosensesupervision}.

\noindent{\small$\blacksquare$}~\textbf{Development.}  Application development requires additional access to the device runtime as well as suitable authoring tools. HoloLens and Magic Leap combine 3D engines with vendor toolkits, such as the Mixed Reality Toolkit (MRTK), linking spatial content to tracking and input~\cite{hl1unity,mrtk2,mrtk,ml1unity,ml1unreal,magicleap2sdk}. OpenXR provides a common application interface across supported devices~\cite{openxr,openxrconformant}, while Spectacles uses Lens Studio~\cite{lensstudio,snap_spectacles2024}.

 Enterprise glasses use generally do not have large displays, so standard Android development tools are commonly used across all platforms in this category~\cite{glassee2dev,vuzixsdk,epsonmoveriosdk,metawearablestoolkit}.

 Host-powered glasses (XREAL and Rokid) provide Unity SDKs~\cite{xrealsdk,rokiduxr}.

 Aria and OCOSense  provide no public SDK, so researchers cannot build their owns applications.

\noindent{\small$\blacksquare$}~\textbf{Models.}  Two commonly used pretrained models allow researchers to reuse learned representations across compatible recordings. EgoVLP supplies video--language representations learned from egocentric data~\cite{lin2022egovlp,grauman2022ego4d}, while the general-purpose Whisper model supports speech transcription~\cite{radford2023whisper}.
 Table~\ref{tab:software-tools-models} lists models that are compatible with platform; Rokid Max, for example, lists Whisper but no camera-based model, since it does not have a camera. In addition to these two models, Aria also incorporates EgoBlur for anonymizing faces and license plates~\cite{raina2023egoblur} and Egocentric Voxel Lifting (EVL) for 3D detection and surface reconstruction~\cite{straub2024efm3d}.

\begin{tcolorbox}[colback=gray!10, colframe=gray!10, boxrule=0pt, arc=0pt, left=5pt, right=5pt, top=4pt, bottom=4pt, before skip=8pt, after skip=4pt]
\textbf{Takeaways.} Across software tools and models, no platform opens everything.
HoloLens 2 and Magic Leap 2 open both raw sensor streams and app development, so they suit studies that build and test interactive systems.
Aria opens rich recordings and the most models but gives no developer access, so it suits studies that collect and analyze egocentric data.
The software around a platform thus decides which questions researchers can answer with it.
\end{tcolorbox}

\subsection{Datasets and Benchmarks}

 This subsection summarizes publicly available 1) datasets, which release recordings for training and analysis, and 2) benchmarks (specify tasks, splits, metrics, and baselines for comparing methods) for smart-glasses research.
 Table~\ref{tab:smartglasses_dataset} summarizes ten datasets spanning low-level perception and higher-level reasoning:

\begin{table*}[t]
\centering
\caption{Summary of representative smart-glasses datasets and benchmarks.}
\label{tab:smartglasses_dataset}

\small

\setlength{\tabcolsep}{2pt}
\renewcommand{\arraystretch}{1.08}

\begin{tabularx}{\textwidth}{
>{\raggedright\arraybackslash\hspace{0pt}}p{3.15cm}
>{\raggedright\arraybackslash\hspace{0pt}}p{3.44cm}
>{\centering\arraybackslash}p{1.17cm}
>{\raggedright\arraybackslash\hspace{0pt}}p{2.75cm}
>{\raggedright\arraybackslash\hspace{0pt}}X
}

\toprule
\textbf{Dataset (Benchmark)} &
\textbf{Modality} &
\textbf{Subjects} &
\textbf{Duration/Samples} &
\textbf{Application Scenarios} \\
\midrule

\rowcolor{gray!10}
Aria Everyday Activities Dataset\newline\cite{lv2024ariaeverydayactivitiesdataset}
& RGB~video, mono~cameras, eye~tracking, IMUs, audio, MPS~outputs
& Multiple
& 143~recordings; 7.3~h; 5~indoor~locations
& Egocentric~perception, 4D~scene~understanding, neural~reconstruction, assistive~AI \\

\rowcolor{gray!10}
Aria Digital Twin\newline\textbf{(Public challenge)}\newline\cite{pan2023ariadigitaltwin}
& RGB~video, mono~cameras, IMUs
& N/A
& 200~sequences; 2~indoor~scenes; 398~object~instances
& Egocentric~3D~machine~perception, object~tracking, scene~reconstruction \\

\rowcolor{gray!10}
UCA-EHAR\newline\textbf{(Fixed split)}\newline\cite{novac2022uca}
& Accelerometer, gyroscope, barometer
& 20
& 8~activities; 64-sample~windows
& Human~activity~recognition~(HAR), elderly-care~monitoring, embedded~AI on~smart~glasses \\

\rowcolor{gray!10}
EgoMusic\newline\cite{ragano2025egomusic}
& RGB~video, SLAM~cameras, 360$^\circ$~video, eye~tracking, IMU, 7-channel~audio, reference~audio
& 5+
& 58~min~20~s; 4~sessions
& Egocentric~music~AAR, personalized~music~enhancement, audio-visual~correspondence \\

\rowcolor{gray!10}
CHiME-8 MMCSG\newline\textbf{(Public challenge)}\newline\cite{vzmolikova2024chime}
& RGB~video, IMU, 7-channel~audio
& 138
& Train/dev/eval:\newline 8.5~/~8.4~/~9.4~h
& Speaker-attributed~ASR, diarization, multimodal~conversation understanding, live~captioning \\

\rowcolor{gray!10}
SoniSpeech\newline\textbf{(Fixed split)}\newline\cite{sonispeech2026}
& Frontal~video, voiced~audio, ultrasound~echo~profiles
& Multiple
& 34~h; 18,000~utterances; voiced and silent modes
& Wearable~silent~speech~interfaces, open-vocabulary~speech~recognition, multimodal~speech~understanding \\

HoloAssist\newline\textbf{(Public challenge)}\newline\cite{wang2023holoassist}
& RGB~video, depth, IMU (HoloLens)
& Multiple
& Large-scale; physical~manipulation~tasks
& Egocentric~human~interaction, interactive~AI~assistance, procedural~task~understanding \\

Ego-Exo4D\newline\textbf{(Public challenge)}\newline\cite{grauman2024egoexo4d}
& Egocentric~video (Project~Aria), exocentric~video, IMU
& Multiple
& Large-scale; skilled~activities; multi-view
& Ego-exo~correspondence, skill~assessment, action~understanding, embodied~AI \\

SUPERGLASSES\newline\textbf{(Fixed split)}\newline\cite{jiang2026superglasses}
& Egocentric~images, questions, search~trajectories, reasoning~annotations
& N/A
& 2,422~QA~pairs; 14~domains; 8~query~categories
& VLM-agent~benchmarking, VQA, object-of-interest~identification, retrieval, multi-hop~reasoning \\

XRF V2\newline\textbf{(Fixed split)}\newline\cite{lan2025xrf}
& RGB-D-IR~video, IMUs (phone/\allowbreak watch/\allowbreak earbuds/\allowbreak glasses), Wi-Fi~CSI
& 16
& 853~sequences; 16~h~16~min; 30~actions; 3~scenes
& Temporal~action~localization~(TAL), action~summarization, smart-home~monitoring \\

\bottomrule
\end{tabularx}
\par\smallskip
\begin{minipage}{\textwidth}\small
Shaded rows target low-level perception, and unshaded rows target \textcolor{black}{higher-level reasoning}.
\textcolor{black}{A bold note in parentheses marks a dataset that also defines a benchmark. \textbf{(Public challenge)} means that the organizers keep the test labels, score submitted results, and rank them on a public leaderboard. \textbf{(Fixed split)} means that the test split and its labels are public, so researchers score their own results on it.}
\end{minipage}
\end{table*}

\noindent{\small$\blacksquare$}~\textbf{Low-level perception.}  These datasets use low-level sensor signals to recognize what the wearer sees, how the wearer moves, and what the wearer hears and says. UCA-EHAR supports inertial activity recognition with lightweight embedded inference, while Aria Everyday Activities combines synchronized vision, audio, motion, gaze, and perception outputs~\cite{novac2022uca,lv2024ariaeverydayactivitiesdataset}. Aria Digital Twin adds annotated spatial references for evaluating 3D perception~\cite{pan2023ariadigitaltwin}. These collections support different questions: whether a compact sensor set can recognize an activity, or how multiple observations relate to a scene. EgoMusic extends egocentric sensing to audio augmented reality, including attention, spatial audio, and personalized enhancement~\cite{ragano2025egomusic}.
 CHiME-8 MMCSG targets speaker-attributed conversational transcription; SoniSpeech pairs video, voiced audio, and ultrasound echoes to study voiced and silent speech~\cite{vzmolikova2024chime,sonispeech2026}.

\noindent{\small$\blacksquare$}~\textbf{\textcolor{black}{Higher-level reasoning.}}  These datasets build on perception to understand the wearer's tasks, skills, questions, and activities. HoloAssist records an instructor guiding a wearer through physical tasks, while Ego-Exo4D pairs egocentric and external views of skilled activities~\cite{wang2023holoassist,grauman2024egoexo4d}. Their complementary perspectives support procedural assistance, skill assessment, and action understanding. SUPERGLASSES evaluates egocentric questions and knowledge retrieval~\cite{jiang2026superglasses}, while XRF V2 combines wearable and environmental observations for activity understanding~\cite{lan2025xrf}. These tasks depend on relationships among observations and activities beyond identifying isolated objects or signals.

\noindent{\small$\blacksquare$}~\textbf{Benchmarks.} \textcolor{black}{Eight of the ten datasets in Table~\ref{tab:smartglasses_dataset} also define benchmarks, four for perception and four for reasoning.}
 Perception benchmarks use outcomes such as object detection, activity recognition, and speaker-attributed or silent-speech word error rates~\cite{pan2023ariadigitaltwin,novac2022uca,vzmolikova2024chime,sonispeech2026}.

 Reasoning benchmarks assess procedural mistakes and interventions, skills, egocentric questions, and action understanding~\cite{wang2023holoassist,grauman2024egoexo4d,jiang2026superglasses,lan2025xrf}.

 Some benchmarks also consider system costs: UCA-EHAR measures latency and energy on the glasses, CHiME-8 MMCSG evaluates algorithmic latency, and Ego-Exo4D estimates power consumption for recognition.

 These resources connect sensing to reasoning, but coverage of long-term, in-the-wild, privacy-sensitive, and socially situated use remains underrepresented. Future benchmarks should assess robustness, latency, energy, privacy risk, and real-world usability alongside accuracy.

\section{Sensing and Perception}
\label{sec:sensing}
\begingroup

Sensing and perception give smart glasses information about the wearer and the surrounding world. The head-worn position provides access to eye and facial movements, speech, bodily signals, and an egocentric view of ongoing activities. Following Section \mbox{\ref{sec:taxonomy-revised-layers}} and Figure \mbox{\ref{fig:taxonomy}}, we review six areas of research. Figure \mbox{\ref{fig:sensing}} maps these fields to representative sensing targets, inputs, and mechanisms. Within each field, we introduce the sensing inputs and explain how they support the capabilities being studied, followed by a discussion of the main design choices and limitations. Tables \mbox{\ref{tab:sensing-body-evidence}}--\mbox{\ref{tab:sensing-scene-evidence}} in the appendix collectively cover the studies discussed in this section, detailing their sensing modalities and setups, tasks and outputs, applications, contributions and mechanisms, evaluation results, and demonstrated reuse.

\begin{figure}[t]
    \centering
    \includegraphics[width=\linewidth]{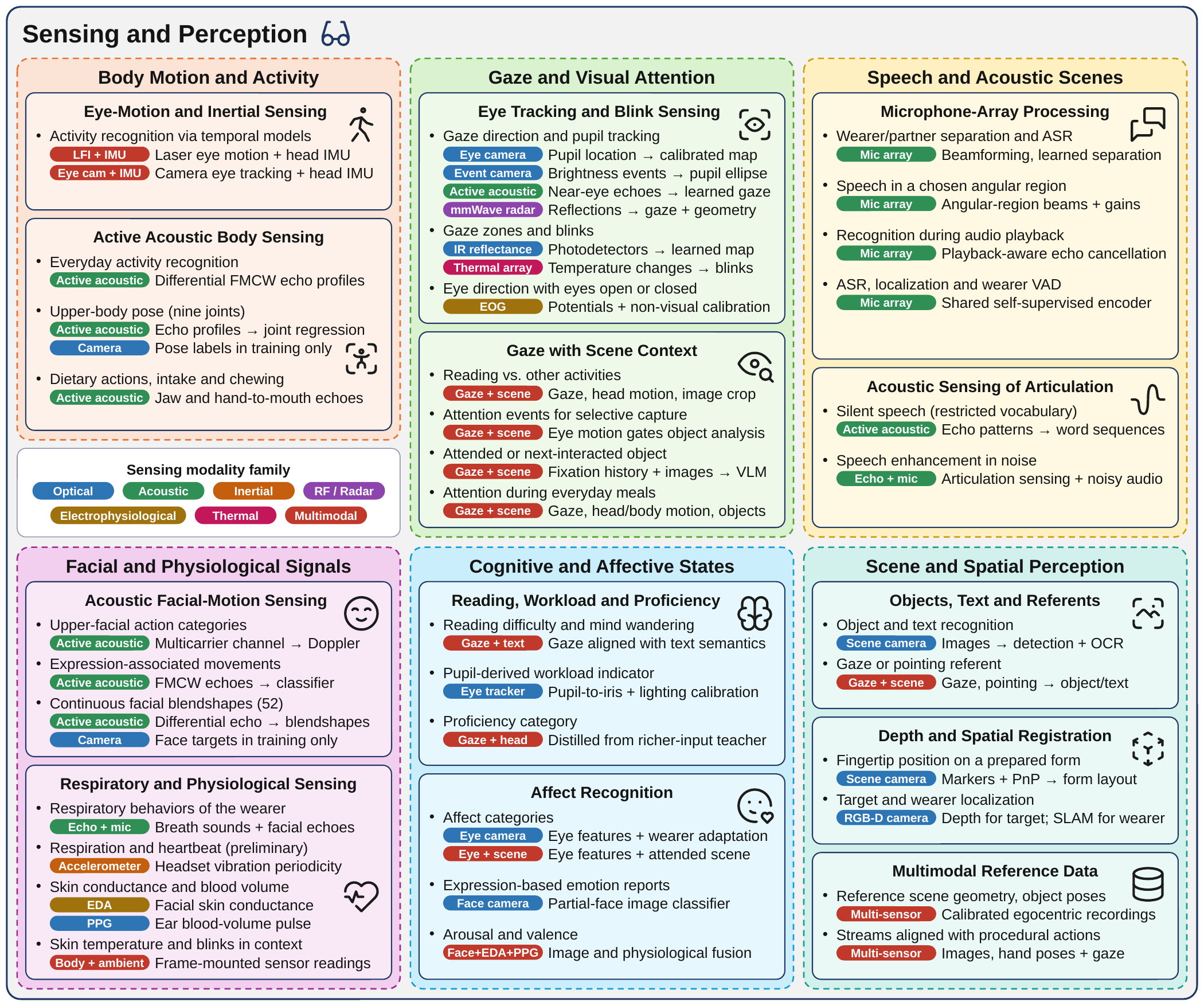}
    \caption{Overview of sensing and perception across six research fields, showing representative sensing targets, inputs, and processing or inference mechanisms. Colored labels indicate sensing modality families.}
    \Description{Six panels summarize body motion and activity, gaze and visual attention, speech and acoustic scenes, facial and physiological signals, cognitive and affective states, and scene and spatial perception. Each panel groups sensing targets with their inputs and mechanisms. Colored labels distinguish sensing modality families, and camera inputs used only for training are explicitly identified.}
    \label{fig:sensing}
\end{figure}

\subsection{Body Motion and Activity}
\label{sec:sensing-body}
Body sensing uses measurements taken at the head to recognize activities or estimate body movement. Existing approaches combine eye and head motion or observe the body through acoustic reflections (Table \mbox{\ref{tab:sensing-body-evidence}}).

\noindent{\small$\blacksquare$}~\textbf{Eye-motion and inertial sensing.}
Frame-mounted accelerometers and gyroscopes measure head movement, while eye sensors capture scanning and fixation. Laser feedback interferometry (LFI) measures interference between emitted and returning light as the eye moves; LFI--IMU activity recognition combines this signal with inertial measurements~\cite{aktaSemi2021cnn}. U-HAR instead combines camera-based eye tracking with an IMU~\cite{bernal2022uhar}. Temporal models identify activities such as walking, cycling, and reading from these complementary streams.

\noindent{\small$\blacksquare$}~\textbf{Active acoustic body sensing.}
Speakers on the glasses emit known frequency-modulated signals, and microphones record their reflections from the face and upper body. Processing these returns produces echo profiles that describe reflections across distance and time. ActSonic uses changes between successive profiles to suppress static reflections and recognize everyday activities~\cite{chen2024actsonic}. PoseSonic uses both original and differential profiles to estimate upper-body joint coordinates, with synchronized visual pose estimates supplying training targets~\cite{wang2023posesonic}. MunchSonic focuses the same sensing approach on jaw movement and hand-to-mouth actions, producing dietary-action labels that can be aggregated into intake and chewing summaries~\cite{munchsonic2024}.

\noindent\textbf{Discussion.}
The useful observation range depends on the task. Eye and head motion provide complementary cues for activities such as reading and locomotion, while acoustic reflections make movement beyond the head observable. MunchSonic illustrates this choice within acoustic sensing: its adaptation of ActSonic reduces gain and selects a nearer range to limit reflections from movements unrelated to eating~\cite{munchsonic2024,chen2024actsonic}. Output requirements also influence representation. For example, ActSonic emphasizes changing echoes for activity recognition, whereas PoseSonic retains original and differential profiles for continuous geometry~\cite{chen2024actsonic,wang2023posesonic}. These examples connect application needs to the information acquired and preserved by the sensing pipeline.

\subsection{Gaze and Visual Attention}
\label{sec:sensing-gaze}
Eye sensing estimates pupil movement, gaze direction, or blink events. Combining these measurements with time and scene content supports further inferences about visual attention (Table \mbox{\ref{tab:sensing-gaze-evidence}}).

\noindent{\small$\blacksquare$}~\textbf{Eye imaging and event-based tracking.}
Inward-facing cameras locate the pupil and estimate gaze through calibrated geometric or learned mappings~\cite{biosensor_glasses_2024}. Event cameras instead report pixel-level brightness changes around moving eye features. Event-based tracking uses predicted movement to trigger pupil detection and ellipse fitting~\cite{zhao2024eventbased}; EyeTrAES adapts event accumulation to slow and rapid eye movement~\cite{sen2024eyetraes}.

\noindent{\small$\blacksquare$}~\textbf{Infrared reflectance and thermal sensing.}
Infrared emitters and lens-edge photodetectors measure how reflected light is distributed as the eye and eyelid move. Learned mappings convert these sparse measurements into eye-related outputs without constructing an image~\cite{li2025lowpower,chen2025hierarchical}. Passive thermal sensing observes temperature patterns instead. ThermalEye uses a low-resolution thermal array and temporal processing to detect blink-related changes around the eye~\cite{thermaleye_2025}.

\noindent{\small$\blacksquare$}~\textbf{Acoustic, radar, and electrical sensing.}
GazeTrak estimates gaze from acoustic reflections near the eyes, while mmET uses millimeter-wave reflections to jointly estimate gaze and eye geometry~\cite{li2024gazetrak,lu2025mmet}. Eye movement changes the geometry observed by these sensors, providing patterns from which direction can be learned. Electro-oculography (EOG) measures electrical potential differences associated with eye rotation. Tactile and auditory targets can establish a calibration mapping without visual fixation, allowing directional estimation with closed eyes in laboratory experiments~\cite{wei2025eog}.

\noindent{\small$\blacksquare$}~\textbf{Gaze combined with scene context.}
Eye measurements can be associated with objects, text, and activities in the outward view. Reading recognition combines gaze patterns, head motion, and small image regions to distinguish reading from other activities~\cite{min2025reading}. MemX relates eye movement to scene objects to identify attention events and select moments for recording~\cite{liu2021memx}. Fixation history has also been used to infer attended or next-interacted objects, while wearable gaze combined with object recognition supports the study of attention during everyday meals~\cite{li2025eyegaze,ladouce2025dinner}. These methods turn eye measurements into information about what the wearer is attending to.

\noindent\textbf{Discussion.}
The appropriate sensing route depends first on the required output: blink detection, gaze-zone classification, and precise gaze estimation are different capabilities. For continuous tracking, maintaining the signal-to-gaze mapping across wearing sessions is a recurring challenge. GazeTrak, for example, uses wearer adaptation and short session calibration~\cite{li2024gazetrak}; mmET uses repeated-remounting training and eye-geometry estimation to reduce recalibration after remounting, while retaining new-user adaptation~\cite{lu2025mmet}. These differences connect sensing performance to frame placement and calibration resources. Combining gaze with scene content introduces another requirement: the objects and events needed by the task must be identifiable and aligned with the eye measurements.

\subsection{Speech and Acoustic Scenes}
\label{sec:sensing-speech}
Acoustic perception captures speech and surrounding sounds, selects relevant sources, and recovers information from articulatory movement. The main inputs are passive microphone recordings and active acoustic measurements near the face (Table \mbox{\ref{tab:sensing-speech-evidence}}).

\noindent{\small$\blacksquare$}~\textbf{Microphone arrays and spatial processing.}
Differences in when and how strongly sound reaches microphones distributed around the frame provide spatial information. Beamforming and learned separation use these differences to retain desired sources from a mixture. Directional speech processing distinguishes wearer and partner speech and combines separation with automatic speech recognition \mbox{\cite{zhang2025directional}}. FoVNet enhances sources within a configurable angular region, allowing several talkers in that region to be retained \mbox{\cite{xu2024fovnet}}. During simultaneous playback and recording, the glasses' own audio also enters the microphones. Full-duplex directional recognition uses the known playback signal for acoustic echo cancellation and trains the recognizer to tolerate the remaining echo \mbox{\cite{peng2025fullduplex}}.

\noindent{\small$\blacksquare$}~\textbf{Acoustic sensing of articulation.}
Near-face acoustic echoes change as the wearer mouths words or speaks. EchoSpeech learns temporal patterns in differential echoes to decode silent articulation into word sequences within a restricted vocabulary, using personal training data~\cite{wang2023echospeech}. WearSE combines articulation-related sensing with noisy audio to reconstruct an enhanced speech waveform~\cite{sun2025wearse}. Its processing suppresses unrelated movement interference before using the complementary channels for enhancement, and a causal implementation supports streaming operation. These approaches provide speech information when audible input is absent or degraded.

\noindent{\small$\blacksquare$}~\textbf{Shared multichannel representations.}
Self-supervised learning uses multichannel recordings to develop representations that can be adapted to several audio tasks. M-BEST-RQ applies this approach to conversational recognition, source localization, and wearer voice-activity detection, including evaluation on a different microphone-array dataset~\cite{shon2025mbestrq}.

\noindent\textbf{Discussion.}
Audio quality and recognition quality do not always improve together. Directional source separation illustrates this distinction: providing only separated signals to the recognizer improves wearer recognition but degrades partner recognition; combining separated and beamformed signals and training them jointly improves overall recognition while changing the balance between speakers~\cite{zhang2025directional}. Full-duplex experiments likewise show different rankings under echo-suppression and word-error measures~\cite{peng2025fullduplex}. The appropriate comparison therefore depends on the output needed by the interaction, including whose speech should be retained. Shared pretraining can support several outputs, but task-specific adaptation remains important to their performance~\cite{shon2025mbestrq}.

\subsection{Facial and Physiological Signals}
\label{sec:sensing-facial}
The proximity of glasses to the face supports measurements of facial movement, respiratory behavior, bodily rhythms, and autonomic responses. These signals can provide input for expression-based interaction, physiological observation, and subsequent user-state inference (Table \mbox{\ref{tab:sensing-facial-evidence}}).

\noindent{\small$\blacksquare$}~\textbf{Acoustic facial-motion sensing.}
Small facial movements change nearby acoustic propagation paths. Upper-facial action recognition extracts Doppler features from multicarrier channel measurements, while EchoExpress classifies expression-associated movements using swept-frequency echoes~\cite{li2021acoustic,zhang2025echoexpress}. EyeEcho uses differential echoes to estimate 52 facial blendshape parameters for continuous avatar animation~\cite{li2024eyeecho}. Synchronized camera-derived targets supervise its model during training, while acoustic signals supply the runtime input.

\noindent{\small$\blacksquare$}~\textbf{Respiratory and physiological measurements.}
EchoBreath combines passive respiratory sounds with active echoes of nearby facial changes to identify respiratory behaviors and attribute them to the wearer~\cite{echobreath_2025}. A preliminary headset study uses small accelerometer variations to estimate respiration and heartbeat periodicity in short, predominantly static trials~\cite{vitalmotion2022}. Other channels include facial electrodermal activity, which measures skin-conductance changes, and ear photoplethysmography, which records optical changes related to blood-volume variation~\cite{kim2021emotion}. AirSpecs measures skin temperature and blinks alongside environmental conditions to support awareness of bodily experiences and their context~\cite{ramsamy2025sensors}.

\noindent\textbf{Discussion.}
Different modalities can resolve different parts of the same sensing problem. In EchoBreath, for instance, sound helps identify respiratory behavior, while local echoes help determine whether it comes from the wearer~\cite{echobreath_2025}. At the same time, speech and facial movement can introduce signals that overlap with the target behavior. The required representation also varies: expression categories support discrete events, whereas continuous facial parameters preserve motion for animation. Interpreting either as evidence of an internal emotional state requires the additional modeling and reference labels considered next.

\subsection{Cognitive and Affective States}
\label{sec:sensing-states}
This field uses observable behavior and context to estimate internal user states such as reading difficulty, mind wandering, affect, and proficiency. Its mechanisms center on combining and interpreting sensed information, with user state definitions with reference labels to build systems and models for estimating states (Table \mbox{\ref{tab:sensing-states-evidence}}).

\noindent{\small$\blacksquare$}~\textbf{Gaze, text, and reading-state inference.}
CASES aligns temporal gaze with text semantics to estimate reading difficulty and mind wandering and locate word explanations, sentence simplification, or rereading reminders~\cite{huang2023cases}. A pupillometry study stabilizes AR pupil measurements through pupil-to-iris correction and illumination calibration, and separately relates pupil-derived workload to wayfinding performance in VR~\cite{xu2026cognitivewayfinding}.

\noindent{\small$\blacksquare$}~\textbf{Eye, facial, and physiological cues for affect.}
EMO extracts features from eye images and adapts classification to the wearer~\cite{guo2020emo}. EMOShip adds attended scene content to eye features, supplying visual context for affective inference~\cite{smartglasses_dream_2022}. EmoGlass learns expression categories from partial-face imagery and pairs emotion reports with contextual diaries for reflection~\cite{lim2022emoglass}. Multimodal work further combines facial imagery with electrodermal activity and photoplethysmography, using visual and physiological features to estimate affective categories~\cite{kim2021emotion}.

\noindent{\small$\blacksquare$}~\textbf{Gaze and head motion for proficiency assessment.}
SkillSight transfers task information from a teacher with richer visual or motion inputs to a student that estimates proficiency from gaze and head motion~\cite{li2026skillsight}. Knowledge distillation and action supervision help the student learn relationships between observed behavior and expert proficiency labels.

\noindent\textbf{Discussion.}
Context helps resolve ambiguity in behavioral signals. For example, CASES improves on gaze-only and text-only alternatives~\cite{huang2023cases}, while fusing eye and scene features improves average affect classification in EMOShip~\cite{smartglasses_dream_2022}. Such gains remain tied to the states and labels studied. These systems also show how existing capabilities can be combined. EMOShip adopts and adapts EMO's eye-feature extraction method and MemX's attention-triggering approach to activate more expensive scene processing when needed~\cite{smartglasses_dream_2022,guo2020emo,liu2021memx}. This provides a concrete instance of cross-study reuse supporting a richer inference task.

\subsection{Scene and Spatial Perception}
\label{sec:sensing-scene}
Outward-facing sensing identifies objects and text and estimates their spatial relationships to the wearer. These representations provide the entities and locations that interaction and assistance can refer to (Table \mbox{\ref{tab:sensing-scene-evidence}}).

\noindent{\small$\blacksquare$}~\textbf{Scene imagery, object recognition, and text extraction.}
Egocentric images provide input for object detection and optical character recognition. GazePointAR organizes detected objects and text into a hierarchy, then projects gaze and pointing coordinates into the same image to identify the user's referent~\cite{zhao2024gazepointar}. This combines scene recognition with the tracking interfaces of an AR platform, associating visual content with user input.

\noindent{\small$\blacksquare$}~\textbf{Depth and spatial registration.}
Depth sensing and pose estimation relate recognized entities to locations where the user can act. IFFA registers fingertip positions to a prepared form layout through marker-based pose estimation, associating pointing locations with text and writing fields~\cite{formfill_2025}. For navigation, visual--auditory E-Glasses combine detected targets with RGB-D depth and localize the wearer within a previously constructed map~\cite{multisensory_nav_2023}. These pipelines establish spatial correspondence at the scale of a page or a room.

\noindent{\small$\blacksquare$}~\textbf{Multimodal correspondence and reference data.}
Aligned research data support the development and evaluation of these perception capabilities. Aria Digital Twin supplies calibrated egocentric recordings with registered scene geometry and object poses \mbox{\cite{pan2023ariadigitaltwin}}. HoloAssist aligns images, hand poses, gaze, and other streams with procedural actions and human interventions \mbox{\cite{wang2023holoassist}}. Such resources make spatial and temporal relationships available as reference information for evaluating what a system can recover.

\noindent\textbf{Discussion.}
Scene representations differ in the questions and actions they can support. Generic object categories may be insufficient for queries about a particular product~\cite{zhao2024gazepointar}, while accurate registration to a prepared form or map supports precise tasks within those environments~\cite{formfill_2025,multisensory_nav_2023}. Temporal coverage creates another boundary. GazePointAR illustrates this limitation: its single captured moment cannot resolve references to objects seen earlier~\cite{zhao2024gazepointar}. Connecting current observations with a history of encountered entities would extend the information available to assistance.

\begin{table*}[t]
\centering
\small
\setlength{\tabcolsep}{4pt}
\renewcommand{\arraystretch}{1.14}
\caption{Recurring approaches to building sensing and perception capabilities. The examples show related design choices; documented reuse is identified in the text and appendix.}
\label{tab:sensing-shared-approaches}
\begin{tabular}{@{}>{\raggedright\arraybackslash}p{\dimexpr.19\textwidth-2\tabcolsep\relax}>{\raggedright\arraybackslash}p{\dimexpr.30\textwidth-2\tabcolsep\relax}>{\raggedright\arraybackslash}p{\dimexpr.31\textwidth-2\tabcolsep\relax}>{\raggedright\arraybackslash}p{\dimexpr.20\textwidth\relax}@{}}
\toprule
\rowcolor{gray!15}
\textbf{Shared approach} & \textbf{What it enables} & \textbf{Central design choice} & \textbf{Examples} \\
\midrule
\textbf{Task-directed observation} & A sensing principle supports different activity, pose, or facial outputs. & Choose the observation range and representation to preserve task-relevant changes. & ActSonic, MunchSonic, PoseSonic, EyeEcho~\cite{chen2024actsonic,munchsonic2024,wang2023posesonic,li2024eyeecho}. \\
\addlinespace
\rowcolor{gray!5}
\textbf{Contextual interpretation} & Measurements become reading states, attended entities, or wearer-attributed events. & Align signals with the content, coordinates, or source information needed to interpret them. & CASES, GazePointAR, EchoBreath~\cite{huang2023cases,zhao2024gazepointar,echobreath_2025}. \\
\addlinespace
\textbf{Rich training, selective runtime inputs} & Richer observations supervise capabilities that use fewer inputs during operation. & Trade runtime sensing demands against training data, supervision, and adaptation requirements. & PoseSonic, mmET, SkillSight~\cite{wang2023posesonic,lu2025mmet,li2026skillsight}. \\
\addlinespace
\rowcolor{gray!5}
\textbf{Selective temporal processing} & Eye motion, attention, or activity determines which information is processed. & Balance processing cost against temporal detail and the coverage of relevant events. & EyeTrAES, MemX, EMOShip, EchoGuide~\cite{sen2024eyetraes,liu2021memx,smartglasses_dream_2022,echoguide2024}. \\
\bottomrule
\end{tabular}
\end{table*}

\begin{tcolorbox}[colback=gray!10, colframe=gray!10, boxrule=0pt, arc=0pt, left=5pt, right=5pt, top=4pt, bottom=4pt, before skip=8pt, after skip=4pt]
\noindent\textbf{Shared Mechanisms and Reuse.} Several approaches recur across the six fields, as summarized in Table~\ref{tab:sensing-shared-approaches}: selecting task-relevant observations, interpreting signals in context, using richer training resources, and processing information selectively over time. These common approaches also create opportunities for reuse. EchoGuide, for example, runs ActSonic's pretrained model to select intervals from recorded video for description and question answering~\cite{echoguide2024}. This offline pipeline demonstrates how an activity-recognition output can direct later semantic analysis. Such examples give concrete form to the layered perspective: sensing and perception supply information that interaction and assistance can use, subject to the conditions under which it was obtained.
\end{tcolorbox}

\endgroup

\section{Interaction and Assistance}
\label{sec:interaction}
\begingroup
Interaction and assistance concern how users act through smart glasses and how systems respond. Eye movements, speech, gestures, and scene information can supply inputs, but an interaction also needs ways to specify a goal, confirm a choice, and understand the response. Following Section \mbox{\ref{sec:taxonomy-revised-layers}}, we organize this work into five research fields.
Figure \mbox{\ref{fig:interaction-assistance}} maps these fields to representative functions, modalities, contextual information, and supporting mechanisms. Within each field, we explain the inputs and mechanisms that support these functions, followed by a discussion of their design implications. Tables \mbox{\ref{tab:interaction-input-evidence}}--\mbox{\ref{tab:interaction-timing-evidence}} in the appendix provide study-level tasks, outputs, and applications alongside inputs, mechanisms, control and feedback, evaluation findings, and demonstrated reuse.

\begin{figure}[t]
    \centering
    \includegraphics[width=\linewidth]{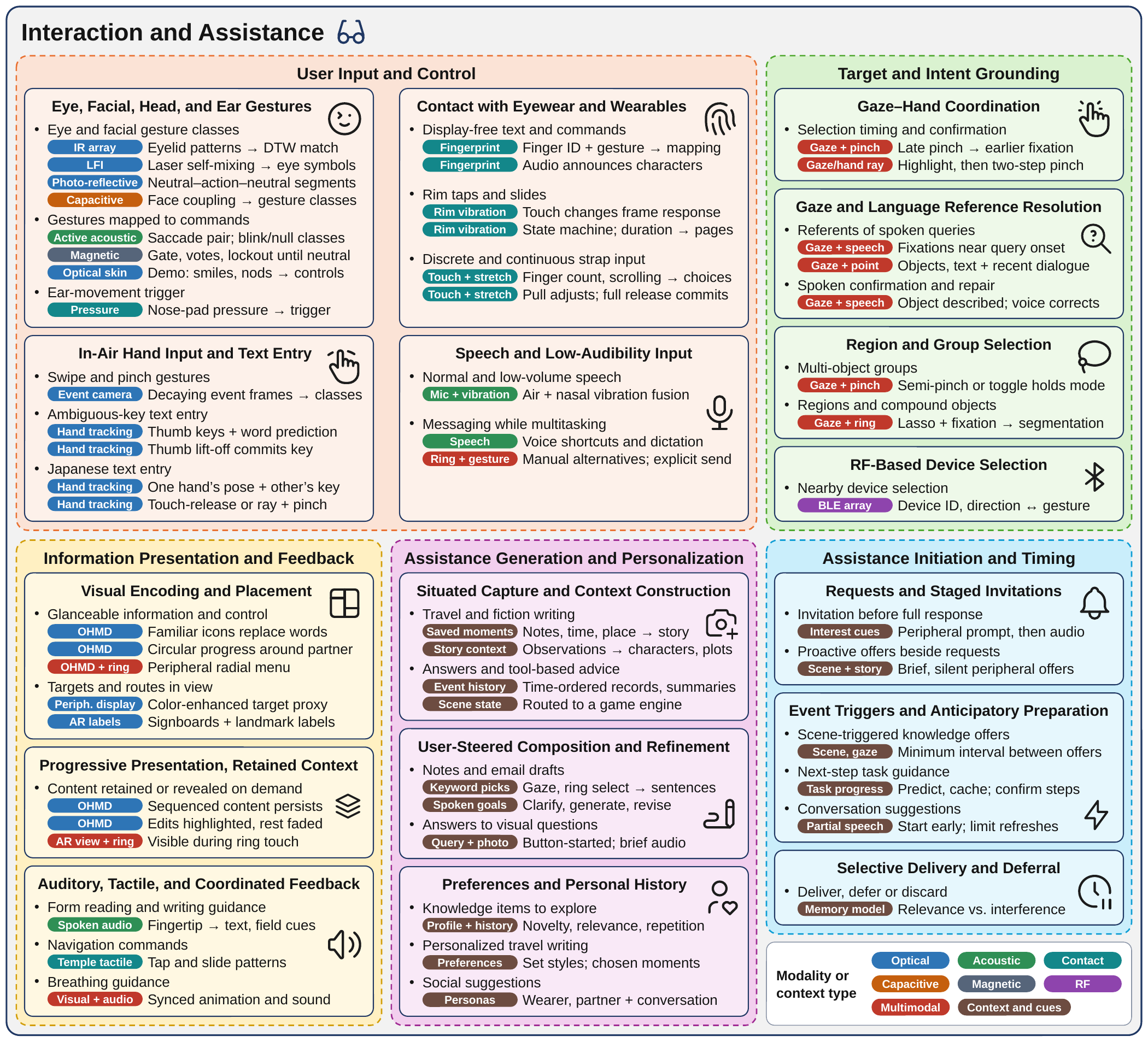}
    \caption{Overview of interaction and assistance across five research fields, showing representative functions and supporting mechanisms. Colored labels identify input and output modalities or contextual information.}
    \Description{Five panels summarize user input and control, target and intent grounding, information presentation and feedback, assistance generation and personalization, and assistance initiation and timing. Each panel groups representative functions with the modalities, contextual information, and mechanisms that support them. Colored labels distinguish optical, acoustic, contact, capacitive, magnetic, radio-frequency, and multimodal approaches, as well as context and cues.}
    \label{fig:interaction-assistance}
\end{figure}

\subsection{User Input and Control}
\label{sec:interaction-input}
Input methods turn deliberate actions into commands, selections, text, or continuous control. The glasses frame provides access to small eye and facial movements and can itself become an input surface. Hand tracking, speech, and companion devices extend the available vocabulary beyond the frame.

\noindent{\small$\blacksquare$}~\textbf{Eye, facial, head, and ear gestures.}
Optical sensors near the face observe tissue displacement or eye movement. Infrared arrays and photo-reflective sensors recognize temporal eye and facial patterns, while laser feedback interferometry extracts ocular motion from returning light~\cite{kaneko2021irarray,McDuff2020facecommands,peng2024fewshot,aktas2021lfigaze}. OCOsense uses optical skin-displacement measurements in demonstrations of gesture-based control~\cite{schneegass2023ocosense}. Other approaches exploit different consequences of movement: CapGlasses measures capacitive coupling to the face, MagFace detects deformation of magnetic inserts within frame pads, and EarWiggl'In senses the pressure that ear movement transmits to the nose pads~\cite{yang2021capglasses,zhang2026magface,kim2025earwigglin}. These channels make small actions observable; the interaction design determines how they become commands. EyeGesener, for example, uses coded acoustic reflections to recognize an outward-and-return pair of saccades, with blink and non-gesture classes helping distinguish the command vocabulary~\cite{li2024eyegesener}. MagFace adds gesture gating, temporal voting, and a lockout until the wearer returns to a neutral state, reducing repeated command firing.

\noindent{\small$\blacksquare$}~\textbf{Contact with eyewear and wearable surfaces.}
Contact provides a physical reference for locating and repeating an action. FingerGlass reads fingerprint images at the temple, using ridge-pattern motion to distinguish gestures and enrolled finger identity to expand their command mappings~\cite{ye2025fingerglass}. Rim-based vibration sensing actively excites the frame and measures how touch changes its response~\cite{sun2022rimsense,xu2024rimsense}. RimSense converts the resulting predictions into bounded interaction events through a state machine; short and long taps map to different commands, while slide duration determines how many pages to turn. StretchAR combines textile touch sensing with resistive stretch sensing on a wearable strap, using contact and deformation for discrete and continuous input~\cite{zhang2022stretchar}. Across these designs, identity, location, motion, and duration allow a small surface to support more than a single button-like action.

\noindent{\small$\blacksquare$}~\textbf{In-air hand input and text entry.}
Event cameras observe brightness changes caused by moving fingers, while conventional camera-based trackers estimate hand geometry. Helios uses event representations for short gestures such as swipes and pinches~\cite{xu2024helios}. Text-entry systems build a larger vocabulary on top of tracked movements. ThumbAir places a small set of reachable virtual keys around the thumbs and combines ambiguous key sequences with word prediction; releasing the thumb from the virtual surface commits the selection~\cite{gupta2023thumbair}. FlickPose instead combines one hand's pose with the other hand's key selection, using touch-and-release or ray pointing with pinch confirmation to enter Japanese characters~\cite{yuasa2025flickpose}. The contribution in these systems includes the arrangement of targets, division of roles between the hands, and confirmation rule, alongside the tracking that makes the actions observable.

\noindent{\small$\blacksquare$}~\textbf{Speech and low-audibility input.}
Speech can express commands and longer text without a large gesture vocabulary. NasoVoce combines an air microphone with a nasal vibration sensor to enhance normal and low-volume speech for listening or recognition~\cite{wu2026nasovoce}. The two channels provide complementary information under noise, although whispering weakens the vibration component. At the interface level, GlassMessaging combines speech commands and dictation with ring and mid-air input, allowing users to choose how they select contacts, compose, and send messages~\cite{li2023glassmessaging}. These examples address different parts of voice interaction: acquiring usable speech and organizing the operations performed with it.

\noindent\textbf{Discussion.}
A useful input method combines a repeatable action with segmentation, command mapping, and feedback. Small vocabularies suit immediate commands; text entry expands them through combinations and prediction, while adding learning and correction demands. The surrounding activity also matters. Facial and eye gestures free the hands but overlap with natural behavior; contact supplies a tangible reference but requires reaching a surface; in-air actions depend on visibility and tracking. Studies of subtle interaction further show that preferences depend on what the hands and attention are already occupied with~\cite{orlosky2021ubiquitous}. Table~\ref{tab:interaction-input-evidence} separates gesture-recognition experiments from evaluated interfaces so that these design choices can be related to the evidence supporting them.

\subsection{Target and Intent Grounding}
\label{sec:interaction-grounding}
Grounding uses spatial context, timing, language, and confirmation to connect an input to the intended object, region, group, or device.

\noindent{\small$\blacksquare$}~\textbf{Gaze and hand-based target selection.}
Gaze can locate a target while a hand action confirms it, but the two signals need not arrive at the same moment. GazeHandSync identifies late pinch triggers and associates them with an earlier fixation when the eyes have already moved on~\cite{park2025gazehandsync}. Selecting several objects introduces another requirement: the system must preserve a group while accepting individual choices. PinchCatcher uses a semi-pinch to maintain selection mode, whereas Eyes on Many examines sustained gestures and explicit toggles for that role~\cite{10.1145/3706598.3713530,bashar2026eyes}. Their mechanisms separate target location, membership changes, and completion of the group. In robot-assisted assembly, target highlighting and two-step pinch confirmation similarly turn a gaze or hand ray into a selected physical part for the robot to retrieve~\cite{robot_assembly_2024}.

\noindent{\small$\blacksquare$}~\textbf{Gaze and language-based reference resolution.}
Language provides semantic information that a location alone may leave ambiguous. G-VOILA selects gaze-associated scene evidence around a spoken query to interpret what the wearer is asking about~\cite{zhang2024gvoila}. GazePointAR combines gaze and pointing with detected objects, text, and recent dialogue to resolve references such as ``this'' or ``that''~\cite{zhao2024gazepointar}. Feedback also makes an interpretation repairable. In Gazeify Then Voiceify, the system describes the gaze-selected object aloud, allowing the user to hear a mismatch and correct it through speech~\cite{li2026gazeifyvoiceify}. The auditory description supplies a confirmation opportunity when a visual selection indicator is unavailable.

\noindent{\small$\blacksquare$}~\textbf{Gaze and gesture-based region selection.}
Some intentions concern a region or a group rather than one point. Squiggle lets users draw an invisible lasso in the air while gaze helps place the traced region within the scene~\cite{yu2025squiggle}. Its temporal combination of fixation and hand trajectory supplies prompts for image segmentation, allowing the selection to encompass compound objects or several objects together. The gesture conveys extent, complementing the location supplied by gaze.

\noindent{\small$\blacksquare$}~\textbf{RF-based device selection.}
An electronic device can also be identified through its radio signal.
BLEselect associates Bluetooth identities with estimated directions and matches them to head or hand gestures~\cite{chen2023bleselect}. Its candidate set comes from communicating devices, allowing device selection without visual object recognition.

\noindent\textbf{Discussion.}
Grounding mechanisms address different ambiguities: a temporal offset between looking and confirming, several objects within one region, an unclear linguistic reference, or the identity of a nearby device. Extra input is useful when it resolves the ambiguity that matters to the task. For repeated selections, the stability of the selection mode and the cost of recovering a lost group can matter as much as pointing speed. For example, Eyes on Many uses these differences to help explain why its results differ from those reported by PinchCatcher~\cite{bashar2026eyes,10.1145/3706598.3713530}. Table~\ref{tab:interaction-grounding-evidence} records the target sets and confirmation rules alongside the results. These distinctions help identify which grounding mechanisms can support a new task and which assumptions about its targets and feedback must also be carried over.

\subsection{Information Presentation and Feedback}
\label{sec:interaction-presentation}
Presentation determines how users receive information while remaining engaged with their surroundings. Feedback also makes a system's interpretation visible or audible, allowing users to confirm, correct, or act on it. Relevant mechanisms concern the encoding, placement, persistence, and sensory channel of the response.

\noindent{\small$\blacksquare$}~\textbf{Visual encoding and placement.}
Compact graphical encodings can convey recurring information with little reading. Research on icon-augmented notifications shows how familiar pictograms can replace several words while retaining numerical or textual detail~\cite{smith2023icons}. For progress information during conversation, circular graphics placed around the central visual target allow users to estimate change while maintaining attention on their partner~\cite{tu2022paracentral}. ParaGlassMenu extends this idea to a radial menu controlled by a ring, pairing peripheral options with selection and continuous-control feedback~\cite{tu2023paraglassmenu}. Placement can also make a target easier to inspect: PeriphAR presents a visual proxy of the gaze-targeted object in a small peripheral display, using color processing to distinguish it from its surroundings~\cite{10.1145/3772318.3791637}. VisiMark combines overview signboards with labels at physical landmarks, providing both a preview of a route and information about nearby entities~\cite{visimark_2025}.

\noindent{\small$\blacksquare$}~\textbf{Progressive presentation and retained context.}
Users who switch attention away from a display need a way to resume. LSVP introduces learning content sequentially while preserving previously presented material, reducing visual search and the need to remember content that has disappeared~\cite{10.1145/3448118}. Its design also adapts backgrounds, contrast, and typography to optical see-through viewing. Different tasks need different amounts of retained information. GlassMail highlights edits while fading unchanged text, making a revised draft easier to inspect without removing its context~\cite{li2024glassmail}. GazeNoter reveals suggestions when the user touches a ring and hides them when contact ends, leaving the user in control of when the visual material occupies the view~\cite{li2025gazenoter}.

\noindent{\small$\blacksquare$}~\textbf{Auditory, tactile, and coordinated feedback.}
Responses can be delivered through channels that leave the visual scene available. IFFA maps a tracked fingertip to text and fields on a prepared form, then provides spoken reading and directional guidance as the user moves~\cite{formfill_2025}. LLM-Glasses encodes navigation commands as temple-mounted tapping and sliding patterns, with separate evaluations of tactile recognition and path following~\cite{llmglasses_2026}. Multiple channels can serve different moments of a task. In Mindful Moments, synchronized animation and audio guide breathing; field reports describe using the visuals to establish or recover the rhythm and the audio to maintain it while looking at the surroundings~\cite{10.1145/3563657.3596030}.

\noindent\textbf{Discussion.} Effective presentation depends on the information users need and the attention available to interpret it. Pictograms, for example, offer advantages that depend on familiarity and how much meaning they encode; precise unfamiliar content may still require text \mbox{\cite{smith2023icons}}. Persistence supports resumption but also occupies the display, while a brief cue may be sufficient for a direction or a progress update. Feedback must likewise match the action being supported. IFFA illustrates this requirement: its users could locate a writing field yet still struggle to remain within its boundaries, indicating a need for feedback during the action as well as at its start \mbox{\cite{formfill_2025}}. Table \mbox{\ref{tab:interaction-presentation-evidence}} relates these design choices to comprehension, interaction, and task-use evidence. This field provides a direct connection between perceptual capabilities and whether their outputs can guide useful action.

\subsection{Assistance Generation and Personalization}
\label{sec:interaction-generation}
Assistance systems organize observations and user-provided information into answers, drafts, explanations, or guidance. The resulting content depends on which context is retained, how the user's purpose is specified, and how the output can be refined.

\noindent{\small$\blacksquare$}~\textbf{Situated capture and context construction.}
A useful context can extend beyond the current camera image. PANDALens accumulates selected travel moments with descriptions, comments, and time and location information, then uses this material to support a narrative~\cite{qian2024pandalens}. CRAFT maintains an evolving fiction context so that environmental observations can become characters, settings, or plot ideas consistent with the writer's project~\cite{bai2026craft}. Egocentric Co-Pilot organizes egocentric descriptions and speech into time-ordered records and query-relevant summaries, supporting questions about recent or earlier events~\cite{wang2026egopilot}. It can also route a structured observation to a specialized tool, such as a game engine, before presenting advice. These mechanisms determine what information is available to generation and how it remains connected across time.

\noindent{\small$\blacksquare$}~\textbf{User-steered composition and refinement.}
Lightweight choices can constrain a broad generative space. GazeNoter extracts candidate keywords from speech and lets users select keywords and sentence suggestions through gaze and ring confirmation~\cite{li2025gazenoter}. The selected items specify what should enter a note. GlassMail instead elicits requirements through speech, offers optional clarification, and supports inspection and revision of the generated email~\cite{li2024glassmail}. An explicit question can provide a similarly focused goal: LifeInsight combines a button-initiated spoken request with a camera snapshot to produce a concise auditory answer~\cite{zhao2025lifeinsight}. Across these interfaces, selection and correction supply information about the user's intended result that the observed scene alone cannot provide.

\noindent{\small$\blacksquare$}~\textbf{Preferences and personal history.}
Personalization changes which content is selected and how it is expressed. AiGet combines a user profile with scene interpretation, interaction history, and rules for novelty, relevance, and repetition when selecting knowledge items~\cite{li2025aiget}. PANDALens collects writing and questioning preferences and allows the user to choose moments for the final narrative~\cite{qian2024pandalens}. SocialMind retains wearer and partner personas and combines them with conversational context to produce short social suggestions~\cite{yang2025socialmind}. These systems illustrate distinct forms of personalization: explicitly configured preferences, choices made during use, and information accumulated from prior interactions.

\noindent\textbf{Discussion.}
The central design problem is to preserve the information that makes assistance relevant while keeping the result open to user correction. A scene description, a set of selected keywords, and a longer event history support different kinds of help. AiGet illustrates the importance of selecting relevant context: its comparison suggests that adding a profile alone is insufficient, and that contextual prioritization and filtering affect the value of the selected content~\cite{li2025aiget}. GazeNoter and GlassMail illustrate the complementary role of user correction, allowing users to shape candidate outputs before accepting them~\cite{li2025gazenoter,li2024glassmail}. Table~\ref{tab:interaction-generation-evidence} distinguishes these mechanisms and the evidence obtained from benchmarks, controlled interaction studies, and situated use. Producing a candidate also leaves a separate question: when, and with whose agreement, should it be presented?

\subsection{Assistance Initiation and Timing}
\label{sec:interaction-timing}
Assistance timing governs when a system prepares content, invites attention, delivers a response, or waits, combining user requests with inferred opportunities.

\noindent{\small$\blacksquare$}~\textbf{Requests and staged invitations.}
An explicit request gives the system a clear starting point.
PANDALens stages an inferred opportunity as a peripheral prompt, enabling audio after the user indicates interest; wearers can also initiate capture or dismiss suggestions~\cite{qian2024pandalens}. CRAFT similarly separates peripheral proactive suggestions from requested transformations~\cite{bai2026craft}. This lets users assess the invitation before engaging with the full response.

\noindent{\small$\blacksquare$}~\textbf{Event triggers and anticipatory preparation.}
Triggers can depend on a scene change, sustained attention, or progress in an ongoing task. AiGet uses scene changes, gaze fixation, and explicit queries, with an interval constraint between scene-triggered suggestions~\cite{li2025aiget}. Satori uses a task plan and observations of progress to predict and cache upcoming guidance, then checks completion and seeks confirmation before advancing~\cite{shen2025satori}. SocialMind begins preparing suggestions from partial conversational utterances and limits updates when successive inputs are too similar~\cite{yang2025socialmind}. Anticipation can reduce waiting, while confirmation and update rules determine when a prepared response becomes visible.

\noindent{\small$\blacksquare$}~\textbf{Selective delivery and deferral.}
ProMemAssist explicitly separates candidate generation from the decision to deliver it~\cite{zhou2025prememassist}. Camera and speech observations form an operational model of recent information, organized into memory items and episodes. The system evaluates a candidate message against its estimated relevance and importance and the interference it could introduce, then delivers, defers, or discards it. The model provides a way to reason about assistance timing; its memory states are computational estimates rather than direct measurements of the wearer's internal state.

\noindent\textbf{Discussion.}
These approaches address different timing problems: recognizing an opportunity, preparing an answer before it is needed, and deciding whether to interrupt. Their evaluation therefore needs to consider delivery decisions and user response alongside the quality of the content. ProMemAssist, for example, provides evidence of more selective delivery and lower frustration in its studied tasks~\cite{zhou2025prememassist}. Satori also shows why perception matters for timing: missed hand actions can lead to step-tracking errors and repeated confirmations~\cite{shen2025satori}. User controls for accepting, postponing, or dismissing help remain consequential even when a system predicts the next step. Table~\ref{tab:interaction-timing-evidence} records the implemented triggers, delivery rules, and evaluation settings.

\begin{table*}[t]
\centering
\small
\setlength{\tabcolsep}{4pt}
\renewcommand{\arraystretch}{1.12}
\caption{Recurring interaction and assistance mechanisms. The examples illustrate related design choices; specific instances of reuse are discussed in the text.}
\label{tab:interaction-shared-mechanisms}
\begin{tabular}{@{}>{\raggedright\arraybackslash}p{\dimexpr.22\textwidth-2\tabcolsep\relax}>{\raggedright\arraybackslash}p{\dimexpr.38\textwidth-2\tabcolsep\relax}>{\raggedright\arraybackslash}p{\dimexpr.40\textwidth\relax}@{}}
\toprule
\rowcolor{gray!15}
\textbf{Shared mechanism} & \textbf{Role in the interaction} & \textbf{Examples} \\
\midrule
\textbf{Explicit commitment and correction} & Let users delimit or commit an action, and inspect or revise an inferred target. & Gesture boundaries in RimSense; spoken target repair in Gazeify Then Voiceify; candidate-note confirmation in GazeNoter~\cite{xu2024rimsense,li2026gazeifyvoiceify,li2025gazenoter}. \\
\addlinespace
\rowcolor{gray!5}
\textbf{Selective presentation} & Control how much information appears, how long it remains, and when users engage with it. & Persistent learning content in LSVP; on-demand ParaGlassMenu; staged invitations in PANDALens~\cite{10.1145/3448118,tu2023paraglassmenu,qian2024pandalens}. \\
\addlinespace
\textbf{Task-relevant context and history} & Preserve the information needed to interpret a request or continue an activity. & Dialogue context in GazePointAR; story context in CRAFT; event histories in Egocentric Co-Pilot~\cite{zhao2024gazepointar,bai2026craft,wang2026egopilot}. \\
\addlinespace
\rowcolor{gray!5}
\textbf{Preparation separated from delivery} & Prepare candidate content while retaining a later decision about whether and when to provide it. & Predicted guidance in Satori; anticipatory conversational suggestions in SocialMind; delivery or deferral in ProMemAssist~\cite{shen2025satori,yang2025socialmind,zhou2025prememassist}. \\
\bottomrule
\end{tabular}
\end{table*}

\begin{tcolorbox}[colback=gray!10, colframe=gray!10, boxrule=0pt, arc=0pt, left=5pt, right=5pt, top=4pt, bottom=4pt, before skip=8pt, after skip=4pt]
\noindent\textbf{Shared Mechanisms and Reuse.}
Confirmation, selective presentation, and retained context recur across these fields (Table~\ref{tab:interaction-shared-mechanisms}), with some designs carried across studies. ParaGlassMenu extends paracentral and near-peripheral visualization into bidirectional control~\cite{tu2022paracentral,tu2023paraglassmenu}; PANDALens adopts its attention-maintaining menu design for situated writing~\cite{qian2024pandalens}. Few-shot facial-gesture work builds on Face Commands data and preprocessing~\cite{McDuff2020facecommands,peng2024fewshot}. These examples show how presentation designs and research resources support capabilities beyond their original study.
\end{tcolorbox}
\endgroup

\section{Applications}
\label{sec:applications}
\label{sec: Applications}
\begingroup
Applications bring smart-glasses resources and capabilities into particular tasks and settings.
Sections~\ref{sec:sensing} and~\ref{sec:interaction} examine sensing and interaction functions; here we examine their benefits, limitations, and requirements for particular users and tasks.
Following Section~\ref{sec:taxonomy-revised-layers}, Figure~\ref{fig:applications} organizes five fields by task and the roles glasses play. These roles may recur across fields or combine within a system. Tables~\ref{tab:applications-everyday-evidence}--\ref{tab:applications-robot-evidence} in the appendix provide study-level tasks, platforms, contributions, evaluation findings, and demonstrated reuse.

\begin{figure}[t]
    \centering
    \includegraphics[width=\linewidth]{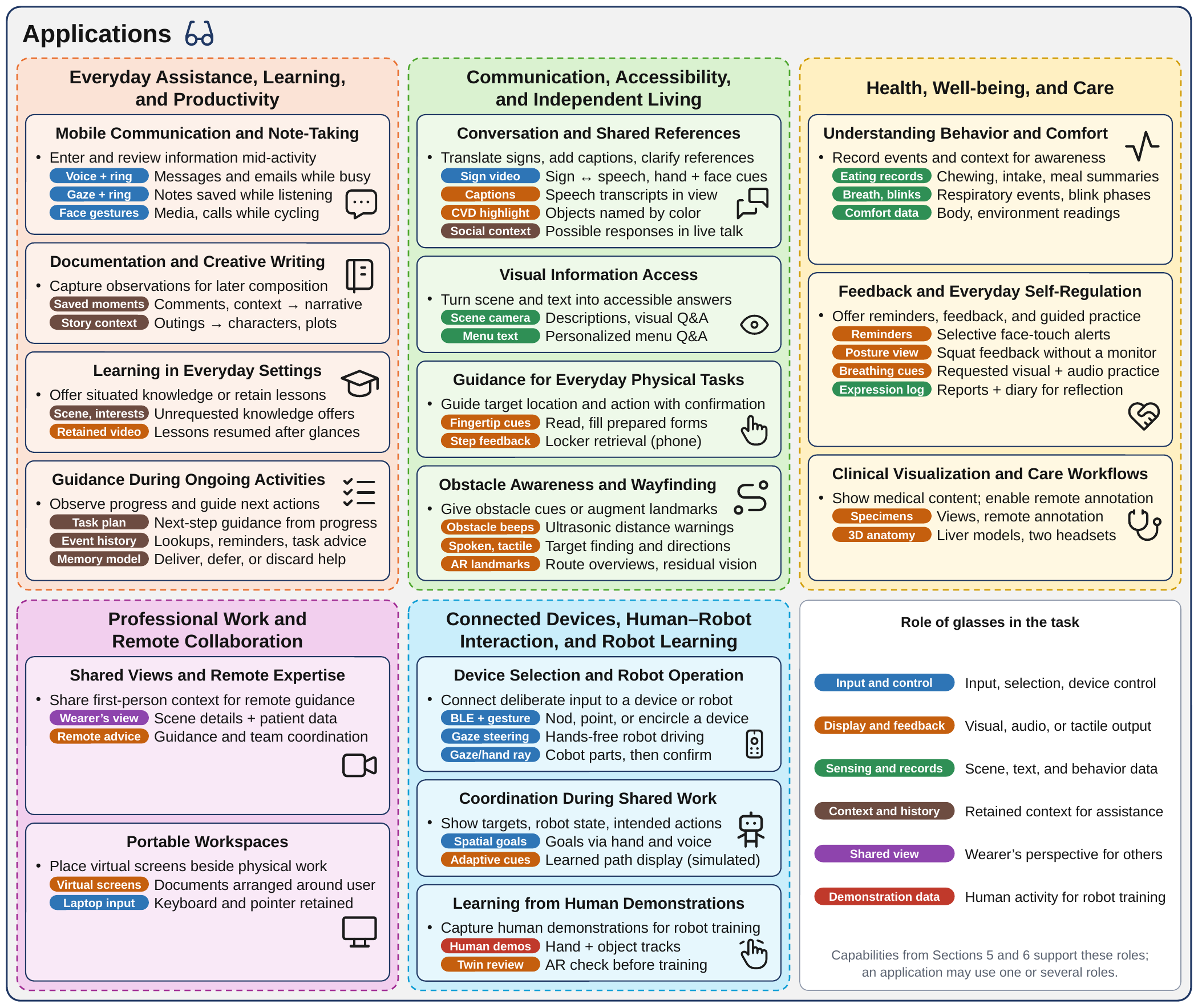}
    \caption{Applications of smart glasses across five research fields, showing representative tasks and the roles of glasses in supporting them. \textcolor{black}{Colored labels indicate how capabilities contribute to these tasks, for example through input and control, feedback, sensing and recording, or sharing the wearer's view. A role can support different applications, and an application can combine several roles.}}
    \Description{Five panels organize applications into everyday assistance, learning, and productivity; communication, accessibility, and independent living; health, well-being, and care; professional work and remote collaboration; and connected devices, human--robot interaction, and robot learning. Each panel pairs task groups with ways in which glasses support the activity. The groups describe complementary uses rather than stages in a single processing pipeline.}
    \label{fig:applications}
    \label{fig:application}
\end{figure}

\subsection{Everyday Assistance, Learning, and Productivity}
\label{sec:applications-everyday}
Everyday applications make information and assistance available while people are occupied with another activity. Their value depends on fitting communication, recording, learning, or guidance into those moments (Table \mbox{\ref{tab:applications-everyday-evidence}}).

\noindent{\small$\blacksquare$}~\textbf{Mobile communication and note-taking.}
 GlassMessaging supports messages during food preparation, eating, and transitions between tasks, increasing messaging opportunities and response speed~\cite{li2023glassmessaging}. GlassMail supports composing and revising email while walking~\cite{li2024glassmail}, while GazeNoter enables detailed, personally relevant notes during speeches or walking meetings~\cite{li2025gazenoter}. MagFace supports media and calling commands during cycling, although traffic-related head movements can trigger unwanted actions~\cite{zhang2026magface}.

\noindent{\small$\blacksquare$}~\textbf{Situated documentation and creative writing.}
 PANDALens connects selected travel moments to later narrative writing; museum participants captured more material and spent less time editing than with phone-based recording~\cite{qian2024pandalens}. CRAFT supports fiction across outings, with field trials showing how writers combine environmental observations and evolving intentions~\cite{bai2026craft}. Both require continuity between brief encounters and later composition.

\noindent{\small$\blacksquare$}~\textbf{Learning in everyday settings.}
 AiGet examines opportunistic learning through context-sensitive knowledge suggestions in a field study~\cite{li2025aiget}. LSVP supports chosen video-learning material while walking; indoor experiments assess immediate and delayed recall and walking performance~\cite{10.1145/3448118}. These applications respectively address discovering learning opportunities and sustaining attention to selected content.

\noindent{\small$\blacksquare$}~\textbf{Guidance during ongoing activities.}
 Satori guides activities such as preparing coffee or setting up a game console~\cite{shen2025satori}, while Egocentric Co-Pilot supports lookup, reminders, and task advice~\cite{wang2026egopilot}. During interrupted tabletop tasks, ProMemAssist delivered assistance more selectively, received a higher proportion of positive responses, and reduced frustration relative to its baseline~\cite{zhou2025prememassist}. Task knowledge and interaction history bound the activities each assistant supports.

\noindent\textbf{Discussion.}
These applications make support available within an ongoing activity. Its usefulness depends on the effort of checking and acting on that support. For example, faster messaging came with lower accuracy~\cite{li2023glassmessaging}, richer notes sometimes took longer to compose~\cite{li2025gazenoter}, and speaking aloud proved inconvenient in some public writing settings~\cite{bai2026craft}. Application findings therefore explain why the input, presentation, and timing choices in Section~\ref{sec:interaction} matter together. Extending these systems to new activities also requires suitable task knowledge and continuity across interruptions, beyond recognizing the immediate scene.

\subsection{Communication, Accessibility, and Independent Living}
\label{sec:applications-accessibility}
\label{sec:l3-comm}
\label{sec:l3-vi}
 Accessibility applications help users communicate, obtain information, and act on their surroundings (Table~\ref{tab:applications-accessibility-evidence}).

\noindent{\small$\blacksquare$}~\textbf{Accessible conversation and shared references.}
Sign-language interfaces connect signing with spoken or written communication. Sign-to-911 supports two-way emergency dialogue, while SiHear incorporates user correction into sign--speech exchange~\cite{sign911_2023,sihear_2025}. SignGlass focuses on ASL-to-English translation and captures facial as well as hand cues, preserving grammatical information that hand trajectories alone can miss~\cite{signglass_2025}. Captioning offers another route: ToozKit makes speech transcripts available on a head-worn display~\cite{10.1145/3594739.3610715}. CC-Glasses addresses a more specific communication barrier by highlighting objects identified through color-based descriptions for users with color vision deficiency~\cite{ccglasses_2023}. SocialMind extends conversational support to context-sensitive suggestions that wearers may consult during live exchanges~\cite{yang2025socialmind}.

\noindent{\small$\blacksquare$}~\textbf{Visual information access.}
Camera-based assistance turns scene and text information into an accessible description or answer. Google Glass scene analysis provides spoken descriptions, while LifeInsight lets blind and low-vision users ask targeted visual questions~\cite{googleglass_vi_2021,zhao2025lifeinsight}. The latter was evaluated primarily with a collar-worn or handheld camera, offering relevant evidence about query formulation and camera coverage. TEXT2TASTE combines captured menu text with personal preferences to support questions and recommendations; its initial evaluation used sighted participants~\cite{text2taste_2024}.

\noindent{\small$\blacksquare$}~\textbf{Guidance for everyday physical tasks.}
 Completing a physical task requires information to guide an action. The Interactive Form Filling Assistant reads text near a fingertip and guides it to fields on prepared paper forms~\cite{formfill_2025}. The related smartphone-based FetchAid supports parcel retrieval through touchscreen guidance, page-change feedback, error recovery, and locker-compartment directions~\cite{10.1145/3613904.3642213}. Both organize scene interpretation into sequential actions and confirmations.

\noindent{\small$\blacksquare$}~\textbf{Obstacle awareness and wayfinding.}
Navigation support ranges from proximity warnings to route guidance. Ultrasonic glasses signal nearby obstacles, E-Glasses combines target recognition with spoken directions, and LLM-Glasses explores temple-based tactile guidance~\cite{ultrasonic_vi_2024,multisensory_nav_2023,llmglasses_2026}. VisiMark instead uses residual vision, augmenting landmarks with visible labels and route overviews~\cite{visimark_2025}. Its low-vision study showed how augmentation changes the landmarks available to users, although retracing time and correctness did not significantly improve.

\noindent\textbf{Discussion.}
Useful assistance must connect the information available to the system with the action the user wants to take. For example, a LifeInsight participant searching for a watch was repeatedly told that no watch was visible. The watch was outside the camera's view, but the participant concluded that it was absent from the table~\cite{zhao2025lifeinsight}. This illustrates the need to help users check whether the camera has captured the relevant area. In the Form Filling Assistant, spoken directions helped blind users find where to begin writing on a paper form, but their writing could still extend beyond the answer box~\cite{formfill_2025}. Finding the starting point therefore did not remove the need for guidance during writing. Confirmation, correction, and feedback during action therefore deserve attention alongside recognition performance. The appropriate output also depends on whether users need an alternative sensory channel or clearer access through residual vision. These task requirements explain why a shared perception capability can lead to different interfaces and different forms of application value.

\subsection{Health, Well-being, and Care}
\label{sec:applications-health}
\label{sec:l3-health}
\label{sec:l3-clinical}
 Health applications turn measurements into monitoring, self-regulation, and clinical support. Table~\ref{tab:applications-health-evidence} relates these tasks to evaluation settings and outcomes.

\noindent{\small$\blacksquare$}~\textbf{Understanding everyday behavior and comfort.}
 Behavioral summaries can make recurring actions easier to review. Chewing detection aggregates sound classifications into eating events; MunchSonic produces dietary-action and intake summaries, while EchoGuide adds descriptions and answers from meal videos~\cite{chewing_2022,munchsonic2024,echoguide2024}. Respiratory-event recognition and blink measurements provide further health-relevant observations~\cite{echobreath_2025,blink_mmwave_2025}. AirSpecs combines bodily and environmental readings for personal-comfort exploration: across 30 participants at three sites, readings helped users investigate uncomfortable spaces and consider actions such as opening a window~\cite{ramsamy2025sensors}.

\noindent{\small$\blacksquare$}~\textbf{Feedback and everyday self-regulation.}
 Feedback connects an observation or goal to an action. EchoTouch targets reminders to intrusive face touches~\cite{echotouch_2025}; GymSoles++ presents pressure-insole posture feedback on Google Glass during squats~\cite{gymsoles_2021}. Mindful Moments supports user-chosen breathing practice through coordinated visual and auditory cues, evaluated in short sessions and a three-day study~\cite{10.1145/3563657.3596030}. EmoGlass supports later reflection through expression reports and contextual diaries~\cite{lim2022emoglass}.

\noindent{\small$\blacksquare$}~\textbf{Clinical visualization and care workflows.}
Clinical applications bring existing medical information into the practitioner's workspace. HoloLens demonstrations in anatomic pathology combined specimen viewing, remote annotation, and access to clinical windows during examination and consultation~\cite{hanna2018hololens}. A comparison of Magic Leap 1 and HoloLens 2 used the same medical-model viewer to examine clinicians' experiences of inspecting liver anatomy~\cite{zari2023magic}. Display quality, manipulation, and wearing comfort affected the usefulness of the models. These studies establish concrete workflows and visualization experiences, while diagnostic accuracy and patient outcomes remain separate questions.

\noindent\textbf{Discussion.}
The required output changes with the health-related task: an eating log needs meaningful episodes, a reminder needs an actionable event, and clinical inspection needs legible information that fits the work. Application studies therefore complement measurement accuracy with evidence about interpretation, feedback, and use. For example, AirSpecs participants differed in how they wanted environmental readings presented: some explored individual measurements, while others preferred a familiar summary because they found gas-concentration values difficult to interpret~\cite{ramsamy2025sensors}. In EchoTouch's reminder study, participants compared reminders for all face touches with reminders only for intrusive touches such as eye rubbing. When considering work scenarios, they were less willing to receive a reminder after every touch, describing frequent vibrations as distracting or irritating~\cite{echotouch_2025}. Extending these systems requires both reliable measurements and feedback whose benefit can be assessed over the intended duration of use.

\subsection{Professional Work and Remote Collaboration}
\label{sec:applications-work}
 Professional applications use glasses to support equipment use, collaboration, and mobile work, with benefits assessed at both wearer and workflow levels (Table~\ref{tab:applications-work-evidence}).

\noindent{\small$\blacksquare$}~\textbf{Shared views and remote expertise.}
A first-person video stream lets remote colleagues inspect details that would otherwise require a verbal description. In simulated emergency medical scenarios, a smart-glasses system combined this view with patient information and returned text or voice guidance~\cite{seewhatisee_2025}. The study documented richer consultation and fewer communication breakdowns than radio communication, but also revealed changes in teamwork. The camera sometimes missed what the wearer was looking at, particularly during hands-on care. Nearby team members could no longer overhear instructions as they could with a radio. Supporting remote expertise therefore involves coordinating whose view and whose instructions are available throughout the team.

\noindent{\small$\blacksquare$}~\textbf{Portable workspaces.}
Virtual displays offer additional screen space when a physical monitor is unavailable. Research on working in extended reality examined personal and structured tasks in public settings, with head-worn displays extending a laptop workspace~\cite{pavanatto2024working}. Users arranged reference material and task windows around them while retaining physical keyboard and pointing input. The controlled comparison found fewer window switches with XR, without a significant improvement in task productivity, and greater visual fatigue than the laptop condition. The findings connect workspace layout to readability, comfort, and the ability to remain aware of nearby people.

\noindent\textbf{Discussion.} Both uses distribute information beyond a conventional screen, but they introduce different coordination needs. Remote guidance must preserve a useful shared view and keep local colleagues informed \mbox{\cite{seewhatisee_2025}}; virtual workspaces must fit the worker's posture, attention, and physical surroundings \mbox{\cite{pavanatto2024working}}. These needs help explain why more available information can change how work is performed without automatically improving its overall outcome.

\subsection{Connected Devices, Human--Robot Interaction, and Robot Learning}
\label{sec:applications-robot}
\label{sec:l3-robot}
 Glasses support device and robot operation or capture demonstrations for robot learning, translating human observations and actions into machine-usable information (Table~\ref{tab:applications-robot-evidence}).

\noindent{\small$\blacksquare$}~\textbf{Device selection and robot operation.}
Selecting the intended device is a first step toward acting through a connected environment. BLEselect associates gestures with nearby Bluetooth transmitters, allowing users to select devices by nodding, pointing, or encircling~\cite{chen2023bleselect}. For mobile robots, gaze-directed control allows the operator to steer while leaving the hands available~\cite{robot_gaze_iros_2023}. In collaborative assembly, gaze or hand pointing instead identifies a part for a robot to retrieve, with explicit confirmation before execution~\cite{robot_assembly_2024}. The latter study found faster part selection with gaze.

\noindent{\small$\blacksquare$}~\textbf{Coordination during shared work.}
Spatial interfaces let users place robot goals in their surroundings and inspect the resulting actions. An interface for a legged manipulator combines spatially registered controls with hand and voice input~\cite{legged_robot_2023}. Its study illustrates the importance of how goals are specified: placing and rotating virtual markers added effort, and the original AR interface was slower than a conventional controller despite providing greater immersion. Other work addresses the information needed during coordination. VARIL learns when to display robot paths, positions, and destinations, reducing waiting and improving reported experience in a simulated warehouse study~\cite{ar_viz_policy_2022}. Its tablet demonstration provides related-platform evidence for selecting useful coordination cues.

\noindent{\small$\blacksquare$}~\textbf{Learning from human demonstrations.}
The wearer's activity can also become a training resource. ARCADE overlays a robot's digital twin on a human demonstration, allowing users to inspect generated trajectories before they train a robot policy~\cite{arcade_2024}. EgoMimic combines Aria-recorded human activity with robot demonstrations, aligning hand and robot actions so that both contribute to policy learning~\cite{egomimic_2025}. AINA extends human demonstration capture to multi-fingered manipulation, using estimated object and fingertip geometry together with a demonstration in the robot's deployment scene~\cite{dexterity_2026}. Here, the value of glasses lies in acquiring useful training observations; the resulting robot behavior does not require a live display for the wearer.

\noindent\textbf{Discussion.} These applications make the translation between human and machine activity consequential. For example, gaze-based robot control depends on suitable motion dynamics \mbox{\cite{robot_gaze_iros_2023}}, while learning from human demonstrations requires action representations that the robot can use \mbox{\cite{egomimic_2025}}. Available measurements also bound what can be learned: hand trajectories supply geometry but omit contact forces, a limitation identified by AINA \mbox{\cite{dexterity_2026}}. The requirements of the downstream task determine the types of control, alignment or sensing required.

\begin{table*}[t]
\centering
\small
\setlength{\tabcolsep}{3pt}
\renewcommand{\arraystretch}{1.02}
\caption{\textcolor{black}{Concrete capabilities used in recurring application patterns. Each row groups related uses across representative studies and identifies requirements arising from their tasks. Individual systems use different subsets of the listed functions.}}
\label{tab:applications-capability-patterns}
\begin{tabular}{@{}>{\color{black}\raggedright\arraybackslash}p{\dimexpr.15\textwidth-2\tabcolsep\relax}>{\color{black}\raggedright\arraybackslash}p{\dimexpr.35\textwidth-2\tabcolsep\relax}>{\color{black}\raggedright\arraybackslash}p{\dimexpr.28\textwidth-2\tabcolsep\relax}>{\color{black}\raggedright\arraybackslash}p{\dimexpr.22\textwidth\relax}@{}}
\toprule
\rowcolor{gray!15}
\textbf{Application pattern} & \textbf{\textcolor{black}{Capabilities used}} & \textbf{Task-specific requirements} & \textbf{Representative studies} \\
\midrule
\textbf{Guiding task completion} & \textcolor{black}{\textit{Perception:} task-progress estimation, or form registration and fingertip localization. \textit{Assistance:} instructions for the next step or spoken guidance to a writing field.} & Relate guidance to task steps and support execution, confirmation, or recovery when needed. & Satori; Form Filling Assistant~\cite{shen2025satori,formfill_2025}. \\
\addlinespace[2pt]
\rowcolor{gray!5}
\textbf{Recording for composition and reflection} & \textcolor{black}{\textit{Sensing/perception:} scene capture or facial-expression recognition. \textit{Assistance:} generating and revising text from captured context, or presenting expression records for reflection.} & Preserve meaningful details and continuity between brief encounters and later writing or reflection. & PANDALens; CRAFT; EmoGlass~\cite{qian2024pandalens,bai2026craft,lim2022emoglass}. \\
\addlinespace[2pt]
\textbf{Supporting self-regulation and practice} & \textcolor{black}{Face-touch classification with reminders; insole pressure measurement with visual posture feedback; or audiovisual breathing cues for a user-chosen practice, without automatic behavior recognition.} & Fit feedback to the activity and desired practice; assess its usefulness over the intended duration of use. & EchoTouch; GymSoles++; Mindful Moments~\cite{echotouch_2025,gymsoles_2021,10.1145/3563657.3596030}. \\
\addlinespace[2pt]
\rowcolor{gray!5}
\textbf{Bringing remote expertise into local work} & \textcolor{black}{Capturing and transmitting the wearer's view; presenting remote instructions, annotations, or clinical information during local work.} & Provide relevant views and understandable guidance while keeping local participants informed. & Remote emergency consultation; HoloLens in anatomic pathology~\cite{seewhatisee_2025,hanna2018hololens}. \\
\addlinespace[2pt]
\textbf{Learning from human demonstrations} & \textcolor{black}{\textit{Perception:} hand tracking and camera or object pose estimation. \textit{Feedback:} visualizing mapped robot motion for demonstration review in ARCADE.} & Align observations with the robot's body and scene; account for task information, such as contact forces, that motion trajectories omit. & ARCADE; EgoMimic; AINA~\cite{arcade_2024,egomimic_2025,dexterity_2026}. \\
\bottomrule
\end{tabular}
\end{table*}

\begin{tcolorbox}[colback=gray!10, colframe=gray!10, boxrule=0pt, arc=0pt, left=5pt, right=5pt, top=4pt, bottom=4pt, before skip=8pt, after skip=4pt]
\noindent\textbf{From Sensing and Interaction Capabilities to Applications.}
\textcolor{black}{Table~\ref{tab:applications-capability-patterns} relates the sensing, perception, interaction, and assistance functions in Sections~\ref{sec:sensing} and~\ref{sec:interaction} to application goals and requirements. Two examples illustrate these relationships.}\par
\textcolor{black}{In the \emph{Guiding task completion} row, the Form Filling Assistant combines form registration and fingertip localization with voice commands and spoken feedback~\cite{formfill_2025}. These capabilities let blind users select a printed line, hear its contents, and locate the corresponding writing field. Their application value comes from connecting perception, target grounding, and feedback into a reading-and-writing workflow. This integration relies on prepared form layouts and keeping the page within the camera's view.}\par
\textcolor{black}{In the \emph{Learning from human demonstrations} row, EgoMimic uses Project Aria video and hand and camera pose estimates to capture human demonstrations~\cite{egomimic_2025}. These sensing and perception outputs become useful for robot learning after human and robot actions are aligned and a policy is trained on human and robot demonstrations. The application value is assessed through the robot's manipulation performance. This example draws directly on sensing and perception, without a live assistance interface for the wearer.}
\end{tcolorbox}
\endgroup

\section[Cross-cutting Deployment Constraints]{Cross-cutting Deployment Constraints}
\label{sec:deployment}
\label{sec:privacy_security_social}
\begin{figure}[t]
    \centering
    \includegraphics[width=\textwidth]{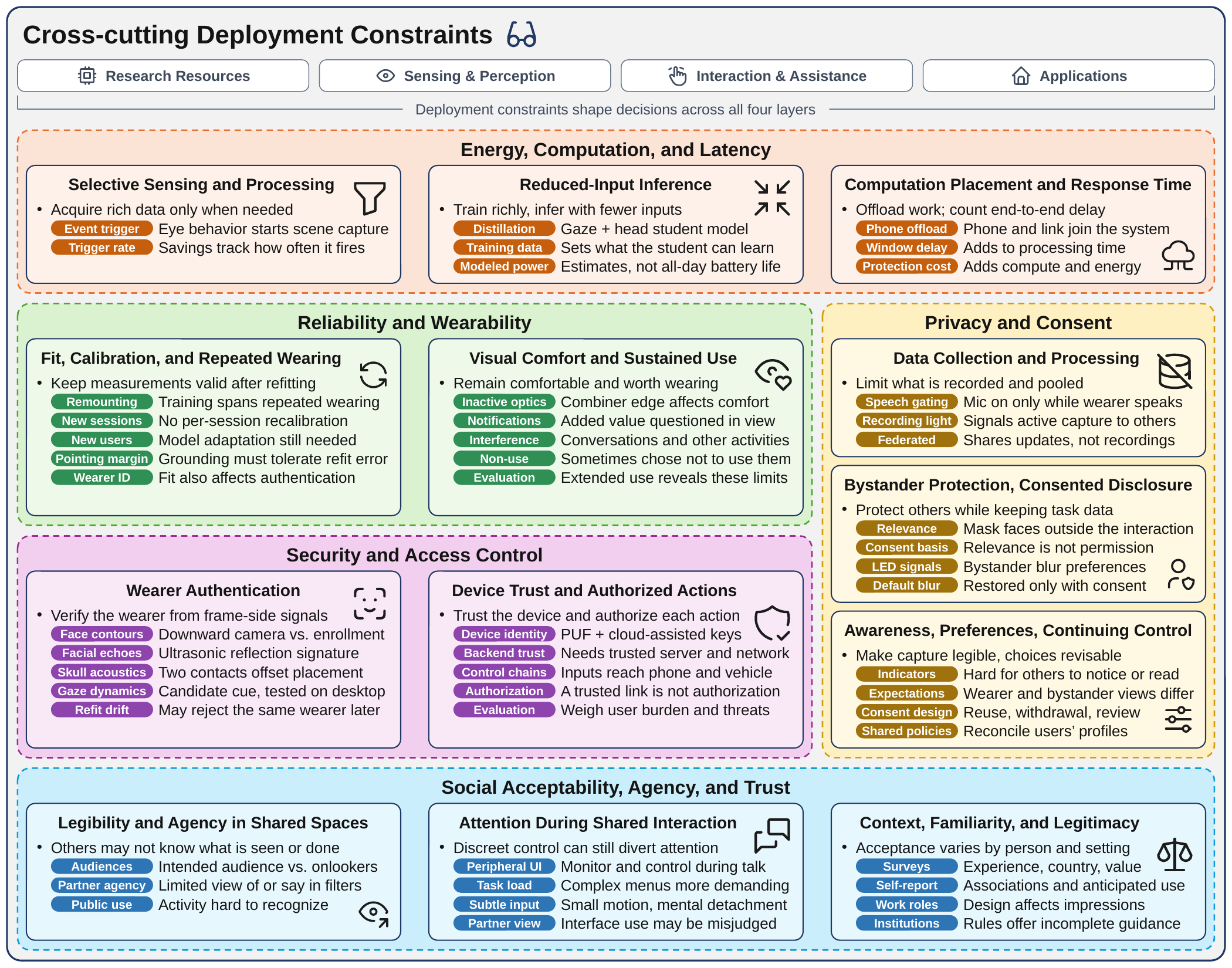}
    \caption{\textcolor{black}{Cross-cutting deployment constraints for smart glasses. The five colored groups summarize recurring concerns, approaches, and conditions of use discussed in this section. The upper strip indicates that these constraints shape decisions across research resources, sensing and perception, interaction and assistance, and applications.}}
    \Description{Five colored groups organize deployment concerns into energy, computation, and latency; reliability and wearability; privacy and consent; security and access control; and social acceptability, agency, and trust. Subgroups summarize approaches and conditions illustrated by the reviewed studies. A shared bracket below the four-layer strip indicates that deployment constraints affect decisions across research resources, sensing and perception, interaction and assistance, and applications.}
    \label{fig:deployment-constraints}
    \label{fig:privacy_security_social}
\end{figure}

 Deployment constraints shape how resources and capabilities can support everyday use across the four layers (Section~\ref{sec:taxonomy}). Figure~\ref{fig:deployment-constraints} groups these five constraints examined in this section: energy, computation, and latency; reliability and wearability; privacy and consent; security and access control; and social acceptability, agency, and trust.

\subsection[Energy, Computation, and Latency]{Energy, Computation, and Latency}
\label{sec:deployment-operation}
Continuously acquiring and processing rich sensor streams can exceed a wearable power budget. The platforms in Section~\ref{sec: Resources} distribute this work differently between the frame, a companion device, and external services. Research systems further manage the cost by deciding which observations to acquire, which computations to run, and when a result is needed. Table~\ref{tab:deployment-operation-evidence} records representative approaches and their evaluation conditions.

\noindent{\small$\blacksquare$}~\textbf{Reducing work during use.}
 Selective processing uses an inexpensive observation to trigger richer sensing: MemX activates outward capture after detecting a possible attention event from eye behavior~\cite{liu2021memx}. SkillSight instead transfers knowledge from richer training observations to a student using gaze and head-related signals at inference~\cite{li2026skillsight}.

\noindent{\small$\blacksquare$}~\textbf{Distributing computation and meeting response times.}
 External computation expands what a lightweight frame can support at the cost of communication latency. WearSE performs streaming speech enhancement on a smartphone, reusing temporal computation~\cite{sun2025wearse}. Its wired prototype reports 130.1\,ms delay: a 100\,ms window plus 30.1\,ms processing. Privacy mechanisms add costs too: SITARA blurs faces and encrypts their original images, with replacement-face generation requiring further processing~\cite{fan2026nowyousee}.

\noindent\textbf{Discussion.} Runtime depends on both the required capability and how often the application invokes it. Selective sensing, reduced-input models, and external processing each preserve useful functions under different assumptions about events, training data, and connectivity. Combining capabilities therefore requires considering their shared sensing, processing, and communication costs.

\subsection[Reliability and Wearability]{Reliability and Wearability}
\label{sec:deployment-reliability}
 Repeated wearing, changing surroundings, and competing activity affect both measurement and sustained use (Sections~\ref{sec:sensing} and~\ref{sec:applications}). Table~\ref{tab:deployment-operation-evidence} records evidence about these conditions, including visual comfort and everyday use.

\noindent{\small$\blacksquare$}~\textbf{Maintaining useful measurements across wearing sessions.}
Sensors close to the eyes or face can observe small movements, but their measurements also change when the frame shifts. mmET addresses this with training across repeated remounting and joint estimation of gaze and eye geometry~\cite{lu2025mmet}. It evaluates new wearing sessions without recalibrating each time, while still adapting the model to new users. The remaining difference between within-session and cross-session errors shows why a capability's operating conditions matter to subsequent interaction: a pointing or grounding method needs to accommodate the accuracy available after ordinary adjustments to the glasses. The same placement issue reappears in wearer authentication, discussed in Section~\ref{sec:deployment-security}.

\noindent{\small$\blacksquare$}~\textbf{Comfort and attention over time.}
Wearability includes what the device does to vision even when it is not presenting information. Going Blank Comfortably examines inactive monocular display optics and finds that the position of the visible combiner boundary affects visual comfort~\cite{lyons2023going}. When information is presented, its convenience must also be considered against distraction. In the three-week Focals user study, participants questioned the added value of notifications in their field of view, reported interference with conversations and other activities, and sometimes chose not to use the glasses~\cite{pyae2022focus}. These findings explain why successful access to a feature does not by itself establish that the feature fits repeated everyday use.

\noindent\textbf{Discussion.} Reliable use depends on maintaining both useful measurements and a tolerable experience. Evaluation across repeated wearing and extended use reveals these conditions more directly than a single successful interaction. Social expectations introduce a further set of conditions, considered in Section \mbox{\ref{sec:deployment-social}}.

\subsection[Privacy and Consent]{Privacy and Consent}
\label{sec:capture_recording_privacy}
\label{sec:bystander_privacy_protection}
\label{sec:consent_permission}

\label{sec:deployment-privacy}
Privacy depends on what smart glasses capture, what they infer, and who can access the resulting information. These decisions affect wearers and people around them. A survey of everyday AR scenarios found that bystanders' awareness and consent preferences varied across recording, identification, and inference activities~\cite{guo2022privacy}. Permission to collect an image therefore needs to be understood in relation to its intended use. Table~\ref{tab:deployment-privacy-evidence} summarizes the mechanisms, user studies, and design frameworks reviewed here.

\noindent{\small$\blacksquare$}~\textbf{Data collection and processing.}
Reducing unnecessary collection changes what a system can retain or disclose. MeMic uses speech-related vibration detected by an IMU to activate its microphone when the wearer speaks, with a light indicating recording~\cite{zulfikar2024memic}. This limits recording between utterances, although simultaneous bystander speech can still be captured. Its technical evaluation used a neck-worn prototype; eyewear was among the form factors considered in its acceptability study. The mechanism illustrates how control at acquisition can complement the recording indicators discussed in responsible-adoption work on camera glasses~\cite{iqbal2023adopting}.

Data handling introduces another control point. Privacy-aware activity recognition compares centralized and federated training using inertial recordings from OCOsense glasses~\cite{leite2023privacyaware}. Federated averaging combines model updates from participant-specific data partitions, reducing the need to pool recordings during training. This changes the training arrangement while leaving the sensing task intact.

\noindent{\small$\blacksquare$}~\textbf{Bystander protection and consent-based disclosure.}
Selective protection aims to preserve information needed by an application while restricting information about other people, a tradeoff emphasized in mixed-reality privacy research~\cite{corbett2023securing}. BystandAR combines gaze history and voice to infer whom the wearer is interacting with, preserving those faces while masking other faces in color and depth data~\cite{corbett2023bystandar}. Its threshold determines the balance between bystander protection and access to the interaction subject. This inference supplies a relevance judgment; permission still requires a separate basis.

 BystandARIA uses LED signals to control blurring in a processed Aria camera feed; matching each sender to a face remains proposed for individual control~\cite{al2025bystandaria}. SITARA instead blurs faces by default and encrypts original regions for consent-mediated restoration through a trusted service~\cite{fan2026nowyousee}. Demonstrations and interviews in the \textit{See Me If You Can} study show how default protection, requester identity and purpose, and disclosure control shape wearers' and bystanders' responses~\cite{li2026seemeifyoucan}.

\noindent{\small$\blacksquare$}~\textbf{Communicating and managing privacy choices.}
People need to understand a privacy choice before they can exercise it. Wearers in \textit{In Focus, Out of Privacy} reported that recording indicators were difficult for others to notice or interpret, leaving wearers responsible for explaining and managing capture~\cite{bhardwaj2024focus}. \textit{Mind the Gap} finds differences between wearer willingness and bystander expectations across situations~\cite{wang2026mind}. These findings make the meaning, timing, and audience of a privacy cue consequential.

Consent designs address the effort of repeated decisions. Co-design work on spontaneous AR interactions proposes mechanisms that reuse preferences, respond to context changes, and permit withdrawal~\cite{windl2025designing}. Privacy Equilibrium explores how multiple users' sensing requirements and privacy profiles can be reconciled through constrained optimization, with simulations and expert feedback exposing dependencies on task needs~\cite{rajaram2025privacy}. A broader XR permission framework organizes consent across onboarding, contextual prompts, ongoing control, and later review~\cite{mansour2025multi}. These approaches give different forms to continuing control over data use, from individual interface choices to policies for shared environments.

\noindent\textbf{Discussion.}
Protection depends on reliable sensing and consumes processing resources. For example, BystandAR reduces computation by spacing out inference, which can leave rapidly moving faces exposed between updates~\cite{corbett2023bystandar}. SITARA's Raspberry Pi evaluation measured energy costs of 1.68 times the recording baseline for its core protection mechanism and 2.8 times for the full pipeline~\cite{fan2026nowyousee}. Privacy decisions therefore change the information available to applications and the resources needed to provide it. A useful design must connect a person's privacy choice to the correct data, enforce that choice, and preserve the information needed for the intended task.

\subsection[Security and Access Control]{Security and Access Control}
\label{sec:authentication_identity}

\label{sec:deployment-security}
Security concerns both access to the glasses and the actions that they can initiate. Their position on the head provides biometric signals for recognizing the wearer, while connections to phones, services, and other devices extend the consequences of an accepted input. Table~\ref{tab:deployment-security-evidence} summarizes the authentication methods, protected operations, evaluation settings, and remaining dependencies.

\noindent{\small$\blacksquare$}~\textbf{Wearer authentication.}
Biometric methods use the relationship between the glasses and the wearer's face or head. C-Auth extracts nose and cheek contours from a downward-facing camera and compares them with enrolled references~\cite{lim2023c}. SonicID identifies the wearer from ultrasonic facial reflections, while SkullID measures how sound changes as it travels through the skull~\cite{li2024sonicid,shin2024skullid}. These approaches obtain identity information from signals available near the frame, reducing reliance on typed credentials or a separate handheld interface.

Their practical difficulty is separating differences between people from changes in how the same person wears the device. C-Auth benefited from adding reference contours from another session, and SonicID's genuine-user acceptance declined across days despite strong same-day performance~\cite{lim2023c,li2024sonicid}. SkullID uses measurements from two contact locations to compensate for placement variation, and evaluates remounting, noise, and repeated visits~\cite{shin2024skullid}. These results connect physical fit and measurement repeatability to the effort of gaining access. Gaze dynamics offer another biometric cue: a study using symmetric moving stimuli evaluated recognition across sessions with a desktop eye tracker~\cite{dondi2026gaze}. Participants viewed dedicated animations, so its relevance to glasses concerns the candidate biometric method; unobtrusive authentication during everyday eyewear use remains to be established.

\noindent{\small$\blacksquare$}~\textbf{Device trust and authorized actions.}
A trusted session requires assurance about the communicating device and the keys protecting its messages. PXRA combines a physically unclonable function with cloud-assisted cryptography to reduce computation on the XR client~\cite{cha2025lightweight}. Its evaluation uses proxy hardware and an emulated hardware-identity response. The design illustrates how shifting security computation off the glasses introduces dependencies on a trusted backend and network availability.

A trusted connection can still carry an unauthorized action. \textit{From Virtual Touch to Tesla Command} demonstrates this problem in configured control chains linking paired glasses, a smartphone, and vehicle services~\cite{yu2024virtualtouch}. Inputs accepted through the glasses could propagate into consequential operations without adequate verification of user authorization. The connection to Sections~\ref{sec:interaction} and~\ref{sec:applications} is direct: recognizing an input, selecting its target, and forwarding a command must preserve the distinction between a detected event and an action the user has authorized.

\noindent\textbf{Discussion.}
 Security evaluation must address both reliable access for legitimate users and authorization of downstream actions. This requires considering multiple dimensions, ranging from sensor placement and enrollment to interface decisions, communication conditions, and the operations that connected devices permit.

\subsection[Social Acceptability, Agency, and Trust]{Social Acceptability, Agency, and Trust}
\label{sec:social_acceptance}

\label{sec:deployment-social}
Social acceptability concerns how smart-glasses use fits the expectations and relationships of people sharing an activity. A useful application for the wearer can still leave a conversation partner unsure what is being seen, recorded, or attended to. Table~\ref{tab:deployment-social-evidence} summarizes the devices, participants, study methods, and findings underlying this discussion.

\noindent{\small$\blacksquare$}~\textbf{Legibility and agency in shared spaces.}
Wearable-display design distinguishes the intended audience for information from the people who can see the device or its use~\cite{lyons2021design}. With glasses, these audiences can have different experiences of the same interaction. In a study of dyadic AR use, wearers could alter their view of a conversation partner through face filters, while partners had limited knowledge of that view or control over their own representation~\cite{chung2023negotiating}. The resulting imbalance concerned participation and self-presentation as well as recording privacy. Making an augmentation available to the wearer therefore introduces a question about what the other person can understand and influence.

Related headset studies show how uncertainty extends to public activity. Bystanders in an XR-work study had more difficulty recognizing that someone was working than when observing laptop use~\cite{pavanatto2024working}. Outdoor tasks with a Quest 3 elicited curiosity, discomfort, and indifference, with eye visibility and unfamiliar behavior shaping encounters~\cite{zhang2025arkward}. A digital ethnography also found tensions around public use and bystander privacy, although most of its 112 videos concerned larger VR/MR headsets~\cite{fleck2025wearable}. These findings identify questions about legibility and familiarity; their device and setting differences matter when interpreting them for everyday eyewear.

\noindent{\small$\blacksquare$}~\textbf{Attention during shared interaction.}
The input and presentation mechanisms in Section~\ref{sec:interaction} can help wearers remain engaged with another person. Paracentral progress displays and ParaGlassMenu support monitoring and control during the conversations studied~\cite{tu2022paracentral,tu2023paraglassmenu}. Their benefits depend on the information and effort involved. ParaGlassMenu participants found complex menu tasks more demanding, and a partner could misinterpret how much the wearer was interacting with the interface. Similarly, a field study of subtle ring input reported mental detachment from conversation despite the small physical action required~\cite{orlosky2021ubiquitous}. Low movement, low attentional demand, and understandable behavior are therefore distinct design goals. Keeping a gesture discreet can support a social activity while still leaving cognitive effort or uncertainty for the other person.

\noindent{\small$\blacksquare$}~\textbf{Context, familiarity, and legitimacy.}
 Surveys show that acceptance varies with experience and perceived value. For eyewear, acceptance was further related to perceived intelligence, naturalness, aesthetics, usefulness, and privacy concerns~\cite{mcgregor2024exploring,kim2025cultural,niu2025exploring}.

 In a photo-based healthcare experiment, glasses design affected impressions of professionals' warmth and competence~\cite{sobieraj2023can}. A healthcare survey found that social-interaction variables helped explain anticipated adoption~\cite{zuidhof2025social}. Classroom policy analysis separately identifies gaps in institutional responsibility and governance~\cite{arantes2026too}. Personal acceptance and institutional readiness therefore require different evidence.

\noindent\textbf{Discussion.}
 Social acceptability depends on how assistance, attention, information, and control are distributed among participants involved. It therefore shapes sensing, feedback, permissions, and application design, with benefits assessed for the shared activity as well as the wearer.

\begin{tcolorbox}[colback=gray!10, colframe=gray!10, boxrule=0pt, arc=0pt, left=5pt, right=5pt, top=4pt, bottom=4pt, before skip=8pt, after skip=4pt]
\noindent\textbf{Deployment Choices Across the Framework.}
\textcolor{black}{Deployment constraints shape how resources and capabilities can be combined. For example, SITARA restricts access to original face images through blurring, encryption, and consent-based restoration, but adds processing time and energy costs \mbox{\cite{fan2026nowyousee}}. Section \mbox{\ref{sec: Future}} builds on these dependencies to develop directions for broader integration and validation.}
\end{tcolorbox}

\section{Future Research Directions}
\label{sec: Future}
\label{sec:future}
\begingroup
\color{black}
 The studies reviewed in this survey show how smart glasses can perceive aspects of the wearer and the world, support interaction, and assist with particular activities. Following Section~\ref{sec:taxonomy-revised-perspective}, we identify seven diverse research directions for making these capabilities useful across activities and time.\par

\subsection{Generalizable Egocentric AI}
\label{sec:future-ai}
\label{sec:future-reuse}
 Egocentric foundation models could support interpretation of scenes, actions, and user requests across applications. EgoVLP and Ego-Exo4D provide video--language and complementary-view resources~\cite{lin2022egovlp,grauman2024egoexo4d}, while M-BEST-RQ demonstrates reuse across audio tasks and microphone arrays~\cite{shon2025mbestrq}. Future work should establish what information and capabilities can be shared across tasks and devices during continued use.\par
\noindent\textbf{Learning representations that support reuse.}
First-person observations bring together scene content, speech, movement, and, on some platforms, gaze or physiological measurements. A reusable model needs to relate these inputs to the wearer and the ongoing activity while accommodating differences in sensor availability and placement. M-BEST-RQ's transfer, for example, still involves device-specific signal processing and task adaptation~\cite{shon2025mbestrq}. Future research should examine whether shared representations preserve information needed for different outputs, from identifying an attended object to interpreting an action or selecting a speaker. Comparing shared models with task-specific alternatives should account for the labels, calibration, and computation needed to adapt them. This would show when pretraining reduces development effort and where specialized methods remain necessary.\par
\noindent\textbf{Learning from richer data than a device can continuously collect.}
Multimodal training can exploit relationships among synchronized observations even when only a subset is available during use. SkillSight provides a concrete example: richer training inputs supervise proficiency estimation from gaze and head motion~\cite{li2026skillsight}. Extending this approach could allow compact models to draw on richer research datasets without requiring every application to collect every modality. The resulting resource need is specific: training and evaluation data must expose the sensor combinations, activities, and missing observations for which transfer is possible. Task performance should be assessed alongside response time on the intended hardware, connecting model development to the systems questions in Section~\ref{sec:future-energy}.\par

\subsection{Understanding Users and Context over Time}
\label{sec:future-context}
 Experience histories and user-state inference provide complementary foundations for adapting assistance strategies over time~\cite{qian2024pandalens,wang2026egopilot,huang2023cases,smartglasses_dream_2022}. Extending them across activities requires preserving the context in which each observation or inference was meaningful.\par
\noindent\textbf{Remembering what was observed and what was missed.}
The head-worn viewpoint provides only a partial account of an activity. Objects leave the camera's view, hands become occluded, and recording can pause to save energy or respect a privacy choice. Missing information can then be mistaken for evidence. In the related collar-worn and handheld LifeInsight study, a participant concluded that a watch was absent from a table after repeated negative answers, although it was outside the camera's view~\cite{zhao2025lifeinsight}. Future memory systems should retain enough information about time, coverage, and subsequent changes to distinguish an observed absence from an unobserved part of the scene. This would allow an assistant to request a new observation or explain why its record cannot answer a question.\par
\noindent\textbf{Interpreting changes in the person.}
Longer records could also reveal changes in attention, proficiency, routines, or well-being. Their interpretation is more demanding than accumulating predictions. A change in gaze may reflect a harder task, a different environment, or a change in the wearer; a shift in frame placement can change the measurement itself. Research should investigate representations that relate individual baselines to task and environmental context, while allowing meaningful personal changes to remain visible. Adaptation should not automatically remove a persistent change as if it were sensor error. Repeated observations, task outcomes, and appropriate reports from participants can help test whether a model's interpretation corresponds to the state it claims to describe.\par
These questions call for evaluation across changes and interruptions, including returns to earlier activities, missing recording intervals, and changes in task demands. Temporal models should be tested for when history helps, when old information misleads, and when a new observation is necessary. The aim is a usable account of the person's activity and its uncertainty, with sufficient evidence to support the assistance that follows.\par

\subsection{Proactive Assistance under Human Control}
\label{sec:future-control}
 Existing proactive assistants differ in when they offer help and how much initiative they take~\cite{qian2024pandalens,li2025aiget,shen2025satori}. Satori exposes a central tension: actions outside of the view of the camera can cause guidance mismatches, while frequent confirmation can interrupt the task~\cite{shen2025satori}.\par
Future assistants should make initiative adjustable as an activity unfolds. A wearer may welcome detailed guidance while learning, prefer occasional reminders later, or temporarily suspend assistance. Systems need ways to learn from these choices without treating every dismissal as a permanent preference or every accepted suggestion as evidence of benefit. Clarification should also reflect the uncertainty of an interpretation and the consequences of acting on it. An ambiguous object may require another look; an operation on a connected device may require explicit authorization. Research should examine whether people can notice, correct, cancel, and recover from mistaken assistance while continuing their activity. These outcomes would reveal more than the rate at which suggestions are accepted.\par
 As the connected-control and consent studies in Section~\ref{sec:deployment} demonstrate, control must extend beyond the glasses~\cite{yu2024virtualtouch,fan2026nowyousee,li2026seemeifyoucan}. Research should establish whether confirmations, cancellations, and restrictions on data use remain effective as assistance crosses services and shared activities.\par

\subsection{Open Research Platforms and Shared Infrastructure}
\label{sec:future-open-platforms}
 Section~\ref{sec: Resources} distinguishes hardware adaptability, sensor-data access, and application development, supported by resources such as Captivates, hl2ss, Aria Tools, and OpenXR~\cite{chwalek2021captivates,dibene2022hl2ss,ariatools,openxr}. However, many of these platforms do not provide access to changing acquisition timing, deploying models onboard, or measuring system costs.\par
Open or well-documented reference platforms could connect these forms of access. Modular hardware would let researchers select sensors and feedback channels, while shared software could support synchronization, logging, privacy filtering, model deployment, and interaction management. Such infrastructure would make it possible to replace a component or processing stage without rebuilding the full system. Its value should be demonstrated through independent reproduction and controlled comparisons, supported by build instructions, calibration procedures, stable interfaces, and measurements of the infrastructure's own overhead.\par
The opportunity also extends across devices. Glasses could combine their first-person view with complementary information from watches, earables, or smart-home devices, and use this context in tasks involving robots or other connected systems. This requires more than moving computation off the frame. Shared infrastructure should help devices relate observations to the same activity, identify the intended target of an action, and communicate whether that action succeeded. Common descriptions of timing, coordinates, available functions, and data-access permissions would support such coordination. Research could test whether an application continues to work when a device is replaced or becomes unavailable, making interoperability a demonstrated property of a task rather than only an interface specification.\par

\subsection{Energy-Efficient Operation across Devices and the Cloud}
\label{sec:future-energy}
 Selective sensing, reduced-input models, and companion processing offer different ways to reduce energy on resource-constrained glasses (Section~\ref{sec:deployment-operation})~\cite{liu2021memx,li2026skillsight,sun2025wearse}. Alongside earlier Android power-management measurements~\cite{chen2020powermanagement}, these studies motivate evaluation of complete applications, including the processing demands of the models reviewed in Section~\ref{subsec:software-tools-models}.\par
\noindent\textbf{Coordinating sensing and computation.}
When several capabilities operate together, they may share observations and processing but also compete for resources. Future systems could coordinate acquisition and inference according to the information needed by the current task. A low-cost signal might be sufficient to detect a relevant event before activating richer sensing or a larger model. Research should establish when such coordination preserves useful information and when it causes missed events or delayed assistance. Energy savings need to be interpreted together with these consequences, particularly when a decision affects later perception or guidance.\par
\noindent\textbf{Adapting where work runs.}
Partitioning inference across devices and servers is an established systems approach~\cite{kang2017neurosurgeon}. Smart-glasses applications add changing observation needs, wearer-facing response deadlines, and constraints on where captured data may go. Future systems could select a processing location for each request or model stage according to available computation, wireless conditions, temperature, and permission to transfer the required information. Evaluations should include sustained operation and connection failures: a configuration that briefly meets a deadline may later throttle, while an unavailable remote service may interrupt assistance. Systems need useful fallback behavior, such as a smaller local capability, a deferred response, or a clear request for user action. Keeping sensitive raw observations on the glasses or phone, where required by the participants' choices, should constrain these decisions.\par
Comparable measurements are essential to developing these policies. Studies should report the device configuration, sensors, processing locations, and workload associated with battery life or energy results. Workload traces combining background sensing, requests, feedback, and idle periods could support comparisons of fixed and adaptive designs. Measurements should include required companion devices and communication, alongside temperature, sustained performance, and task outcomes. The open infrastructure in Section~\ref{sec:future-open-platforms} could make these measurements reproducible and expose the costs of combining capabilities that appear efficient in isolation.\par

\subsection{Robustness and Adaptation in Everyday Use}
\label{sec:future-adaptation}
 Remounting and across-day experiments show that within-session performance cannot establish continued reliability~\cite{lu2025mmet,li2024sonicid}. Research must determine which changes in wearing position, user, or environment require adaptation and how much effort that adaptation demands.\par
 Placement changes may require compensation or recalibration, while unfamiliar activities may require new training data. mmET and SkullID illustrate training and measurement approaches to remounting~\cite{lu2025mmet,shin2024skullid}. Continual learning should be compared with fixed models and periodic calibration, using reliable observations or user corrections to avoid reinforcing inference errors. Evaluation must also test whether adaptation preserves previously useful behavior.\par
Evaluation should separate new-user transfer, repeated wearing by the same person, environmental change, and missing sensor inputs. Data divided by person and session can support these comparisons, while reporting calibration effort, additional labels, computation, and recovery time makes the cost of adaptation visible. For personalized state inference, evaluations must also distinguish changes in measurement from genuine changes in the wearer, as discussed in Section~\ref{sec:future-context}. The aim is to maintain useful capabilities through ordinary changes in use without making continuous maintenance another task for the user.\par

\subsection{Sustained Application Value and Behavioral Change}
\label{sec:future-validation}
 Technical gains must be evaluated through task outcomes, errors, and recovery effort (Section~\ref{sec:applications}). The XR workspace study, for example, found fewer window switches without significantly improving productivity and with greater visual fatigue~\cite{pavanatto2024working}. Such results motivate evaluating what a capability enables people to accomplish.\par
Longer-term evaluation can also ask whether assistance changes what the user can do beyond the immediate interaction. In learning, training, or habit support, researchers could examine whether gains persist as guidance is reduced and whether they transfer to activities not practiced with the system. Mindful Moments explicitly identifies the distinction between immediate mindfulness effects and lasting change as a question for further study~\cite{10.1145/3563657.3596030}. Such studies should connect the assistance provided to subsequent behavior and participant goals, rather than infer improvement from repeated use alone. Continued support can itself be the desired outcome in accessibility and everyday assistance; independence from the device is not a universal criterion of success.\par
 Sustained value also includes the decision to keep using a system. In the Focals study, participants sometimes decided to forego using the glasses because its disruption outweighed its utility~\cite{pyae2022focus}. Studies should assess these outcomes alongside task benefits and the burdens of calibration, charging, interruptions, and attention, including partners' and bystanders' experiences in shared activities.\par
\endgroup

\section{Conclusion}
\label{sec:conclusion}
This survey examines 163 publications on smart glasses for ubiquitous computing, focusing on work published from 2020 to October 2026 and retaining selected earlier studies for historical context. We organize the literature through four complementary layers: Research Resources, Sensing \& Perception, Interaction \& Assistance, and Applications, with deployment constraints considered across all four. The framework brings together hardware platforms, software tools and models, datasets, technical capabilities, and evidence of use to explain both current progress and opportunities for future development.\par
Across the reviewed work, the usefulness of a capability depends on the task it supports and the conditions under which it operates. Sensing choices determine what information is available; interaction and assistance mechanisms determine how that information reaches users or supports action; application studies establish whether these combinations provide meaningful benefits. Shared mechanisms create opportunities to adapt advances across tasks, while differences in hardware access, evaluation conditions, and deployment requirements shape the limits of that reuse. This perspective helps identify what has been established, what could be combined or extended, and what still requires validation.\par
Advancing smart glasses as a ubiquitous computing platform therefore requires coordinated progress in reusable resources, generalizable intelligence, understanding users and contexts over time, and assistance that remains under human control. Open research platforms and efficient operation can support this progress, while evaluation across users, environments, and extended periods must establish reliability, acceptability, and sustained value. Together, these directions provide a path from demonstrated capabilities toward practical and trustworthy systems that support everyday activities.

\bibliographystyle{ACM-Reference-Format}

\bibliography{rfs-chapter5-revised,taxonomy-additions}

\newpage
\appendix
\clearpage

\section{Appendix}
\label{sec:supplementary}
\begingroup
Tables~\ref{tab:sensing-body-evidence}--\ref{tab:sensing-scene-evidence} bring together the studies discussed in Section~\ref{sec:sensing} and additional related entries from the literature inventory. They identify the sensed inputs and device setup, the task and output together with its application, the contribution and its mechanism, the conditions supporting the reported findings, and demonstrated reuse across studies. A study may appear in more than one field when it provides distinct contributions; proposed uses are distinguished from demonstrated applications. Results refer to each study's own task and protocol. The reuse column highlights documented use of methods, models, platforms, software, or datasets across studies; --- indicates that no specific link is recorded here.\par
\small
\setlength{\tabcolsep}{4pt}
\renewcommand{\arraystretch}{1.10}
\setlength{\LTpre}{8pt}
\setlength{\LTpost}{8pt}
\begin{longtable}{@{}>{\raggedright\arraybackslash}p{\dimexpr.12\textwidth-2\tabcolsep\relax}>{\raggedright\arraybackslash}p{\dimexpr.16\textwidth-2\tabcolsep\relax}>{\raggedright\arraybackslash}p{\dimexpr.15\textwidth-2\tabcolsep\relax}>{\raggedright\arraybackslash}p{\dimexpr.20\textwidth-2\tabcolsep\relax}>{\raggedright\arraybackslash}p{\dimexpr.23\textwidth-2\tabcolsep\relax}>{\raggedright\arraybackslash}p{\dimexpr.14\textwidth\relax}@{}}
\caption{Sensing and perception evidence: body motion and activity.}\label{tab:sensing-body-evidence}\label{tab:sensing_activity_reading_speech}\label{tab:l1_body_motion}\\
\toprule
\textbf{Study} & \textbf{Sensing modality and setup} & \textbf{Task, output, and application} & \textbf{Key contribution and mechanism} & \textbf{Evaluation and main finding} & \textbf{Demonstrated reuse} \\
\midrule
\endfirsthead
\caption[]{Sensing and perception evidence: body motion and activity. (continued)}\\
\toprule
\textbf{Study} & \textbf{Sensing modality and setup} & \textbf{Task, output, and application} & \textbf{Key contribution and mechanism} & \textbf{Evaluation and main finding} & \textbf{Demonstrated reuse} \\
\midrule
\endhead
\midrule
\multicolumn{6}{r}{\footnotesize Continued on the next page}\\
\endfoot
\bottomrule
\endlastfoot
\textbf{LFI--IMU activity recognition}\newline\cite{aktaSemi2021cnn} & Laser feedback interferometry measures eye motion; an IMU measures head motion on prototype eyewear. & Physical/cognitive activity recognition, producing seven activity labels. & Combines eye and head features in a temporal CNN with transfer learning and user adaptation. Context-aware device behavior is a proposed downstream use, not an evaluated application. & Fifteen participants; held-out-user models personalized with three 30-second labeled windows per class achieved 88.15\% macro-F1. & --- \\
\addlinespace[4pt]
\textbf{ActSonic}\newline\cite{chen2024actsonic} & Glasses-mounted speakers and microphones emit FMCW probes and receive reflections from the face and upper body. & Everyday activity logging, producing activity labels including null. & Differential echo profiles and self-supervised pretraining support activity classification without the test user's labeled training data. & Seven participants' unconstrained home sessions: 86.6\% macro-F1 at 1 Hz; a separate twelve-person prompted home study achieved 93.4\%. & EchoGuide reuses its pretrained model~\cite{echoguide2024}; MunchSonic adapts its hardware design~\cite{munchsonic2024}. \\
\addlinespace[4pt]
\textbf{PoseSonic}\newline\cite{wang2023posesonic} & Glasses-mounted FMCW speakers/microphones; synchronized camera-derived poses provide supervision, not runtime input. & Continuous 3D upper-body pose estimation, producing nine joint positions. & Original and differential echo profiles feed coordinate regression, extending egocentric acoustic sensing from movement categories to body geometry. Downstream health or interaction applications are proposed rather than evaluated. & Leave-one-participant-out mean per-joint position error: 6.17 cm in the twelve-person laboratory study and 14.12 cm in the ten-person semi-in-the-wild study with prescribed actions. & --- \\
\addlinespace[4pt]
\textbf{U-HAR}\newline\cite{bernal2022uhar} & Commercial video-based eye tracking and a head-mounted IMU provide eye/head movement features. & Physical/cognitive activity recognition, producing seven activity labels. & A temporal convolutional encoder--decoder and personalized classifier combine complementary eye/head cues. The collected dataset supports further research; context-dependent display control remains a proposed use. & Twenty participants; three 30-second labeled windows per class adapt the held-out user's classifier, yielding 86.59\% macro-F1. This result includes personal calibration. & --- \\
\addlinespace[4pt]
\textbf{Smartphone and smart-glasses activity analysis}\newline\cite{smSurvey2024har} & Smartphone and smart-glasses data; individual sensor channels are not specified in the available abstract. & Human activity recognition. & Develops separate ensemble/deep-learning models for data from each device, rather than combining their features. & The abstract reports improved recognition performance but does not provide numerical results or enough protocol detail for comparison. & --- \\
\addlinespace[4pt]
\textbf{MunchSonic}\newline\cite{munchsonic2024} & FMCW speakers/microphones on glasses; reduced gain and selected near-face ranges emphasize jaw and hand-to-mouth motion. & Dietary-action logging: six dietary/non-dietary classes, intake counts, and chewing summaries. & A separately trained MobileNetV2 classifier and temporal aggregation recognize dietary actions. Restricting the sensed region helps distinguish fine dietary actions from unrelated body motion. & Twelve participants in 45-minute meal-containing sessions; leave-one-participant-out evaluation achieved 93.5\% macro-F1 on two-second windows. & Adapts ActSonic's hardware design~\cite{chen2024actsonic}; trains a separate dietary-action classifier. \\
\addlinespace[4pt]
\textbf{EchoGuide}\newline\cite{echoguide2024} & Glasses-mounted active acoustic sensing plus synchronized, separately head-mounted GoPro video. & Eating-event records: activity intervals, timestamped descriptions, and retrieved answers. & Acoustic activity predictions select recorded-video intervals; a pretrained captioner and retrieval-based question answering add semantic detail. & Nine participants, 40-minute meal-containing sessions: offline selection reduced caption-record length by 68\% on average versus dense processing while retaining useful summary content; food identity remained difficult. Online camera control and energy savings were not demonstrated. & Reuses the pretrained ActSonic model~\cite{chen2024actsonic} and a LaViLa captioner trained on Ego4D~\cite{grauman2022ego4d}. \\
\end{longtable}

\begin{longtable}{@{}>{\raggedright\arraybackslash}p{\dimexpr.12\textwidth-2\tabcolsep\relax}>{\raggedright\arraybackslash}p{\dimexpr.16\textwidth-2\tabcolsep\relax}>{\raggedright\arraybackslash}p{\dimexpr.15\textwidth-2\tabcolsep\relax}>{\raggedright\arraybackslash}p{\dimexpr.20\textwidth-2\tabcolsep\relax}>{\raggedright\arraybackslash}p{\dimexpr.23\textwidth-2\tabcolsep\relax}>{\raggedright\arraybackslash}p{\dimexpr.14\textwidth\relax}@{}}
\caption{Sensing and perception evidence: gaze and visual attention.}\label{tab:sensing-gaze-evidence}\label{tab:sensing_eye_tracking}\label{tab:l1_gaze_eye}\\
\toprule
\textbf{Study} & \textbf{Sensing modality and setup} & \textbf{Task, output, and application} & \textbf{Key contribution and mechanism} & \textbf{Evaluation and main finding} & \textbf{Demonstrated reuse} \\
\midrule
\endfirsthead
\caption[]{Sensing and perception evidence: gaze and visual attention. (continued)}\\
\toprule
\textbf{Study} & \textbf{Sensing modality and setup} & \textbf{Task, output, and application} & \textbf{Key contribution and mechanism} & \textbf{Evaluation and main finding} & \textbf{Demonstrated reuse} \\
\midrule
\endhead
\midrule
\multicolumn{6}{r}{\footnotesize Continued on the next page}\\
\endfoot
\bottomrule
\endlastfoot
\textbf{Lens-edge photodetector tracker}\newline\cite{li2025lowpower} & Active near-infrared illumination and photodetectors hidden along the lens edges; custom glasses with Bluetooth acquisition. & Gaze angles, five gaze-zone labels, and blink events. & Uses transparent lenses as light guides. Sequential LED illumination and ambient-light subtraction produce eye-reflection features for regression/classification without an eye camera. & Motorized-eye mean errors: 3.79 degrees yaw and 3.63 degrees pitch. Preliminary human five-zone accuracy: 87\% for the same wearer and glasses position; these are different evaluation tasks. & --- \\
\addlinespace[4pt]
\textbf{Nonvisual EOG calibration}\newline\cite{wei2025eog} & Skin EOG electrodes around the eyes, with a mastoid reference; Pupil Core supplies the eyes-open comparison. & Horizontal and vertical eye-direction estimation with eyes open or closed. & Maps eye-rotation-related potential differences to direction using tactile and auditory calibration targets, avoiding the need to see a fixation target. & 11 participants, chin-rest laboratory trials. Offline closed-eye estimates outperform shuffled references against instructed trajectories. Calibration is user/session specific; no camera ground truth is available with eyes closed. & --- \\
\addlinespace[4pt]
\textbf{Event-based eye tracking}\newline\cite{zhao2024eventbased} & Near-eye DAVIS event-camera recordings; a grayscale-frame-derived calibration maps pupil coordinates to a screen. & Pupil-center tracking and screen point-of-regard estimation. & An event-burst predictor activates pupil detection and ellipse-based tracking around eye motion, reducing processing during still-eye intervals. & Public event-eye dataset: 9.31-degree mean angular error with frame-derived calibration. The movement predictor has F1 0.60; its activation performance and the tracker update rate are distinct from gaze accuracy. & --- \\
\addlinespace[4pt]
\textbf{Low-Power Hierarchical Network}\newline\cite{chen2025hierarchical} & Sequential infrared LEDs and frame-mounted photodiodes; robotic eyes provide labels, with separate microcontroller inference tests. & Five gaze-zone classification. & Organizes photodiode features by sensor and eye in a hierarchical network and removes redundant features to reduce model size. & Robot-eye sessions with glasses repositioning: best HNN achieves 98.9\% intra-session and 89.7\% inter-session accuracy. Smaller hierarchical models run slower than the tested dense model on the measured MCUs; these are not cross-user human results. & --- \\
\addlinespace[4pt]
\textbf{Everyday-meal attention study}\newline\cite{ladouce2025dinner} & Video-based gaze tracking, scene video, and head IMU on Tobii Pro Glasses 3, supplemented by body-motion recording. & Gaze eccentricity and object-related fixation measures during meals and tablet use. & Combines gaze and movement measurements with scene-object recognition to relate spatial attention to the task and surrounding objects. & 24 participants with manipulated phone placement. Gaze shifts toward the phone and attention patterns depend on task context; the smartphone-related spatial bias did not change meal duration. & --- \\
\addlinespace[4pt]
\textbf{GazeTrak}\newline\cite{li2024gazetrak} & Active near-eye FMCW acoustic sensing using miniature speakers and microphones mounted on glasses. & Continuous gaze-coordinate estimation. & Maps acoustic echo profiles to gaze through a learned regressor, with wearer/session adaptation. Optimized preprocessing and inference support a real-time processing prototype. & 12-user study: 4.9-degree mean error with session fine-tuning. A separate 10-user final-prototype test reports 3.6 degrees within-session. MCU performance is assessed separately; these figures are not one common deployment condition. & --- \\
\addlinespace[4pt]
\textbf{mmET}\newline\cite{lu2025mmet} & Near-eye 60 GHz FMCW radar modules on prototype glasses; eye images provide training supervision. & Gaze direction with auxiliary eye-geometry estimates. & Learned beamforming and joint geometry/gaze prediction address mounting changes. An eye-image teacher guides radar-feature pretraining while runtime inference uses radar. & Personalized models yield 1.49-degree within-session and 4.47-degree cross-session mean error after remounting. New-user experiments include adaptation; the results do not establish calibration-free tracking for unseen wearers. & --- \\
\addlinespace[4pt]
\textbf{Reading Recognition in the Wild}\newline\cite{min2025reading} & Head-worn gaze, egocentric RGB image crops, and head-motion signals. & Reading/non-reading states and reading-event boundaries. & A flexible transformer fuses behavioral and visual evidence; modality dropout supports unavailable inputs. Detected reading events can provide intervals for selective OCR. & 111-person task-directed naturalistic dataset with spoken boundary labels. Combined accuracy is 86.9\% in Seattle and 82.9\% in Columbus; writing and typing remain difficult cases. This measures reading activity, not comprehension. & --- \\
\addlinespace[4pt]
\textbf{Eye Gaze for Contextual AI}\newline\cite{li2025eyegaze} & Aria Digital Twin egocentric images and supplied prior object-fixation information. & Prediction of the currently fixated or next-interacted object. & Adds structured object-fixation history to vision-language queries, linking visual context with temporal evidence of attention. & Offline ADT evaluation with supplied visible-object candidates: 24.8\% fixation-object and 49.5\% next-object accuracy. Next-object samples precede an interaction; the result is not action recognition or deployed assistance performance. & Uses Aria Digital Twin recordings and annotations~\cite{pan2023ariadigitaltwin} for offline gaze-context evaluation. \\
\addlinespace[4pt]
\textbf{MemX}\newline\cite{liu2021memx} & Inward eye camera and outward scene camera on wearable glasses. & Object-directed attention events and automatic capture of moments of interest. & Eye-motion candidates selectively activate scene analysis; object-aware temporal fusion determines capture. & Prescribed-target laboratory study: 93.65\% average precision. Eleven field pilots compare capture with retrospective interest labels. Reported energy savings and endurance are model-based estimates, rather than measured continuous battery life. & EMOShip adapts its attention-triggering method~\cite{smartglasses_dream_2022}. \\
\addlinespace[4pt]
\textbf{ThermalEye}\newline\cite{thermaleye_2025} & Passive 16-by-12 thermal array positioned near the eye on a glasses frame; an external RGB camera supplies annotations only. & Eye-open/closed probabilities and blink intervals. & Uses blink-related eyelid/eye temperature changes; spatial modeling and temporal filtering turn low-resolution thermal measurements into blink events. & Four participants, fixed forward-facing posture, prescribed blinking and temperature conditions. Table 2 reports overall event-level F1 0.80 in the held-out split. Fatigue assessment and command input are proposed applications. & --- \\
\addlinespace[4pt]
\textbf{Biosensor smart-glasses eye tracker}\newline\cite{biosensor_glasses_2024} & Binocular inward near-infrared cameras and an outward scene camera on a Huawei Vision Glass-based prototype. & Pupil localization and calibrated gaze-point estimation. & Combines pupil segmentation with a six-point geometric/polynomial mapping from eye-image features to gaze in the scene. & Pupil-dataset comparisons and six screen targets at 1 m; the paper reports two-stage gaze RMSE of 1.0 degree. Participant count is not specified for the described gaze experiment, limiting comparison with other trackers. & --- \\
\addlinespace[4pt]
\textbf{EyeTrAES}\newline\cite{sen2024eyetraes} & Near-eye event-camera streams, including a DAVIS346-equipped custom headset. & Pupil trajectories and motion features used for biometric authentication. & Adapts event accumulation to motion, avoiding sparse pupil representations or overlapping motion. Geometric segmentation and filtering recover temporal features; adaptive slicing is also tested with an existing baseline. & Two datasets and a 40-participant authentication study. Adaptive slicing improves fixed-window baselines; mean pupil-segmentation computation is 4.7 ms. This is not a complete gaze-estimation latency or angular-accuracy result. & --- \\
\end{longtable}

\begin{longtable}{@{}>{\raggedright\arraybackslash}p{\dimexpr.12\textwidth-2\tabcolsep\relax}>{\raggedright\arraybackslash}p{\dimexpr.16\textwidth-2\tabcolsep\relax}>{\raggedright\arraybackslash}p{\dimexpr.15\textwidth-2\tabcolsep\relax}>{\raggedright\arraybackslash}p{\dimexpr.20\textwidth-2\tabcolsep\relax}>{\raggedright\arraybackslash}p{\dimexpr.23\textwidth-2\tabcolsep\relax}>{\raggedright\arraybackslash}p{\dimexpr.14\textwidth\relax}@{}}
\caption{Sensing and perception evidence: speech and acoustic scenes.}\label{tab:sensing-speech-evidence}\label{tab:l1_acoustic_speech}\\
\toprule
\textbf{Study} & \textbf{Sensing modality and setup} & \textbf{Task, output, and application} & \textbf{Key contribution and mechanism} & \textbf{Evaluation and main finding} & \textbf{Demonstrated reuse} \\
\midrule
\endfirsthead
\caption[]{Sensing and perception evidence: speech and acoustic scenes. (continued)}\\
\toprule
\textbf{Study} & \textbf{Sensing modality and setup} & \textbf{Task, output, and application} & \textbf{Key contribution and mechanism} & \textbf{Evaluation and main finding} & \textbf{Demonstrated reuse} \\
\midrule
\endhead
\midrule
\multicolumn{6}{r}{\footnotesize Continued on the next page}\\
\endfoot
\bottomrule
\endlastfoot
\textbf{Directional source separation}\newline\cite{zhang2025directional} & Seven-channel speech mixtures simulated for the Project Aria microphone-array geometry. & Conversational speech recognition amid bystanders: separated wearer/partner speech and attributed transcripts. & Directional/neural beamforming supplies spatial information to separation; joint ASR training fuses separated and beamformed signals so the enhancement front end supports recognition. & On simulated smart-glasses mixtures, joint fusion achieved 13.25\% overall WER versus 14.14\% for directional ASR; the comparison concerns this recognition pipeline and dataset. & --- \\
\addlinespace[4pt]
\textbf{Full-duplex directional recognition}\newline\cite{peng2025fullduplex} & Simulated five-microphone glasses recordings with loudspeaker playback and its known reference signal. & Simultaneous speech capture and playback, producing wearer/partner transcripts during device output. & Multichannel acoustic echo cancellation uses the playback reference; AEC-aware directional-ASR training accounts for residual echo and processing distortion. Translation is a motivating use, not an evaluated user benefit. & At simulated $-$5 dB speech-to-echo ratio, AEC-aware STFT-RLS achieved 13.8\% attributed partner WER versus 66.3\% without AEC. & --- \\
\addlinespace[4pt]
\textbf{EchoSpeech}\newline\cite{wang2023echospeech} & Near-face FMCW speakers/microphones observe articulatory tissue motion during silent mouthing. & Silent input: 31 isolated commands and connected sequences of three to six digits. & Differential echoes feed a temporal-preserving CNN and CTC decoder. Phone/Bluetooth demonstrations map recognized commands into application controls; continuous decoding does not imply unrestricted vocabulary. & Twelve seated users; ten personal training sessions (about 30 minutes) yielded 4.5\% WER for isolated commands and 6.1\% for connected digits. & --- \\
\addlinespace[4pt]
\textbf{FoVNet}\newline\cite{xu2024fovnet} & Simulated five-microphone glasses recordings and a specified angular field of view (FoV). & Auditory source selection, producing enhanced speech from within the requested angular region. & FoV-conditioned spatial beams and neural gains avoid requiring each target's precise direction; optional Wiener filtering/post-processing adjusts distortion and quality. Inferring the requested FoV from user intent is not evaluated. & Two-target, three-interferer simulations: base FoVNet achieved 4.26 dB SI-SDR versus $-$1.01 dB for the noisy mixture. & --- \\
\addlinespace[4pt]
\textbf{M-BEST-RQ}\newline\cite{shon2025mbestrq} & Multichannel glasses audio mapped to directional beams using array-specific acoustic transfer information. & Conversational ASR, source localization, and wearer voice-activity detection. & Masked self-supervised pretraining supplies shared acoustic representations for task-specific heads and fine-tuning. Cross-device adaptation is tested on EasyCom; geometry-independent encoder input still requires an appropriate beamforming front end. & MMCSG model C: eight labeled fine-tuning hours after 2.8k hours of real/synthetic pretraining yielded 21.5\% wearer / 29.5\% partner WER. & Uses MMCSG~\cite{vzmolikova2024chime} for conversational ASR and EasyCom for localization and wearer voice-activity detection. \\
\addlinespace[4pt]
\textbf{WearSE}\newline\cite{sun2025wearse} & FMCW probes observe near-face articulatory motion while passive microphones record speech; a wired phone executes enhancement. & Streaming speech enhancement, producing an enhanced waveform for listening or ASR. & Movement-interference suppression improves the sensing channel before acoustic--speech fusion; amplitude/phase enhancement and cached computation support streaming. & Mobile-condition Dataset 2 with generated corruption: 10.14 dB SDR versus 8.02 dB without sensing enhancement. Phone buffering plus processing latency was 130.1 ms; this does not include subsequent communication-network delay. & --- \\
\end{longtable}

\begin{longtable}{@{}>{\raggedright\arraybackslash}p{\dimexpr.12\textwidth-2\tabcolsep\relax}>{\raggedright\arraybackslash}p{\dimexpr.16\textwidth-2\tabcolsep\relax}>{\raggedright\arraybackslash}p{\dimexpr.15\textwidth-2\tabcolsep\relax}>{\raggedright\arraybackslash}p{\dimexpr.20\textwidth-2\tabcolsep\relax}>{\raggedright\arraybackslash}p{\dimexpr.23\textwidth-2\tabcolsep\relax}>{\raggedright\arraybackslash}p{\dimexpr.14\textwidth\relax}@{}}
\caption{Sensing and perception evidence: facial and physiological signals.}\label{tab:sensing-facial-evidence}\\
\toprule
\textbf{Study} & \textbf{Sensing modality and setup} & \textbf{Task, output, and application} & \textbf{Key contribution and mechanism} & \textbf{Evaluation and main finding} & \textbf{Demonstrated reuse} \\
\midrule
\endfirsthead
\caption[]{Sensing and perception evidence: facial and physiological signals. (continued)}\\
\toprule
\textbf{Study} & \textbf{Sensing modality and setup} & \textbf{Task, output, and application} & \textbf{Key contribution and mechanism} & \textbf{Evaluation and main finding} & \textbf{Demonstrated reuse} \\
\midrule
\endhead
\midrule
\multicolumn{6}{r}{\footnotesize Continued on the next page}\\
\endfoot
\bottomrule
\endlastfoot
\textbf{Multichannel facial responses}\newline\cite{kim2021emotion} & Left-eye-region camera, nose/mastoid skin-conductance electrodes, and earlobe-clip optical PPG on a glasses-type wearable. & Local facial images, electrodermal responses, and pulse-related features for affect analysis. & Selects glasses-compatible EDA contact locations and combines local facial appearance with skin-conductance and pulse features. The emotion-classification aspect is covered in Table~\ref{tab:sensing-states-evidence}. & Ten-person placement study compares facial EDA with finger recordings. Nose-to-mastoid placement has the highest mean correlation, r=0.676; subsequent film-viewing recordings support multimodal affect classification. & --- \\
\addlinespace[4pt]
\textbf{Upper facial action recognition}\newline\cite{li2021acoustic} & Active OFDM acoustics from two speakers and four microphones arranged around one side of an eyeglass frame. & Six upper-facial action/neutral categories. & Combines frequency- and link-specific channel-state measurements to reduce fading; Doppler features characterize small facial movements for classification. & 26-person laboratory dataset: average F1 0.92 under held-out-subject pretraining with first-session personalization and second-session testing. Processing is offline; a brief everyday-activity extension identifies running and chewing as difficult conditions. & --- \\
\addlinespace[4pt]
\textbf{EchoExpress}\newline\cite{zhang2025echoexpress} & Active FMCW acoustics using two speakers and two microphones on the glasses frame. & Facial-expression-associated movement categories. & Orthogonal chirps capture complementary facial-motion patterns. Feature attention, non-target-action filtering, and semi-supervised user adaptation support expression classification. & Laboratory, video-elicited, and daily-use studies. Overlapping speech and expressions degrade recognition; daily-use labels include self-reports rather than dense independent annotations. Expression classification is distinct from inferring the wearer's experienced emotion. & --- \\
\addlinespace[4pt]
\textbf{EyeEcho}\newline\cite{li2024eyeecho} & Glasses-mounted FMCW acoustic speakers/microphones; an iPhone TrueDepth camera supplies training/reference facial parameters. & Continuous estimation of 52 facial blendshape coefficients and avatar animation. & Maps differential echo profiles to camera-derived facial geometry; runtime input is acoustic. The implemented phone pipeline transmits estimated coefficients to an avatar. & Tests user-specific, held-out-user, and adapted models, including noise and semi-field conditions. Nine instructed expressions elicit the movements used for evaluation; the output is continuous coefficients, not nine expression labels. & --- \\
\addlinespace[4pt]
\textbf{AirSpecs comfort study}\newline\cite{ramsamy2025sensors} & Frame-mounted environmental sensors, microphone, skin-temperature sensors, and optical blink sensing; phone/watch interfaces display readings. & Personal environmental and bodily readings for comfort awareness and interaction research. & Connects near-body environmental measurements with skin temperature and blinks in situated use. Displayed heart rate comes from HealthKit, not the glasses. & 30 participants across three sites, mostly five-day use, with experience sampling, notification responses, interviews, and co-design. The study does not independently validate a physiological-to-comfort or cognitive-load prediction model. & Builds on the Captivates open-source eyeglass platform~\cite{captivates_2021}. \\
\addlinespace[4pt]
\textbf{EchoBreath}\newline\cite{echobreath_2025} & Passive respiratory audio and active near-face FMCW echoes from a glasses-mounted speaker/microphone, connected to a smartphone. & Six respiratory-behavior categories with non-target rejection. & Fuses sound spectra with echo evidence of wearer activity; Null training and open-set filtering reject unrelated events and help distinguish wearer from bystander sounds. & 25-person laboratory and 10-person semi-wild studies, about 90 hours. User-dependent accuracy is 89.8\% in an apartment, 89.5\% in an office, and 82.8\% in a cafeteria; labels combine participant logs with recording checks. & --- \\
\addlinespace[4pt]
\textbf{Motion-based vital-sign monitoring}\newline\cite{vitalmotion2022} & Built-in Meta Quest accelerometer sensing head/face micro-vibrations; respiration belt and ECG provide references. & Respiration-rate and heartbeat-rate estimation on a related headset. & Removes high-energy movement-contaminated windows, then separates respiratory and cardiac frequency bands and estimates periodicity. & Two-page poster: 20 participants, one minute each of sitting, standing, and seated controller use. Reported average errors are below 3.3\% for respiration and 5.7\% for heartbeat; everyday moving-use coverage on lightweight glasses remains untested. & --- \\
\end{longtable}

\begin{longtable}{@{}>{\raggedright\arraybackslash}p{\dimexpr.12\textwidth-2\tabcolsep\relax}>{\raggedright\arraybackslash}p{\dimexpr.16\textwidth-2\tabcolsep\relax}>{\raggedright\arraybackslash}p{\dimexpr.15\textwidth-2\tabcolsep\relax}>{\raggedright\arraybackslash}p{\dimexpr.20\textwidth-2\tabcolsep\relax}>{\raggedright\arraybackslash}p{\dimexpr.23\textwidth-2\tabcolsep\relax}>{\raggedright\arraybackslash}p{\dimexpr.14\textwidth\relax}@{}}
\caption{Sensing and perception evidence: cognitive and affective states.}\label{tab:sensing-states-evidence}\label{tab:sensing_attention_emotion}\label{tab:l1_affect_emotion}\label{tab:l1_cognitive_attention}\\
\toprule
\textbf{Study} & \textbf{Sensing modality and setup} & \textbf{Task, output, and application} & \textbf{Key contribution and mechanism} & \textbf{Evaluation and main finding} & \textbf{Demonstrated reuse} \\
\midrule
\endfirsthead
\caption[]{Sensing and perception evidence: cognitive and affective states. (continued)}\\
\toprule
\textbf{Study} & \textbf{Sensing modality and setup} & \textbf{Task, output, and application} & \textbf{Key contribution and mechanism} & \textbf{Evaluation and main finding} & \textbf{Demonstrated reuse} \\
\midrule
\endhead
\midrule
\multicolumn{6}{r}{\footnotesize Continued on the next page}\\
\endfoot
\bottomrule
\endlastfoot
\textbf{EMOShip}\newline\cite{smartglasses_dream_2022} & Inward eye images, pupil size and gaze, with outward scene images captured by eyewear. & Emotion-aware moment capture and review: emotion categories, attended-scene descriptions, and model-derived associations between them. & Attention and non-neutral-expression gates select scene processing; learned fusion associates visual semantics with eye/pupil features. & Film-elicited, self-labelled responses and an additional unseen-user test. Reported macro binary accuracy across six non-neutral categories is 80.2\%; scene cues sometimes confuse negative categories. Associations are model explanations, not identified emotional causes. & Adapts EMO's eye-feature extraction~\cite{guo2020emo} and MemX's attention-triggering method~\cite{liu2021memx}. \\
\addlinespace[4pt]
\textbf{EMO}\newline\cite{guo2020emo} & Infrared single-eye images, including periocular appearance, on resource-constrained eyewear computing platforms. & Continuous expression/emotion recognition, producing seven category labels. & Combines a learned eye-feature extractor with per-user reference classification. Stable-frame forwarding and sampling avoid repeated inference. & Film-elicited responses labelled by participants; each test wearer supplies a 70-second initialization. The optimized configuration reports 72.2\% accuracy at 12.8 fps. Runtime evaluation excludes camera acquisition cost. & EMOShip adapts its eye-feature extraction method~\cite{smartglasses_dream_2022}. \\
\addlinespace[4pt]
\textbf{EmoGlass}\newline\cite{lim2022emoglass} & Glasses-mounted camera imaging part of the wearer's face, with embedded inference and a companion phone app. & Emotional self-awareness: seven facial-expression categories presented as emotion reports, contextual diaries, and positive-activity retrieval. & Progressively adapts a per-user local-face classifier across controlled lighting, varied lighting, and elicited expressions; connects recognition with a usable self-monitoring and reflection interface. & 15 participants in a three-day field study: 73.0\% mean accuracy against sparse self-reports. Three participants with low recognition accuracy were excluded from subsequent qualitative analysis. & --- \\
\addlinespace[4pt]
\textbf{Multi-Channel Facial Responses}\newline\cite{kim2021emotion} & Local facial images, facial electrodermal activity (EDA), and ear photoplethysmography (PPG); finger measurements provide physiological references. & Affect inference and illustrative emotion logging: arousal/valence categories and four affective quadrants. & Fuses facial-image features with physiological features over 100-second windows to combine expression-related appearance with autonomic responses. & 20 analyzed participants, stimulus-derived labels, and leave-one-trial-out evaluation. Fusion improves mean classification with participant-dependent gains. & --- \\
\addlinespace[4pt]
\textbf{CASES / EYEReader}\newline\cite{huang2023cases} & Inward eye camera for gaze and outward scene camera for text, with PC/server processing and computer-displayed interventions. & English reading assistance: word/sentence processing difficulty and sentence-level mind wandering trigger explanations, simplification, or rereading reminders. & Aligns temporal gaze features with words and sentences, then combines visual attention with semantic and linguistic features to select targeted reading interventions. & A 25-person self-annotated reading dataset shows gains over gaze-only and text-only baselines. A 13-person pilot reports fewer difficult items after assistance; this measures reported reading difficulty rather than long-term learning gains. & --- \\
\addlinespace[4pt]
\textbf{AR-Supported Wayfinding Load}\newline\cite{xu2026cognitivewayfinding} & HoloLens 2 with an added infrared pupil camera for AR calibration; a separate HTC VIVE Pro/Tobii setup for VR wayfinding. & Workload measurement for wayfinding: stabilized pupil measurements and a pupil-derived workload indicator. & Pupil-to-iris correction reduces motion-related measurement variation; luminance and individual-baseline calibration address non-cognitive effects. A separate VR experiment relates pupil dilation to navigation performance. & AR corridor stability trials and a separate 30-firefighter VR task reporting r = -0.397. The fitted workload threshold is task-specific; integrated workload-adaptive AR guidance is proposed rather than evaluated. & --- \\
\addlinespace[4pt]
\textbf{Heads-Up Multitasker}\newline\cite{iqbal2024headsup} & Behavioral simulation using modeled visual/task state and human-study benchmarks; no new runtime sensing device or state recognizer. & OHMD interface design: predicted attention allocation, reading resumption, and walking behavior during multitasking. & A hierarchical behavioral model coordinates reading, visual scanning, eye movements, and walking under memory uncertainty and attention-switching costs, enabling exploration of display layouts. & Calibration and validation against human data reproduce several aggregate behavioral trends. Results for additional display layouts are model predictions. & --- \\
\addlinespace[4pt]
\textbf{SkillSight}\newline\cite{li2026skillsight} & Gaze and head-motion features at inference; richer egocentric video or body-motion information is available to the training teacher. & Skill assessment, producing proficiency categories from activity recordings. & Distills task-relevant information from the richer teacher into a gaze-plus-head-motion student, reducing inference inputs while retaining action context relevant to proficiency. & Subject-separated testing against expert proficiency labels: 44.4\% student accuracy on Ego-Exo4D. The reported 14--73-fold power reduction is estimated against selected video-based methods, not measured battery life. Fine motor actions remain difficult; teaching feedback is prospective. & Uses the Ego-Exo4D proficiency benchmark~\cite{grauman2024egoexo4d} and two additional sports datasets. \\
\addlinespace[4pt]
\textbf{HQS-AR}\newline\cite{song2025holistic} & Motion, visual, audio, behavioral, and physiological datasets; these inputs should not be read as a single glasses-integrated sensor configuration. & User-state estimation and office-scenario classification for personalized AR. & Organizes user modeling around physical, cognitive-emotional, social, and behavioral dimensions, linking inferred states to proposed interface configurations and user routines. & Publisher-reported hold-out validation: average accuracy of 88\% for state estimation and 99\% for scenario classification. These results concern the modeled office conditions. & --- \\
\end{longtable}

\begin{longtable}{@{}>{\raggedright\arraybackslash}p{\dimexpr.12\textwidth-2\tabcolsep\relax}>{\raggedright\arraybackslash}p{\dimexpr.16\textwidth-2\tabcolsep\relax}>{\raggedright\arraybackslash}p{\dimexpr.15\textwidth-2\tabcolsep\relax}>{\raggedright\arraybackslash}p{\dimexpr.20\textwidth-2\tabcolsep\relax}>{\raggedright\arraybackslash}p{\dimexpr.23\textwidth-2\tabcolsep\relax}>{\raggedright\arraybackslash}p{\dimexpr.14\textwidth\relax}@{}}
\caption{Sensing and perception evidence: scene and spatial perception.}\label{tab:sensing-scene-evidence}\\
\toprule
\textbf{Study} & \textbf{Sensing modality and setup} & \textbf{Task, output, and application} & \textbf{Key contribution and mechanism} & \textbf{Evaluation and main finding} & \textbf{Demonstrated reuse} \\
\midrule
\endfirsthead
\caption[]{Sensing and perception evidence: scene and spatial perception. (continued)}\\
\toprule
\textbf{Study} & \textbf{Sensing modality and setup} & \textbf{Task, output, and application} & \textbf{Key contribution and mechanism} & \textbf{Evaluation and main finding} & \textbf{Demonstrated reuse} \\
\midrule
\endhead
\midrule
\multicolumn{6}{r}{\footnotesize Continued on the next page}\\
\endfoot
\bottomrule
\endlastfoot
\textbf{GazePointAR}\newline\cite{zhao2024gazepointar} & HoloLens 2 scene images, eye gaze, finger pointing, and spoken queries; spatial coordinates register user input to the image. & Contextual voice assistance: scene entities and text associated with expressions such as `this' or `that'. & Combines object detection and OCR in an object/text hierarchy, then projects gaze or pointing coordinates into the image to construct a referent description for a language-model query. & 12-person exploratory laboratory study and one-researcher five-day diary. Recognition coverage, single-moment gaze capture, processing delay, and missing historical referents constrain which queries the implemented assistant can answer. & Customizes MRTK gaze/pointing and spatial-awareness interfaces~\cite{mrtk2}. \\
\addlinespace[4pt]
\textbf{IFFA}\newline\cite{formfill_2025} & Vuzix M4000 scene camera streams frames to a backend; printed forms carry ArUco markers and have prepared layouts. & Form-filling assistance for blind users: fingertip position relative to text and writing fields drives spoken reading/writing guidance. & Hand segmentation and fingertip localization combine with marker-based PnP registration to place the finger in form coordinates, enabling assistance without unrestricted document understanding. & Four blind participants completed practice and three forms. 63.2\% of responses started in the intended field and 43.6\% remained fully inside it; participants also reported effort maintaining the form in view. & --- \\
\addlinespace[4pt]
\textbf{Visual-auditory E-Glasses navigation}\newline\cite{multisensory_nav_2023} & RealSense RGB-D camera/IMU, microphone and headphones, with a wearable computing client and remote server; navigation uses prepared indoor maps. & Object-directed indoor navigation: requested target identity and 3D position support route planning and spoken guidance. & Fuses spoken object-name features with visual detection; depth locates the target and ORB-SLAM2 registers the wearer within an existing map. & 25 object-search trials report 95.46\% detection success and 1.74-second mean processing/transfer response. Route demonstrations use sighted participants, including a blindfolded test; detection success is not an overall navigation or safety measure. & --- \\
\addlinespace[4pt]
\textbf{Aria Digital Twin}\newline\cite{pan2023ariadigitaltwin} & Aria RGB/monochrome cameras, eye tracking, and IMUs, aligned with digitized scenes/objects and external motion capture in instrumented spaces. & Scene-perception development and benchmarking: reference geometry, object poses, depth, and instance annotations. & Common coordinates, calibrated sensors, and synchronized motion capture yield registered per-frame reference annotations. & 200 sequences in two spaces; annotation-pipeline validation and object/3D benchmarks expose egocentric-domain and depth-estimation difficulties. Accurate supplied scene state is a research resource, not demonstrated online recognition in arbitrary environments. & Provides data for fixation/next-object inference in Eye Gaze for Contextual AI~\cite{li2025eyegaze}. \\
\addlinespace[4pt]
\textbf{HoloAssist}\newline\cite{wang2023holoassist} & HoloLens 2 RGB, depth, gaze, 3D hand/head poses, audio, and IMU streams, synchronized and registered to a world anchor. & Assistance-relevant perception benchmarks: action recognition, mistake detection, intervention-type prediction, and hand forecasting. & Pairs aligned user/world observations with manual action, mistake, and intervention annotations from procedural tasks, supporting comparison of which modalities help each prediction problem. & 166 hours from 222 participants. Fusion benefits depend on task and pretraining; guidance during collection is human-provided. Benchmarks evaluate predictions from recorded data, rather than a deployed autonomous assistant. & --- \\
\end{longtable}

\endgroup

\renewcommand{\arraystretch}{1.08}
\renewcommand\tabularxcolumn[1]{m{#1}}
\clearpage
\subsection{Interaction and Assistance Evidence}
\label{app:interaction-evidence}
\begingroup
Tables~\ref{tab:interaction-input-evidence}--\ref{tab:interaction-timing-evidence} summarize the studies discussed in Section~\ref{sec:interaction}. They identify each study's inputs and setup; task, output, and application; contribution through its interaction mechanism, user control, and feedback; evaluation findings; and demonstrated reuse. Results refer to the tasks and conditions tested in each study. When a study appears in more than one table, each entry describes a different contribution, and reuse links are recorded with the most relevant entry. The reuse column highlights documented use of methods, models, platforms, software, or datasets across studies; --- indicates that no specific link is recorded here.\par
\small
\setlength{\tabcolsep}{3pt}
\renewcommand{\arraystretch}{1.10}
\setlength{\LTpre}{8pt}
\setlength{\LTpost}{8pt}
\begin{longtable}{@{}>{\raggedright\arraybackslash}p{\dimexpr.12\textwidth-2\tabcolsep\relax}>{\raggedright\arraybackslash}p{\dimexpr.15\textwidth-2\tabcolsep\relax}>{\raggedright\arraybackslash}p{\dimexpr.14\textwidth-2\tabcolsep\relax}>{\raggedright\arraybackslash}p{\dimexpr.25\textwidth-2\tabcolsep\relax}>{\raggedright\arraybackslash}p{\dimexpr.21\textwidth-2\tabcolsep\relax}>{\raggedright\arraybackslash}p{\dimexpr.13\textwidth\relax}@{}}
\caption{Interaction and assistance evidence: user input and control.}\label{tab:interaction-input-evidence}\label{tab:interaction_handsfree}\label{tab:interaction_input_text}\label{tab:l2_gesture_handsfree}\label{tab:l2_touch_text}\\
\toprule
\textbf{Study} & \textbf{Inputs and setup} & \textbf{Task, output, and application} & \textbf{Mechanism, control, and feedback} & \textbf{Evaluation and main finding} & \textbf{Demonstrated reuse} \\
\midrule
\endfirsthead
\caption[]{Interaction and assistance evidence: user input and control. (continued)}\\
\toprule
\textbf{Study} & \textbf{Inputs and setup} & \textbf{Task, output, and application} & \textbf{Mechanism, control, and feedback} & \textbf{Evaluation and main finding} & \textbf{Demonstrated reuse} \\
\midrule
\endhead
\midrule
\multicolumn{6}{r}{\footnotesize Continued on the next page}\\
\endfoot
\bottomrule
\endlastfoot
\textbf{IR eye gestures}\newline\cite{kaneko2021irarray} & Deliberate eye trajectories; sixteen inward-facing infrared sensors on eyewear. & Eyes-only command input: directional and compound eye-gesture symbols; media mappings are proposed. & Eye movement deforms eyelid/periocular tissue; temporal sensor patterns are matched with dynamic time warping and classified as gesture symbols.\newline\emph{Control and feedback:} Evaluates directional and compound gestures; media-command mappings are proposed rather than tested. & Five seated participants followed marked targets without blinking. Within-person validation reached F1 .94 for twenty gestures; continuous unintended-activation rejection was not evaluated. & --- \\
\addlinespace[4pt]
\textbf{Laser gaze gestures}\newline\cite{aktas2021lfigaze} & Eye movements and blinks; two laser-feedback channels and external laboratory instruments. & Gaze-gesture input: directional eye-motion and blink symbols. & Laser self-mixing yields velocity/distance features; threshold and direction rules identify atomic eye-motion symbols.\newline\emph{Control and feedback:} Directional movements and blink form a potential command vocabulary; no application feedback loop is evaluated. & Two participants with head support; sample-level mean F1 .9344. The study demonstrates symbol recognition, not discrimination of intentional commands during ordinary looking. & --- \\
\addlinespace[4pt]
\textbf{Face Commands}\newline\cite{McDuff2020facecommands} & Eight facial/eye gestures; photo-reflective sensor array and IMU. & Hands-free input: eight facial/eye gesture events and elicited command mappings. & Facial deformation changes reflected intensity; change detection segments neutral-to-action-to-neutral movements, followed by classification and temporal smoothing.\newline\emph{Control and feedback:} Elicited gesture--command associations; technical evaluation recognizes gestures rather than a complete application interface. & Thirteen participants with user-specific retraining: F1 .911. A separate one-hour exploratory test used one experimenter and two gestures; it does not establish whole-vocabulary everyday performance. & Dataset and gesture-segmentation method reused by the few-shot facial-gesture study~\cite{peng2024fewshot}. \\
\addlinespace[4pt]
\textbf{Few-shot facial gestures}\newline\cite{peng2024fewshot} & Eight facial gestures; photo-reflective/IMU data, reusing Face Commands data and adding participants. & Low-calibration facial input: personalized labels for eight presegmented gestures. & Augmentation and few-shot calibration adapt gesture classification to a new wearer and wearing condition.\newline\emph{Control and feedback:} Reduces onboarding examples; continuous intent detection and command feedback are outside the evaluated classifier. & Twenty-two participants across datasets; four examples per class yielded 91.38\% classification accuracy. Inputs were presegmented, so this result does not measure false triggers in continuous use. & Reuses Face Commands data and gesture-segmentation preprocessing~\cite{McDuff2020facecommands}; adds new wearers and calibration data. \\
\addlinespace[4pt]
\textbf{Helios}\newline\cite{xu2024helios} & In-air swipes and pinches; downward-facing event camera with external computation. & In-air command input: swipe and double-pinch events, with background classes. & Temporally decaying event representations support motion localization/classification; action duration determines how much evidence falls within the recognition window.\newline\emph{Control and feedback:} Media-control concepts motivate the gestures; background/no-hand classes distinguish some non-command activity. & Twenty-person live testing produced gesture-dependent recall, including 54\% for double pinch and 86\% for left swipe. The separate 91\% synthetic-sequence validation is not live command accuracy. & --- \\
\addlinespace[4pt]
\textbf{OCOsense}\newline\cite{schneegass2023ocosense} & Facial and head actions; noncontact optical skin-displacement sensing with calibration. & Hands-free application control: facial/head actions mapped to media, calling, and game operations. & Optomyographic signals are interpreted as facial/eye/head actions and assigned to interface operations.\newline\emph{Control and feedback:} Demonstrations map smiles, eyebrow movements, nods, and tilts to media, calling, and game controls; tablet visualization shows signals. & Demonstration paper with implemented interaction examples; no quantified recognition or usability study is reported. Accessibility and broader monitoring uses remain proposed extensions. & --- \\
\addlinespace[4pt]
\textbf{ThumbAir}\newline\cite{gupta2023thumbair} & Bimanual thumb motion tracked by an Oculus Quest in controlled laboratory lighting. & In-air text entry: ambiguous thumb-key sequences decoded into words. & Four reachable virtual targets per hand encode ambiguous character groups; key selection commits when the thumb leaves a virtual surface.\newline\emph{Control and feedback:} Highlighted selections and predicted text; index-finger actions provide space and deletion. & Fourteen-person typing study reached 13.73 WPM in final phrase trials. Test-word frequencies were adjusted to guarantee correct full-sequence predictions; unrestricted candidate selection was not evaluated. & Uses Oculus Quest SDK hand tracking to obtain thumb motion and define virtual input targets. \\
\addlinespace[4pt]
\textbf{EyeGesener}\newline\cite{li2024eyegesener} & Deliberate out-and-back eye gestures; two speakers and two microphones on glasses. & Eyes-only media control: eight directional commands, with blink/null recognition. & Coded acoustic probing captures periocular changes; paired motion peaks segment eight directional gestures, with blink/null classes filtering non-command activity.\newline\emph{Control and feedback:} Directions operate a media interface; repeating a toggle reverses it. A monitor simulates the AR display. & Eight-person live study: ten-class F1 .89 and SUS 86.5. Task-focused execution and accidental commands still affected use despite explicit non-gesture handling. & --- \\
\addlinespace[4pt]
\textbf{CapGlasses}\newline\cite{yang2021capglasses} & Facial/head gestures; four capacitive channels, including a transparent lens electrode. & Hands-free gesture input: twelve facial/head/default-state labels. & A generated nearby electric field supports untethered sensing; gesture-dependent body--electrode coupling changes become classification features.\newline\emph{Control and feedback:} Twelve classes include the default state; application mappings and interactive feedback are proposed uses. & Ten participants, user-dependent ten-fold validation: 89.6\% accuracy. Transfer between wearers and rejection of natural behavior remain unresolved; large and subtle movements require different sensitivity. & --- \\
\addlinespace[4pt]
\textbf{EarWiggl'In}\newline\cite{kim2025earwigglin} & Voluntary ear movements; two resistive pressure sensors in nose pads. & Ear-based input: a discrete trigger; richer pressure patterns are explored. & Ear movement pulls the frame, changing nose-pad pressure; timing, magnitude, and side produce distinguishable signal patterns.\newline\emph{Control and feedback:} Evaluates a simple trigger responding to a visual cue; richer patterns and gaze-plus-ear confirmation are exploratory. & Five preselected ear-wigglers: mean trigger response 378.48 ms, slower than the frame button. The eight explored patterns are not an evaluated eight-command classifier. & --- \\
\addlinespace[4pt]
\textbf{MagFace}\newline\cite{zhang2026magface} & Facial/head gestures during cycling; magnetic silicone inserts inside frame pads and magnetometers. & Cycling interaction: facial/head gestures mapped to media and call commands. & Facial contact deforms the inserts; a gesture gate, classifier, temporal votes, and lockout convert magnetic changes into discrete commands.\newline\emph{Control and feedback:} Cycling-oriented media/call operations with audio prompts; null detections release the lockout between commands. & Fifteen-person urban cycling study after fine-tuning: 85.9\% command accuracy, including missed response windows. Across fourteen hours, 4.07 false activations/hour remained, often from traffic-related head turns. & --- \\
\addlinespace[4pt]
\textbf{FlickPose}\newline\cite{yuasa2025flickpose} & Two-hand text entry using built-in HMD hand tracking. & Japanese text entry: hand-pose/key composition with explicit word-conversion candidates. & Left-hand pose chooses a character within a key; the right hand selects the key by touch/release or ray pointing plus pinch.\newline\emph{Control and feedback:} Linked keyboard/panel highlights, explicit Japanese conversion candidates, deletion, and mode switching. & Ten participants typed while standing. Raycasting produced fewer deletions, but the standard keyboard was faster overall. Automatic panel following was implemented; typing benefits while walking were not measured. & Uses Oculus Integration hand tracking and extends NMeCab's mecab-ipadic dictionary for word conversion. \\
\addlinespace[4pt]
\textbf{FingerGlass}\newline\cite{ye2025fingerglass} & Finger contact on a temple-mounted capacitive fingerprint sensor; external PC processing. & Display-free text entry and application control: finger identity and gesture determine characters or commands. & Fingerprint translation, contact shifts, and rotation identify gestures; enrolled finger identity combines with gesture type to expand command and character mappings.\newline\emph{Control and feedback:} Audio announces entered characters; explicit case/numeric modes support display-free text entry and correction. & Twelve-person study: 95.42--97.60\% command accuracy. Text entry rose from 9.61 to 12.72 WPM after a one-week follow-up with training; rapid repeated-phrase performance is a separate condition. & --- \\
\addlinespace[4pt]
\textbf{RimSense}\newline\cite{xu2024rimsense} & Touch gestures on the rim; active piezoelectric excitation/measurement with external audio equipment. & Rim-based photo/ebook control: gesture events with duration-dependent operation variants. & Contact changes the frame's vibration response. Temporal recognition and a finite-state machine produce bounded gesture events; duration adds operation variants.\newline\emph{Control and feedback:} Taps, slides, presses, and zoom gestures control photo/ebook interfaces, shown on a monitor simulating AR. & Thirty-person held-out-user evaluation: event F1 .95; fourteen-person application study: F1 .88. Accidental adjustment touches and mobile operation were not evaluated. & --- \\
\addlinespace[4pt]
\textbf{StretchAR}\newline\cite{zhang2022stretchar} & One-handed touch/stretch on wearable straps at four body locations; HoloLens 2 display. & Wearable AR control: discrete touch choices and continuous stretch values. & Separate capacitive and resistive layers measure contact and stretch; finger count, scrolling, and pull amount compose discrete and continuous controls.\newline\emph{Control and feedback:} Partial release adjusts the value; complete release commits. Visual targets/menus guide control. & Fifteen-person indoor study after calibration: wrist/forearm were more convenient than head placement for stretching. Perfect stretch-task accuracy included target tolerance; twenty-eight configurations are observed interaction variants, not classifier classes. & --- \\
\addlinespace[4pt]
\textbf{Early rim touch panel}\newline\cite{sun2022rimsense} & Taps at five rim locations; active piezoelectric probe and receiver. & Rim touch input: five contact-location labels and a null state. & Finger loading changes the complex frequency response of the frame; temporal features distinguish contact locations and null.\newline\emph{Control and feedback:} Produces location symbols that could support commands; no application mapping or feedback task is evaluated. & Ten participants, leave-one-subject-out evaluation: F1 .91. This establishes a touch-location capability with an external audio interface, preceding richer gesture/event interaction in RimSense. & --- \\
\addlinespace[4pt]
\textbf{Gaze robot control}\newline\cite{robot_gaze_iros_2023} & HoloLens 2 eye tracking, registered virtual robot, and real mobile robot. & Hands-free robot navigation: gaze-derived velocity commands. & Gaze displacement from the virtual robot becomes a virtual force; an admittance controller converts it to velocity, with a no-motion dead zone.\newline\emph{Control and feedback:} Continuous gaze-directed steering; registered AR robot and control display provide visual reference. & Ten-person obstacle-navigation comparison with joystick and three damping settings. The fastest gaze setting was not the most comfortable; joystick was easier on a single-ease measure, while SUS did not differ significantly. & Modifies ARETT for real-time gaze access; uses MRTK eye-tracking interfaces~\cite{mrtk2}. \\
\addlinespace[4pt]
\textbf{GlassMessaging}\newline\cite{li2023glassmessaging} & OHMD messaging through speech, ring-mouse buttons, and mid-air gestures. & Messaging while multitasking: dictated/typed messages and contact/chat navigation. & Voice shortcuts and interchangeable manual actions navigate contacts and dictate/type messages; explicit send and hide commands manage interaction state.\newline\emph{Control and feedback:} Visible transcription/chat history, auditory action cues, incoming alerts, and an on-demand display. & Sixteen-person comparison during everyday tasks used prescribed messages. GlassMessaging enabled faster entry and shorter response times, but lower accuracy and greater workload than phone messaging; free composition was not tested. & Uses MRTK gesture, voice, virtual-keyboard, and stabilization functions~\cite{mrtk2}. \\
\addlinespace[4pt]
\textbf{NasoVoce}\newline\cite{wu2026nasovoce} & Normal/whispered speech; synchronized microphone and vibration sensor at the nasal pad. & Low-audibility speech input: enhanced speech waveforms for listening or recognition. & Air- and skin/bone-conducted signals feed a dual-input enhancement model; recovered audio supports listening or existing speech recognition.\newline\emph{Control and feedback:} Supports intentional low-audibility input; adaptive mode selection and an integrated conversational control loop remain future work. & Held-out speech with added noise and fifty-person listening evaluation support enhancement over noisy microphone input. Severe noise and weak whisper vibration limit fusion; streaming smartphone operation was not demonstrated. & Reuses frozen Whisper encoder/decoder for ASR-guided enhancement training and recognition evaluation~\cite{radford2023whisper}. \\
\addlinespace[4pt]
\textbf{Subtle interactions in everyday settings}\newline\cite{orlosky2021ubiquitous} & Vuzix Blade with foot, arm, jaw-teeth, and thumb-index gestures; Wizard-of-Oz sensing in the comparison, working ring mouse in the follow-up. & Input-method selection for hands-busy use: call responses, menu selections, and scripted SMS replies. & Compares bodily-action demands and primary-task disruption while holding the interface tasks constant.\newline\emph{Control and feedback:} Laboratory actions use a triple-action delimiter. Visual prompts have experimenter-managed feedback in the comparison and live display updates in the follow-up. & Sixteen-person comparison across presenting, folding clothes, and carrying bags: 14/16 preferred thumb-index overall. Eight two-hour field sessions exposed context-dependent access, comfort, and grip constraints. & --- \\
\end{longtable}

\begin{longtable}{@{}>{\raggedright\arraybackslash}p{\dimexpr.12\textwidth-2\tabcolsep\relax}>{\raggedright\arraybackslash}p{\dimexpr.15\textwidth-2\tabcolsep\relax}>{\raggedright\arraybackslash}p{\dimexpr.14\textwidth-2\tabcolsep\relax}>{\raggedright\arraybackslash}p{\dimexpr.25\textwidth-2\tabcolsep\relax}>{\raggedright\arraybackslash}p{\dimexpr.21\textwidth-2\tabcolsep\relax}>{\raggedright\arraybackslash}p{\dimexpr.13\textwidth\relax}@{}}
\caption{Interaction and assistance evidence: target and intent grounding.}\label{tab:interaction-grounding-evidence}\label{tab:interaction_object_grounding}\label{tab:l2_object_grounding}\\
\toprule
\textbf{Study} & \textbf{Inputs and setup} & \textbf{Task, output, and application} & \textbf{Mechanism, control, and feedback} & \textbf{Evaluation and main finding} & \textbf{Demonstrated reuse} \\
\midrule
\endfirsthead
\caption[]{Interaction and assistance evidence: target and intent grounding. (continued)}\\
\toprule
\textbf{Study} & \textbf{Inputs and setup} & \textbf{Task, output, and application} & \textbf{Mechanism, control, and feedback} & \textbf{Evaluation and main finding} & \textbf{Demonstrated reuse} \\
\midrule
\endhead
\midrule
\multicolumn{6}{r}{\footnotesize Continued on the next page}\\
\endfoot
\bottomrule
\endlastfoot
\textbf{GazeHand\-Sync}\newline\cite{park2025gazehandsync} & Virtual targets; gaze and hand-pinch timing on Quest Pro. & Gaze-and-pinch object selection: corrected attribution of a selection action to its intended virtual target. & Detects late pinches from gaze--hand coordination features and reassigns selection to the preceding fixation, rather than the object gazed at when the pinch arrives.\newline\emph{Control and feedback:} Pinch commits selection; gaze highlights the candidate. Correction is inferred from recent interaction history. & 20 participants, controlled target grid. Applying the final model to the full recorded dataset reduced coordination errors from 12.10\% to 5.39\%; this was offline correction, not a new online study. & --- \\
\addlinespace[4pt]
\textbf{PinchCatcher}\newline\cite{10.1145/3706598.3713530} & Groups of virtual objects; gaze, semi-pinch, and alternative hand or dwell triggers. & Serial selection of virtual-object groups for subsequent manipulation. & Separates maintaining multi-selection mode from adding individual gazed targets. Semi-pinch maintains the group state; full pinch concludes grouping and enables manipulation.\newline\emph{Control and feedback:} Dwell, swipe, tilt, or other-hand pinch toggles membership; highlights and movement indicators expose selection state. & 30 participants, Quest Pro, groups of 2--6 marked targets. Semi-pinch methods reduced accidental subselections, but final group errors varied; completion time did not differ significantly across techniques. & Eyes on Many adapts its semi-pinch mode-switching design~\cite{bashar2026eyes}. \\
\addlinespace[4pt]
\textbf{Eyes on Many}\newline\cite{bashar2026eyes} & Groups of virtual objects; gaze with hand gestures, dwell, or voice. & XR multi-object selection: group membership and selection-mode state under alternative confirmation techniques. & Combines four ways of maintaining selection mode with three ways of confirming each gazed target. Persistent toggles avoid the need to sustain a hand posture throughout grouping.\newline\emph{Control and feedback:} Visual mode and item-state indicators; each item is confirmed by dwell, pinch, or a short spoken command. & 30 participants, Quest Pro, 6--10 targets in a flat grid. Persistent modes performed better under the tested recovery rules; repeated voice subselection added latency and was disliked. & Adapts PinchCatcher's semi-pinch mode-switching design~\cite{10.1145/3706598.3713530}, with different hand roles and recovery rules. \\
\addlinespace[4pt]
\textbf{G-VOILA}\newline\cite{zhang2024gvoila} & Objects relevant to spoken questions; gaze, egocentric images, and speech. & Contextual information querying: reconstructed spoken questions and answers about attended objects. & Selects frames using fixation duration and proximity to query onset; gaze-guided segmentation and scene descriptions help reconstruct the query's intended referents.\newline\emph{Control and feedback:} Users initiate queries; reconstructed queries and answers are displayed on a computer for review. & A 21-person enactment informed a separate 16-person evaluation. Response-object recall reached 89.1\%; spatial and temporal gaze cues helped ambiguous queries. Processing used after-round transfers and manually annotated query boundaries. & Uses pretrained Whisper~\cite{radford2023whisper} to transcribe queries before gaze-guided reference resolution. \\
\addlinespace[4pt]
\textbf{GazePointAR}\newline\cite{zhao2024gazepointar} & Physical objects and text; gaze, hand pointing, speech, and recent dialogue. & Scene-grounded voice assistance: object/text referents associated with contextual query answers. & Projects gaze and pointing into a scene photograph, associates them with object detections and nested text, and supplies referents and conversation history to the language model.\newline\emph{Control and feedback:} Wake phrase initiates interaction; an indicator shows estimated gaze, and responses are displayed and spoken. Users can ask follow-up questions. & HoloLens 2: 12-person exploratory laboratory study and a five-day researcher diary. Demonstrated contextual queries; single-moment gaze capture and missing object categories constrained answers. & Customizes MRTK gaze/pointing and spatial-awareness interfaces~\cite{mrtk2}. \\
\addlinespace[4pt]
\textbf{Gazeify Then Voiceify}\newline\cite{li2026gazeifyvoiceify} & Physical-object masks; gaze, selection trigger, and spoken corrections. & Displayless physical-object selection: a provisional object mask that users can verify and correct through speech. & Clustered gaze samples prompt initial segmentation. Object properties and spatial, ordinal, or part--whole relations in corrective speech filter candidate masks and update the selection.\newline\emph{Control and feedback:} The system describes its selected object aloud; users can identify mistakes and request correction. & 23 participants using Quest Pro as a displayless-glasses proxy, with controller triggers. Initial mask success was 53\%; 58.4\% of trials entering voice disambiguation produced a correct mask within two turns. & Uses pretrained EfficientSAM to turn clustered gaze into an initial object mask. \\
\addlinespace[4pt]
\textbf{Squiggle}\newline\cite{yu2025squiggle} & Objects, parts, and groups; gaze, camera images, and ring-sensed pinch and hand motion. & Selecting real-world objects, parts, and groups: an image mask specifying the intended extent. & Gaze places an invisible lasso in the scene; duration-dependent fusion of gaze and drawing trajectories prompts image segmentation, expressing the intended target's extent.\newline\emph{Control and feedback:} Thumb contact starts the stroke; release submits it. No feedback appears during drawing; masks are reviewed on a desktop afterward. & 12 participants using Aria glasses and a ring. Lasso improved mask overlap for compound objects and groups, while requiring longer input than gaze-plus-pinch. & Fine-tunes EfficientSAM's prompt encoder and mask decoder while retaining its image encoder. \\
\addlinespace[4pt]
\textbf{BLEselect}\newline\cite{chen2023bleselect} & BLE-tagged devices; head nods or hand gestures, glasses-mounted RF array, and wrist transmitter. & IoT device selection: the identity of the tagged device indicated by a head or hand action. & Estimates device and wrist directions, recognizes gestures, and matches their angular evidence to a candidate device identity using distance scores or gesture-specific classifiers.\newline\emph{Control and feedback:} Nodding, pointing dwell, or encircling supplies the selection action; hand-down events delimit hand-gesture processing. & 22 participants selected six tagged objects per room while facing targets. Accuracy was 94.8--97.8\% across gestures. Encircling was slower; devices sharing a direction but differing in depth remain ambiguous. & --- \\
\addlinespace[4pt]
\textbf{AR robot assembly}\newline\cite{robot_assembly_2024} & Physical assembly parts; gaze or hand-ray pointing with pinch confirmation on HoloLens 2. & Robot-assisted assembly: confirmed part selection invokes a predefined robot pick-and-place action. & QR registration establishes a shared robot--headset frame. Selecting a part binds the user's pointing to a predefined robot pick-and-place action.\newline\emph{Control and feedback:} The candidate is highlighted; one pinch opens a confirmation prompt and a second confirms before robot movement. & 16-person comparative assembly study following formative refinement. Gaze reduced total part-selection time relative to hand rays; usability did not significantly differ. Overall assembly time also included manual work and robot motion. & Builds the AR interaction on MRTK~\cite{mrtk2} and links the headset to robot control through ROS. \\
\end{longtable}

\begin{longtable}{@{}>{\raggedright\arraybackslash}p{\dimexpr.12\textwidth-2\tabcolsep\relax}>{\raggedright\arraybackslash}p{\dimexpr.15\textwidth-2\tabcolsep\relax}>{\raggedright\arraybackslash}p{\dimexpr.14\textwidth-2\tabcolsep\relax}>{\raggedright\arraybackslash}p{\dimexpr.25\textwidth-2\tabcolsep\relax}>{\raggedright\arraybackslash}p{\dimexpr.21\textwidth-2\tabcolsep\relax}>{\raggedright\arraybackslash}p{\dimexpr.13\textwidth\relax}@{}}
\caption{Interaction and assistance evidence: information presentation and feedback.}\label{tab:interaction-presentation-evidence}\\
\toprule
\textbf{Study} & \textbf{Inputs and setup} & \textbf{Task, output, and application} & \textbf{Mechanism, control, and feedback} & \textbf{Evaluation and main finding} & \textbf{Demonstrated reuse} \\
\midrule
\endfirsthead
\caption[]{Interaction and assistance evidence: information presentation and feedback. (continued)}\\
\toprule
\textbf{Study} & \textbf{Inputs and setup} & \textbf{Task, output, and application} & \textbf{Mechanism, control, and feedback} & \textbf{Evaluation and main finding} & \textbf{Demonstrated reuse} \\
\midrule
\endhead
\midrule
\multicolumn{6}{r}{\footnotesize Continued on the next page}\\
\endfoot
\bottomrule
\endlastfoot
\textbf{Icon-augmented notifications}\newline\cite{smith2023icons} & Calendar notifications on an optical head-mounted display during a primary task. & Calendar notification reading during another task: concise icon/text representations of event information. & Familiar icons encode event meanings; text or numbers retain secondary details. Experiments separate text reduction, icon familiarity, and encoding density.\newline\emph{Control and feedback:} Users interpret notifications while performing vigilance, navigation, or browsing tasks; icon selection can reflect personal familiarity. & Four studies (16, 12, 24, and 12 participants). Benefits depend on familiarity and encoded meaning; replacing equally concise text with researcher-selected icons alone yields no significant advantage. & --- \\
\addlinespace[4pt]
\textbf{Paracentral and near-peripheral visualization}\newline\cite{tu2022paracentral} & Progress information around a conversation partner's face, using an optical head-mounted display. & Progress monitoring during conversation: glanceable indicators positioned around the conversation partner. & Circular and linear graphics encode progress through shape; numerical text supplies precise values. The paired-conversation study uses intermittent updates.\newline\emph{Control and feedback:} Wearers estimate progress while maintaining conversation; the visualization requires no manual input. & Two studies with 12 wearers each, first simulated and then with trained conversation partners. Circular graphics support progress estimation and attention maintenance; results concern these encodings and controlled conversation settings. & Extended into ParaGlassMenu~\cite{tu2023paraglassmenu}; peripheral layout adopted by GazeNoter~\cite{li2025gazenoter}. \\
\addlinespace[4pt]
\textbf{ParaGlassMenu}\newline\cite{tu2023paraglassmenu} & Radial optical-display menu paired with a ring mouse for IoT control during conversation. & IoT control during conversation: an interactive radial menu showing options and continuous device values. & Extends paracentral and near-peripheral visualization. Icons/text surround the central target; circular indicators show continuous values.\newline\emph{Control and feedback:} Ring buttons activate and select; scrolling changes values. The hidden-by-default menu highlights selections and supports returning to previous levels. & Comparative interface and paired-conversation studies support reduced disruption alongside IoT control. More complex searches still compete with conversation; the prototype's four buttons constrain each menu level. & Extends paracentral visualization~\cite{tu2022paracentral}; its layout is adopted by PANDALens and GazeNoter~\cite{qian2024pandalens,li2025gazenoter}. \\
\addlinespace[4pt]
\textbf{PeriphAR}\newline\cite{10.1145/3772318.3791637} & A Quest Pro simulates a small monocular peripheral display showing a proxy of the gaze-targeted object. & Gaze-based object selection: a peripheral visual proxy of the current target before confirmation. & Color quantization/enhancement distinguishes the candidate against its surroundings or a similarly colored neighboring object.\newline\emph{Control and feedback:} The proxy makes the current candidate inspectable; a controller button commits the choice. & Controlled studies use virtual targets. Color proxies are faster than shape proxies in Study 1; Study 2 (12 participants) finds no significant overall accuracy/time differences between enhancement strategies. Real-object tests are separate demonstrations. & --- \\
\addlinespace[4pt]
\textbf{VisiMark}\newline\cite{visimark_2025} & HoloLens 2 landmark augmentations in a pre-scanned environment; a QR code establishes route alignment. & Low-vision indoor navigation: route overviews and in-situ landmark labels. & Intersection signboards preview hallway structures and upcoming landmarks; in-situ icons/text identify nearby landmarks. Labels face the wearer and permit size/color customization.\newline\emph{Control and feedback:} Users inspect overview and local cues while navigating; landmark numbering distinguishes repeated types. & A 16-person low-vision evaluation finds perceived navigation benefits but no significant difference in route-retracing time or correctness. Pre-scanning isolates presentation design from online landmark-recognition errors. & Uses MRTK 3~\cite{mrtk} and HoloLens spatial-mapping/QR APIs to implement registered landmark displays. \\
\addlinespace[4pt]
\textbf{LSVP}\newline\cite{10.1145/3448118} & Educational videos on Epson BT-300 optical displays during indoor walking tasks. & Learning while walking: educational video content arranged for interrupted viewing and later recall. & Sequential disclosure introduces content with narration; persistence retains earlier material. Transparent backgrounds, contrast, and legible typography support see-through viewing.\newline\emph{Control and feedback:} Users can return attention to retained content after checking the environment; playback is preloaded rather than adapted by live sensing. & Two controlled studies with 16 participants each examine video design and display platform. Benefits concern recognition/recall, including seven-day retention, and walking measures; hazard detection and higher-order learning are not established. & --- \\
\addlinespace[4pt]
\textbf{GlassMail: draft review}\newline\cite{li2024glassmail} & Near-eye presentation of voice-generated email drafts on an OHMD with ring control. & Email draft review: visually distinguished edits with optional audio, supporting voice-led revision. & Revised content is highlighted while unchanged context is faded. This retains orientation without giving all text equal visual prominence.\newline\emph{Control and feedback:} Users inspect changes, request further voice edits, and toggle audio; ring input initiates voice interaction. & A Wizard-of-Oz output-design study compares default, faded-context, and changes-only displays. The selected design enters a 12-person prototype study. Findings concern draft review and whole-interface use, not autonomous email authoring. & --- \\
\addlinespace[4pt]
\textbf{GazeNoter: on-demand suggestions}\newline\cite{li2025gazenoter} & Suggested keywords and sentences in a Quest Pro AR view, paired with a custom ring. & Note taking during speeches and walking meetings: on-demand keyword and sentence candidates. & Suggestions become visible during ring touch and disappear on release, allowing users to control their visual presence.\newline\emph{Control and feedback:} Gaze identifies a candidate and a ring button confirms it. A quick-note mode records selected keywords without waiting for full sentences. & Two 12-person studies examine speeches and walking meetings. Evidence concerns note-taking behavior and perceived fit to intent; it does not isolate display visibility as the cause of all system benefits. & Adopts the peripheral, center-clear layout of prior OHMD designs~\cite{tu2022paracentral,tu2023paraglassmenu}. \\
\addlinespace[4pt]
\textbf{IFFA}\newline\cite{formfill_2025} & Vuzix M4000 camera with server processing; prepared forms carry ArUco markers for fingertip/form registration. & Form filling by blind users: spoken text, field identification, and fingertip-position guidance. & Audio reads nearby text and fields, indicates page centering, and guides the fingertip to a writing location. Repeated feedback is suppressed.\newline\emph{Control and feedback:} Spoken commands switch reading/writing modes; finger movement selects a line or field and drives directional cues. & Four-person preliminary evaluation with blind users. Writing starts in the intended field in 63.2\% of cases, but only 43.6\% remains fully within it, motivating feedback during writing. & --- \\
\addlinespace[4pt]
\textbf{LLM-Glasses}\newline\cite{llmglasses_2026} & Camera-based scene interpretation and cloud reasoning supply commands to temple-mounted tactile actuators. & Navigation feedback: tactile directional/event patterns derived from guidance commands. & Taps and sliding contact encode direction and events. A structured scene representation prioritizes nearby obstacles before command generation.\newline\emph{Control and feedback:} Users follow tactile cues after intensity/contact calibration; pattern duration affects responsiveness. & Separate studies evaluate pattern recognition (13 participants), motion-capture-guided paths (6), and video/LLM decisions. These establish component-level guidance evidence, not integrated autonomous navigation with blind participants in everyday environments. & --- \\
\addlinespace[4pt]
\textbf{Mindful Moments}\newline\cite{10.1145/3563657.3596030} & Nreal Light, headphones, and a tethered iPad deliver a breathing-guidance video during walking. & Mindfulness during walking: synchronized visual/audio breathing-rhythm guidance. & Abstract expanding/contracting shapes synchronize with sustained and cue sounds to communicate a breathing rhythm.\newline\emph{Control and feedback:} Participants initiate sessions in field use and follow the rhythm; the presentation is predefined, not adapted from measured breathing. & An 18-person controlled study and 15-person field study examine mindfulness use. Field reports describe visuals establishing/recovering rhythm and audio maintaining it while users attend to their surroundings. & --- \\
\end{longtable}

\begin{longtable}{@{}>{\raggedright\arraybackslash}p{\dimexpr.12\textwidth-2\tabcolsep\relax}>{\raggedright\arraybackslash}p{\dimexpr.15\textwidth-2\tabcolsep\relax}>{\raggedright\arraybackslash}p{\dimexpr.14\textwidth-2\tabcolsep\relax}>{\raggedright\arraybackslash}p{\dimexpr.25\textwidth-2\tabcolsep\relax}>{\raggedright\arraybackslash}p{\dimexpr.21\textwidth-2\tabcolsep\relax}>{\raggedright\arraybackslash}p{\dimexpr.13\textwidth\relax}@{}}
\caption{Interaction and assistance evidence: assistance generation and personalization.}\label{tab:interaction-generation-evidence}\label{tab:interaction_ai_assistance}\label{tab:l2_proactive_ai_1}\\
\toprule
\textbf{Study} & \textbf{Inputs and setup} & \textbf{Task, output, and application} & \textbf{Mechanism, control, and feedback} & \textbf{Evaluation and main finding} & \textbf{Demonstrated reuse} \\
\midrule
\endfirsthead
\caption[]{Interaction and assistance evidence: assistance generation and personalization. (continued)}\\
\toprule
\textbf{Study} & \textbf{Inputs and setup} & \textbf{Task, output, and application} & \textbf{Mechanism, control, and feedback} & \textbf{Evaluation and main finding} & \textbf{Demonstrated reuse} \\
\midrule
\endhead
\midrule
\multicolumn{6}{r}{\footnotesize Continued on the next page}\\
\endfoot
\bottomrule
\endlastfoot
\textbf{AiGet}\newline\cite{li2025aiget} & Gaze-annotated scene frames, optional speech, location, profile, and interaction history; XREAL Air, Pupil Core, laptop, and cloud models. & Informal learning: prioritized contextual knowledge items and answers to follow-up questions. & Context analysis identifies potential learning interests; generation ranks knowledge by novelty, preference alignment, utility, and unexpectedness, filtering repetition and limiting simultaneous items.\newline\emph{Control and feedback:} Ring controls follow-up queries, dismissal, audio, and system activation. & A 12-person comparison favored contextual prioritization over simpler pipelines; adding a profile alone was insufficient. An 18-person field study examined perceived knowledge value. & Follows peripheral-presentation guidance from ParaGlassMenu and paracentral visualization work~\cite{tu2023paraglassmenu,tu2022paracentral}. \\
\addlinespace[4pt]
\textbf{CRAFT}\newline\cite{bai2026craft} & First-person images, location, spoken observations, and evolving fiction context; XREAL Air with camera, ring, laptop, and cloud models. & Situated fiction writing: characters, settings, plot ideas, and drafts assembled from accumulated observations. & Transforms observations into fictional characters, settings, and plots; preserves story context across sessions and organizes accumulated material for subsequent draft generation.\newline\emph{Control and feedback:} Writers set and revise story context, request transformations, dismiss suggestions, and edit drafts. & Eight writers completed 24 supported field sessions. Findings concern perceived creative support and emerging practices; literary quality was not independently compared against a baseline. & --- \\
\addlinespace[4pt]
\textbf{Egocentric Co-Pilot}\newline\cite{wang2026egopilot} & Timestamped video descriptions and speech transcripts; smart-glasses front end connected to cloud or on-premises processing. & Egocentric assistance: context-aware answers and tool-mediated actions, including board-game guidance. & Selects recent temporal evidence and compresses longer histories for query answering. Tool orchestration combines perception with specialized reasoning, including a symbolic board-game engine.\newline\emph{Control and feedback:} Explicit queries, clarification dialogue, and conversational interruption guide responses. & Egolife and HD-EPIC benchmarks and ablations support temporal context management. Human evaluation used blinded ratings of prerecorded interaction logs, rather than live sessions across devices. & Fine-tunes on egocentric data including Ego4D~\cite{grauman2022ego4d}; evaluates on EgoLife and HD-EPIC. \\
\addlinespace[4pt]
\textbf{GazeNoter}\newline\cite{li2025gazenoter} & Recent speech transcripts, selected keywords, and predefined personal keywords; Quest Pro with eye tracking and a confirmation ring. & Speech and meeting note-taking: selected keywords and user-confirmed sentence notes. & Extracts context keywords, derives related alternatives, and composes candidate sentences. Iterative keyword selection constrains generation to the note the user intends to record.\newline\emph{Control and feedback:} Gaze selects; ring confirms. Users can save quick keyword notes or refine notes later. & Speech and indoor walking-meeting studies, each with 12 participants, found favorable intent-match ratings and more detailed notes compared with smartphone methods, with time--detail tradeoffs. & Uses pretrained Whisper transcription~\cite{radford2023whisper} to supply text for keyword extraction. \\
\addlinespace[4pt]
\textbf{GlassMail}\newline\cite{li2024glassmail} & Spoken email requirements and editing requests; Nreal Light, ring control, speech recognition, and an LLM-backed interface. & Mobile email composition: clarified requirements, generated drafts, and voice-requested revisions. & Extracts requirements for optional clarification before drafting, then generates revisions. Personal style is pursued through user editing; learning a persistent style model remains a proposed extension.\newline\emph{Control and feedback:} Users inspect requirements, clarify, request edits, and review highlighted changes with optional audio. & Controlled design studies informed the interface; a 12-person walking observation study demonstrated composition feasibility while revealing persistent difficulties with personal style and precise edits. & --- \\
\addlinespace[4pt]
\textbf{LifeInsight}\newline\cite{zhao2025lifeinsight} & Camera snapshot and spoken visual query; primarily collar-worn or handheld camera/microphone, with Raspberry Pi and cloud-model support. & Visual assistance for blind and low-vision users: concise spoken answers to targeted visual questions. & A multimodal model answers the user's specific question; a task prompt requests brief, immediately relevant spoken information for blind and low-vision users.\newline\emph{Control and feedback:} A button initiates querying; users can reframe questions, follow up, or reposition the camera. & Thirteen blind and low-vision participants completed six structured scenarios; 63 of 78 experiences succeeded. Camera coverage and visual interpretation affected responses. & --- \\
\addlinespace[4pt]
\textbf{PANDALens}\newline\cite{qian2024pandalens} & Confirmed moments combine speech, image descriptions, objects, OCR, time, location, and audio context; XREAL Air, Pupil Core, ring, and laptop. & Travel documentation: selected moments, contextual follow-up questions, and personalized written narratives. & Contextual questions elicit personal detail; summarized session history and user-selected moments feed narrative generation. Configured question and writing preferences shape the output.\newline\emph{Control and feedback:} Users select moments, revise style, and review or edit the resulting narrative. & A 16-person museum comparison with phone-based LiveSnippets found higher self-rated writing quality and lower post-editing effort/time; individual component effects were not isolated. & --- \\
\addlinespace[4pt]
\textbf{SocialMind}\newline\cite{yang2025socialmind} & Conversation, partner nonverbal cues, social factors, and personas inferred from previous conversations; RayNEO X2 with local and server processing. & Live conversational assistance: brief social suggestions and example utterances conditioned on context and personas. & Combines static communication guidelines with current cues and identity-linked persona retrieval; generates concise suggestions and example sentences relevant to the ongoing exchange.\newline\emph{Control and feedback:} Users decide whether to follow suggestions and may explicitly specify social goals beforehand. & Dataset comparisons used model-judged suggestion-quality measures; a separate 20-person live-use study collected perceived usefulness and acceptance. These do not establish clinical or long-term social outcomes. & Uses DailyDialog, Synthetic-Persona-Chat, and SocialDial to construct dataset-based conversation evaluations. \\
\end{longtable}
\begin{longtable}{@{}>{\raggedright\arraybackslash}p{\dimexpr.12\textwidth-2\tabcolsep\relax}>{\raggedright\arraybackslash}p{\dimexpr.15\textwidth-2\tabcolsep\relax}>{\raggedright\arraybackslash}p{\dimexpr.14\textwidth-2\tabcolsep\relax}>{\raggedright\arraybackslash}p{\dimexpr.25\textwidth-2\tabcolsep\relax}>{\raggedright\arraybackslash}p{\dimexpr.21\textwidth-2\tabcolsep\relax}>{\raggedright\arraybackslash}p{\dimexpr.13\textwidth\relax}@{}}
\caption{Interaction and assistance evidence: assistance initiation and timing.}\label{tab:interaction-timing-evidence}\label{tab:l2_proactive_ai_2}\\
\toprule
\textbf{Study} & \textbf{Inputs and setup} & \textbf{Task, output, and application} & \textbf{Mechanism, control, and feedback} & \textbf{Evaluation and main finding} & \textbf{Demonstrated reuse} \\
\midrule
\endfirsthead
\caption[]{Interaction and assistance evidence: assistance initiation and timing. (continued)}\\
\toprule
\textbf{Study} & \textbf{Inputs and setup} & \textbf{Task, output, and application} & \textbf{Mechanism, control, and feedback} & \textbf{Evaluation and main finding} & \textbf{Demonstrated reuse} \\
\midrule
\endhead
\midrule
\multicolumn{6}{r}{\footnotesize Continued on the next page}\\
\endfoot
\bottomrule
\endlastfoot
\textbf{AiGet}\newline\cite{li2025aiget} & Scene change, sustained gaze fixation, or an explicit ring-initiated verbal query can request contextually relevant knowledge. & Knowledge discovery: event-triggered information offers alongside user-initiated queries. & Scene-triggered requests require a minimum interval; knowledge selection suppresses repetitive or low-value content. General interruptibility prediction is proposed future work, not an implemented trigger.\newline\emph{Control and feedback:} Users cancel the current display, mute audio, pause the system, or ask follow-up questions. & Eighteen participants used the system in low-demand activities, including repeated sessions for six. Findings support perceived relevance and manageable distraction in those settings. & --- \\
\addlinespace[4pt]
\textbf{CRAFT}\newline\cite{bai2026craft} & Environmental observations are matched to the current fictional project; writers can also initiate photo- and voice-based transformations. & Fiction writing: proactive inspiration offers and explicitly requested transformations of observations. & A periodic loop filters visually repetitive contexts and semantically redundant suggestions. The refined probe adds model-based contextual interruptibility checks; proactive suggestions appear briefly without audio.\newline\emph{Control and feedback:} Ring controls dismissal and explicit capture; writers retain control over story direction and generated material. & Eight writers used the refined probe across 24 supported sessions. Qualitative findings describe helpful opportunities and interruption concerns; timing mechanisms were not separately evaluated. & Applies peripheral-positioning guidance from PANDALens, ParaGlassMenu, and paracentral visualization work~\cite{qian2024pandalens,tu2023paraglassmenu,tu2022paracentral}. \\
\addlinespace[4pt]
\textbf{PANDALens}\newline\cite{qian2024pandalens} & Interest cues include gaze fixation, approaching an object, positive speech, and preferred object or sound categories; manual capture remains available. & Travel documentation: capture/comment invitations, confirmed recordings, and deferred or dismissed suggestions. & Peripheral visual invitations precede confirmed capture/commenting. Similar invitations are rate-limited, and some gaze-triggered suggestions wait for travel transitions; audio follows confirmed interest.\newline\emph{Control and feedback:} Users confirm verbally, ignore or dismiss invitations, mute feedback, and capture independently. & A 16-person museum comparison found more captured moments with fewer manual initiations. Whole-system evidence does not isolate the contribution of individual triggers or deferral rules. & Adopts ParaGlassMenu's attention-maintaining menu design for user-initiated capture~\cite{tu2023paraglassmenu}. \\
\addlinespace[4pt]
\textbf{ProMemAssist}\newline\cite{zhou2025prememassist} & Camera-derived objects and speech transcripts update a computational working-memory representation of recent context; updates can produce candidate messages. & Proactive task assistance: deliver, defer, or discard decisions for candidate spoken messages. & A separate predictor balances candidate importance/relevance against estimated memory displacement and interference, then delivers, defers for reconsideration, or discards the message.\newline\emph{Control and feedback:} Users can ignore suggestions; adjusting assistance frequency or scope through feedback was not implemented. & Twelve participants performed tabletop tasks. Compared with a prompt-based baseline, delivery was more selective and frustration lower; task time and timing ratings did not significantly differ. & --- \\
\addlinespace[4pt]
\textbf{Satori}\newline\cite{shen2025satori} & Scene, object, and action history inform a belief--desire--intention representation; task plans supply checkpoints for detecting progress. & Procedural task guidance: anticipated next-step support released around confirmed task transitions. & Forecasts and caches upcoming guidance while monitoring step completion. On completion, it retrieves the prediction and asks for confirmation before continuing with task-relevant content.\newline\emph{Control and feedback:} Users confirm goals or steps through voice or pinch; an explicit wake phrase supports requests. & Sixteen participants completed four predefined tasks on HoloLens 2. Several subjective guidance measures were non-inferior to designer-authored Wizard-of-Oz assistance; field of view and latency constrained flow. & Uses GTEA and EgoTaskQA to evaluate the task-state and action-prediction components supporting guidance. \\
\addlinespace[4pt]
\textbf{SocialMind}\newline\cite{yang2025socialmind} & Partial partner utterances, social context, and conversational history provide opportunities to prepare suggestions during an ongoing exchange. & Live conversational assistance: prepared suggestions and selectively refreshed display updates. & Retrieves suitable cached suggestions or starts model inference before an utterance ends. Minimum refresh intervals and semantic-change checks limit repeated display updates.\newline\emph{Control and feedback:} Users choose whether to use displayed suggestions; manual activation can limit continuous operation. & Prototype measurements demonstrate the cache/inference latency distinction; 20-person live-use feedback indicates generally acceptable responsiveness. This does not establish a causal behavioral engagement gain over baselines. & --- \\
\end{longtable}

\endgroup

\renewcommand{\arraystretch}{1.08}
\renewcommand\tabularxcolumn[1]{m{#1}}

\clearpage
\subsection{Application Evidence}
\label{app:applications-evidence}
\begingroup
Tables~\ref{tab:applications-everyday-evidence}--\ref{tab:applications-robot-evidence} summarize the applications discussed in Section~\ref{sec:applications} and relevant enabling studies. They record intended users and tasks, platform capabilities, contributions and workflows, evaluation findings, and demonstrated reuse. Results refer to the participants, tasks, and conditions examined in each study. The reuse column identifies documented adoption of platforms, software, models, data, or designs; --- indicates that no specific link is recorded here.\par
\small
\setlength{\tabcolsep}{3pt}
\renewcommand{\arraystretch}{1.10}
\setlength{\LTpre}{8pt}
\setlength{\LTpost}{8pt}
\begin{longtable}{@{}>{\raggedright\arraybackslash}p{\dimexpr.12\textwidth-2\tabcolsep\relax}>{\raggedright\arraybackslash}p{\dimexpr.14\textwidth-2\tabcolsep\relax}>{\raggedright\arraybackslash}p{\dimexpr.16\textwidth-2\tabcolsep\relax}>{\raggedright\arraybackslash}p{\dimexpr.21\textwidth-2\tabcolsep\relax}>{\raggedright\arraybackslash}p{\dimexpr.24\textwidth-2\tabcolsep\relax}>{\raggedright\arraybackslash}p{\dimexpr.13\textwidth\relax}@{}}
\caption{Application evidence: everyday assistance, learning, and productivity.}\label{tab:applications-everyday-evidence}\label{tab:daily_life}\label{tab:context_aware}\label{tab:education}\\
\toprule
\rowcolor{gray!15}
\textbf{Study} & \textbf{Users, task, and context} & \textbf{Platform and capabilities} & \textbf{Contribution and workflow} & \textbf{Evaluation and main finding} & \textbf{Demonstrated reuse} \\
\midrule
\endfirsthead
\caption[]{Application evidence: everyday assistance, learning, and productivity. (continued)}\\
\toprule
\rowcolor{gray!15}
\textbf{Study} & \textbf{Users, task, and context} & \textbf{Platform and capabilities} & \textbf{Contribution and workflow} & \textbf{Evaluation and main finding} & \textbf{Demonstrated reuse} \\
\midrule
\endhead
\midrule
\multicolumn{6}{r}{\footnotesize Continued on the next page}\\
\endfoot
\bottomrule
\endlastfoot
\textbf{GlassMessaging}\newline\cite{li2023glassmessaging} & Everyday messaging while walking, preparing food, eating, and cleaning. & HoloLens 2; speech, ring mouse, hand input, and visual/audio feedback. & Voice shortcuts and interchangeable manual input support contact selection, dictated messages, chat navigation, and explicit sending. & Sixteen participants re-entered prescribed messages. Faster entry and responses, more messaging opportunities, but lower accuracy and higher workload than phone messaging; no overall platform effect on primary-task time. & MRTK gesture, voice, keyboard, and stabilization functions~\cite{mrtk2}. \\
\addlinespace[4pt]
\rowcolor{gray!5}
\textbf{GlassMail}\newline\cite{li2024glassmail} & Composing and editing personal emails while walking. & Nreal glasses, ring mouse, speech recognition, GPT-3.5, and visual/audio output. & Intent clarification precedes generation; highlighted changes with faded context help users inspect and refine drafts. & Controlled dialog/display studies followed by twelve-person observation with participant-provided scenarios. Formal emails were feasible; precise edits and personal style could require additional effort. & GPT-3.5 for requirement extraction, drafting, and revision. \\
\addlinespace[4pt]
\textbf{GazeNoter}\newline\cite{li2025gazenoter} & Taking personally relevant notes during speeches and walking meetings. & AR headset, gaze and ring selection, speech transcription, and LLM suggestions. & Users select keywords and candidate sentences, confirm notes, and reveal content on demand while listening. & Two studies of twelve participants each compared phone typing, the same AI on a phone, and AR. More detailed notes and better perceived intent match; time per note could increase. Walking meetings occurred in a small room. & Whisper transcription; ParaGlassMenu presentation design~\cite{tu2023paraglassmenu}. \\
\addlinespace[4pt]
\rowcolor{gray!5}
\textbf{MagFace}\newline\cite{zhang2026magface} & Media and call control while cycling. & Magnetic inserts/sensors on eyewear pads; facial gestures and audio prompts. & Movement-related magnetic patterns become commands through gesture gating, voting, and a lockout until neutral posture. & Fifteen-person urban cycling study after adaptation: 85.9\% command accuracy, including missed response windows. Fourteen hours yielded 4.07 false activations/hour, often from traffic-related head turns. & --- \\
\addlinespace[4pt]
\textbf{PANDALens}\newline\cite{qian2024pandalens} & Capturing travel experiences and writing a personal narrative afterward. & XREAL Air, Pupil Core, ring mouse, laptop, multimodal context, and GPT-3.5. & Optional capture invitations and follow-up questions build a history of selected moments; users select material and style for the final narrative. & Sixteen-person museum comparison with phone-based LiveSnippets. More captured moments/comments, less measured post-editing time, and higher self-rated writing quality; whole-system comparison. & Adopts ParaGlassMenu peripheral-menu design~\cite{tu2023paraglassmenu}. \\
\addlinespace[4pt]
\rowcolor{gray!5}
\textbf{CRAFT}\newline\cite{bai2026craft} & Developing fiction from observations during everyday outings. & XREAL Air, Pupil Core, ring mouse, laptop, and Gemini models. & Recorded observations and writer-defined story context support character/plot ideas and later drafts across sessions. & Eight writers completed 24 supported field sessions over one week. Logs and interviews show incorporation of observations and perceived creative support; no baseline writing comparison or independent literary-quality assessment. & Gemini models; peripheral presentation draws on PANDALens and ParaGlassMenu~\cite{qian2024pandalens,tu2023paraglassmenu}. \\
\addlinespace[4pt]
\textbf{AiGet}\newline\cite{li2025aiget} & Discovering relevant knowledge during walking, shopping, and museum visits. & XREAL Air, Pupil Core, ring mouse, laptop, gaze/scene context, and Gemini. & Interests, contextual interpretation, prioritization rules, and history guide knowledge offers; users can query, dismiss, or mute. & Twelve-person pipeline comparison and eighteen-person field study, forty sessions. Contextual prioritization improves perceived knowledge desirability; field findings concern learning opportunities, recalled items, and experience. & Gemini models; sentence embeddings for history retrieval and repetition filtering. \\
\addlinespace[4pt]
\rowcolor{gray!5}
\textbf{LSVP}\newline\cite{10.1145/3448118} & Learning from video while walking. & Optical head-mounted display; transparent video with sequential and persistent content. & Information stays available after appearing so learners can return to it after attending to their surroundings. & Two sixteen-person controlled experiments. Platform, video design, and indoor mobility task affect recall/recognition and walking speed; the second study includes seven-day delayed tests. & Adapts Khan Academy educational videos to the LSVP presentation style. \\
\addlinespace[4pt]
\textbf{Satori}\newline\cite{shen2025satori} & Step guidance for flowers, game-console setup, cleaning, and coffee preparation. & HoloLens 2, scene/action interpretation, server/cloud models, and multimodal feedback. & Task plans and progress observations support anticipated next-step guidance, with confirmation before advancing. & Sixteen participants compared four predefined tasks with designer-authored Wizard-of-Oz guidance. Subjective content/timing measures support feasibility; missed hand actions caused errors and repeated confirmation. & Uses BLIP-2/GPT-4V and GTEA/EgoTaskQA for module evaluation. \\
\addlinespace[4pt]
\rowcolor{gray!5}
\textbf{Egocentric Co-Pilot}\newline\cite{wang2026egopilot} & Information lookup, reminders, visual questions, and tool-based task advice. & Smart-glasses camera/audio, egocentric models, event history, cloud reasoning, and web/local tools. & Queries and temporally organized observations guide tool selection and argument generation; symbolic tools support specialized advice. & Egocentric QA benchmarks, implemented task trials, and ratings of recorded interaction logs. Human comparative ratings are offline, rather than live wearer deployment. & Ego4D training data; EgoLife and HD-EPIC evaluation data; existing models and tool interfaces. \\
\addlinespace[4pt]
\textbf{ProMemAssist}\newline\cite{zhou2025prememassist} & Receiving assistance during interrupted tabletop activities. & Camera and microphone-array glasses stream to laptop; laptop supplies spoken output. & An operational memory model balances relevance against interference to deliver, defer, or discard candidate messages. & Twelve participants completed four tasks with scripted interjections. More selective delivery, a higher proportion of positive responses, and lower frustration; no significant task-time or rated-timeliness advantage. & CLIP representations and language models for context and assistance generation. \\
\addlinespace[4pt]
\end{longtable}

\begin{longtable}{@{}>{\raggedright\arraybackslash}p{\dimexpr.12\textwidth-2\tabcolsep\relax}>{\raggedright\arraybackslash}p{\dimexpr.14\textwidth-2\tabcolsep\relax}>{\raggedright\arraybackslash}p{\dimexpr.16\textwidth-2\tabcolsep\relax}>{\raggedright\arraybackslash}p{\dimexpr.21\textwidth-2\tabcolsep\relax}>{\raggedright\arraybackslash}p{\dimexpr.24\textwidth-2\tabcolsep\relax}>{\raggedright\arraybackslash}p{\dimexpr.13\textwidth\relax}@{}}
\caption{Application evidence: communication, accessibility, and independent living.}\label{tab:applications-accessibility-evidence}\label{tab:comm}\label{tab:vision}\label{tab:navigation}\label{tab:l3_comm_accessibility}\label{tab:l3_visual_assistance}\\
\toprule
\rowcolor{gray!15}
\textbf{Study} & \textbf{Users, task, and context} & \textbf{Platform and capabilities} & \textbf{Contribution and workflow} & \textbf{Evaluation and main finding} & \textbf{Demonstrated reuse} \\
\midrule
\endfirsthead
\caption[]{Application evidence: communication, accessibility, and independent living. (continued)}\\
\toprule
\rowcolor{gray!15}
\textbf{Study} & \textbf{Users, task, and context} & \textbf{Platform and capabilities} & \textbf{Contribution and workflow} & \textbf{Evaluation and main finding} & \textbf{Demonstrated reuse} \\
\midrule
\endhead
\midrule
\multicolumn{6}{r}{\footnotesize Continued on the next page}\\
\endfoot
\bottomrule
\endlastfoot
\textbf{Sign-to-911}\newline\cite{sign911_2023} & ASL users communicating with emergency services through two-way ASL--English translation. & INMO AIR glasses and phone; first-person signing video, gyroscope compensation, sign animations and audio. & Emergency-dialogue context improves sign interpretation; compact language processing and streamed sign animations support a glasses--phone translation workflow. & Six signers supplied emergency-dialogue data; twelve rated experience against text/video-relay approaches. Context increased word accuracy from 84.49\% to 91.37\%; no actual emergency calls were made. & Glasses--phone processing design later informed SocialMind, as reported in the Assistive AR overview~\cite{assistive_ar_2025}. \\
\addlinespace[4pt]
\rowcolor{gray!5}
\textbf{SignGlass}\newline\cite{signglass_2025} & Deaf ASL users; translating manual and non-manual signing into English. & Custom three-camera eyewear with external Raspberry Pi units, power bank, and phone inference. & Captures hands and facial cues together, allowing grammatical information from head and face movements to contribute to sentence translation. & Fourteen native Deaf signers re-signed selected sentences in a controlled room. Leave-one-user-out word error rate was 16.98\% with manual/non-manual cues versus 31.17\% with manual cues only. Reverse translation is future work. & --- \\
\addlinespace[4pt]
\textbf{SiHear}\newline\cite{sihear_2025} & Deaf/hard-of-hearing and hearing interlocutors; sign--speech exchange. & INMO Air2 glasses for both interlocutors, Android phones, and laptop GPU; video, speech and text/sign displays. & Translates observed signs into speech and spoken replies into signed output. Users select among candidate signs and correct transcription before messages are relayed. & Poster prototype tested signs sampled from ASL tutorial material; 87.8\% average accuracy after user feedback. Evaluation concerns corrected sign recognition rather than a conversational deployment. & Uses MediaPipe Holistic and DSTA-SLR pretrained on WLASL2000. \\
\addlinespace[4pt]
\rowcolor{gray!5}
\textbf{ToozKit}\newline\cite{10.1145/3594739.3610715} & Deaf/hard-of-hearing users and captioning researchers; access to speech transcripts. & Tooz DevKit monocular display, Android phone microphone and cloud speech transcription; Bluetooth display updates. & Open-source captioning toolkit with configurable caption size, color, position and update behavior; changed-character updates accommodate limited display bandwidth. & Demonstration paper presents working captioning and customization. The implementation supports experiments with head-worn captions; systematic user testing is future work. & Extends Termux and uses Live Transcribe Speech Engine with Google Cloud Speech-to-Text. \\
\addlinespace[4pt]
\textbf{CC-Glasses}\newline\cite{ccglasses_2023} & People with color vision deficiency; identifying objects in color-dependent communication. & HoloLens 2 camera/display and GPU server; typed color/object descriptions and referring image segmentation. & Highlights the region referred to by a color-based expression, making the intended object identifiable without requiring accurate color discrimination. & Three deutan-CVD participants answered fourteen sets of object/color questions. Mean correct answers rose from 3.67 to 12/14 with assistance. Speech input was not part of the evaluated interface. & Fine-tunes VLT referring segmentation using the authors' color-object noun-pair dataset. \\
\addlinespace[4pt]
\rowcolor{gray!5}
\textbf{SocialMind}\newline\cite{yang2025socialmind} & Wearers seeking suggestions during everyday conversations, including unfamiliar social situations. & RayNEO X2 with local/server processing; conversational audio, partner visual cues, social context and persona history. & Combines ongoing conversation and nonverbal cues with identity-linked profiles to present brief social suggestions and example responses. Wearers choose whether to use them. & Dataset comparisons assess model-judged suggestion quality; twenty participants subsequently converse with a partner and rate usefulness, latency and acceptance. The live study does not measure a comparative behavioral engagement gain. & Uses DailyDialog, Synthetic-Persona-Chat and SocialDial for dataset-based evaluation. \\
\addlinespace[4pt]
\textbf{Google Glass scene analysis}\newline\cite{googleglass_vi_2021} & Blind and visually impaired users; obtaining spoken information about surroundings. & Google Glass camera, smartphone connection, Azure vision processing and spoken output. & Captures the wearer's scene, generates a description through cloud vision, and returns speech. A locally collected image dataset adapts recognition to Indian scenes. & Technical tests and questionnaires from fifty students and five teachers at a school for visually impaired users. Reported subsecond responses use the tested 4G connection; network dependence and voice/noise issues constrain operation. & Uses Azure Custom Vision and Android text-to-speech; adapts the vision service with local images. \\
\addlinespace[4pt]
\rowcolor{gray!5}
\textbf{LifeInsight}\newline\cite{zhao2025lifeinsight} & Blind and low-vision users; goal-directed visual questions in everyday scenarios. & Primarily collar-worn or handheld camera/microphone, click interface, Raspberry Pi and cloud vision-language model. & A button starts a spoken query about a camera snapshot; concise spoken answers allow follow-up questions. Flexible camera placement lets users seek a useful viewpoint. & Thirteen blind/low-vision participants completed six structured scenarios; 63 of 78 experiences succeeded. Camera coverage, occlusion and mismatches between the queried and captured object affected usefulness. & --- \\
\addlinespace[4pt]
\textbf{TEXT2TASTE}\newline\cite{text2taste_2024} & Intended reading assistance; menu questions and personalized dish recommendations. & Project Aria RGB recordings, object detection and OCR, GPT-4 and Gradio chat interface. & Selects a well-centered menu frame, reconstructs its text, and retrieves personal preference information to condition answers and recommendations. & Four sighted participants used four foreign-language menus. The system reproduced 96.77\% of menu items; recommendation ratings averaged 4.87/5. Voice control and target-population evaluation remain further steps. & Uses DETIC, EasyOCR, GPT-4 and Gradio; retrieves user-provided personal data. \\
\addlinespace[4pt]
\rowcolor{gray!5}
\textbf{Meta glasses perspective}\newline\cite{meta_llm_vi_2024} & Individuals with vision impairments; prospective consumer-glasses assistance. & Commentary on camera/audio glasses, language models and possible AR features. & Discusses how consumer hardware could widen access to visual descriptions, questions, reading and navigation, and identifies future possibilities for AR enhancement. & An Eye commentary rather than an implemented-system or user evaluation. Camera-grounded questions and several assistive functions are discussed as prospective capabilities at publication. & --- \\
\addlinespace[4pt]
\textbf{Form Filling Assistant}\newline\cite{formfill_2025} & Blind users reading and completing paper forms. & Vuzix M4000 camera and server; fingertip detection, known form layouts and printed ArUco markers. & Reads text near the fingertip, directs it to a selected field, and gives page-centering cues. Voice commands switch reading/writing modes; repeated audio is suppressed. & Four blind participants completed three forms after practice. Writing began in the intended field in 63.2\% of cases; 43.6\% of responses stayed entirely inside it. Continuous writing guidance was a remaining need. & Adopts an existing hand-segmentation model for fingertip localization. \\
\addlinespace[4pt]
\rowcolor{gray!5}
\textbf{FetchAid}\newline\cite{10.1145/3613904.3642213} & Blind and low-vision users retrieving parcels from public lockers. & Related smartphone platform: iOS camera, fingertip/button recognition, OCR and AR tracking. & Guides touchscreen choices, announces page changes and supports recovery after wrong selections; uses locker information and spatial tracking to direct users to a compartment. & Twelve blind/low-vision participants tested touchscreen and door-finding stages at public lockers, with a simulated open door. Success improved; door-finding became faster, while touchscreen completion time did not differ significantly. & Uses Apple ARKit for plane/pose tracking and its data-scanner API for text. \\
\addlinespace[4pt]
\textbf{Ultrasonic assistive glasses}\newline\cite{ultrasonic_vi_2024} & Visually impaired users; awareness of nearby obstacles. & Frame-mounted ultrasonic ranging, Arduino Nano controller and buzzer. & Maps measured distance to three beep rates within 30 cm, providing progressively more frequent proximity warnings without image interpretation. & Technical prototype demonstrates the distance-to-warning rules. The reported results do not provide a participant protocol or quantitative navigation comparison; the evidence supports near-obstacle signalling. & --- \\
\addlinespace[4pt]
\rowcolor{gray!5}
\textbf{E-Glasses navigation}\newline\cite{multisensory_nav_2023} & Intended assistance for people with impaired vision; locating a named object and following a route. & Wearable camera/depth sensing, spoken target queries, spatial mapping, path planning and audio feedback. & Combines visual recognition with a spoken target, constructs spatial information, and converts a planned route into spoken navigation instructions. & Object-search and indoor-route tests used sighted participants, sometimes blindfolded. Twenty-five object-search trials yielded 95.46\% detection success and 1.74 s mean detection time; these measures concern target finding. & --- \\
\addlinespace[4pt]
\textbf{LLM-Glasses}\newline\cite{llmglasses_2026} & Intended navigation assistance for visually impaired people through tactile commands. & ESP32 camera, object detection, cloud language reasoning and temple-mounted tactile actuators. & Prioritizes objects in a structured scene description and converts generated directions into taps and sliding contact patterns, with adjustable contact and intensity. & Separate studies tested pattern recognition with thirteen participants, guided paths with six, and visual/LLM decisions in scenario trials. The studies characterize components rather than integrated blind-user navigation. & Uses YOLO-World for object detection and GPT-4o for command generation. \\
\addlinespace[4pt]
\rowcolor{gray!5}
\textbf{VisiMark}\newline\cite{visimark_2025} & People with low vision; recognizing landmarks and retracing indoor routes. & HoloLens 2, pre-scanned environment, route-start QR alignment, configurable AR labels and overviews. & Augments functional landmarks with visible icons/text and previews route structure, making spatial cues available through residual vision. & Sixteen low-vision participants used the landmark augmentations. Recall of augmented landmarks changed, with perceived help finding facilities; route-retracing time, correctness and total landmark recall did not significantly improve. & Uses MRTK 3~\cite{mrtk} and HoloLens spatial-mapping/QR APIs. \\
\addlinespace[4pt]
\textbf{Assistive AR overview}\newline\cite{assistive_ar_2025} & Accessibility and everyday human--human/environment interaction; research agenda. & Overview of glasses sensing, phone/cloud processing and situated information presentation across existing projects. & Relates sign translation, social suggestions and environmental-data presentation, and proposes modular coordination for future assistive systems. & Two-page overview draws on existing projects; it does not report a newly evaluated common accessibility pipeline. The modular MASG-MCP architecture is part of its future agenda. & Reports Sign-to-911's glasses--phone processing design informing SocialMind; it synthesizes prior systems. \\
\addlinespace[4pt]
\end{longtable}

\begin{longtable}{@{}>{\raggedright\arraybackslash}p{\dimexpr.12\textwidth-2\tabcolsep\relax}>{\raggedright\arraybackslash}p{\dimexpr.14\textwidth-2\tabcolsep\relax}>{\raggedright\arraybackslash}p{\dimexpr.16\textwidth-2\tabcolsep\relax}>{\raggedright\arraybackslash}p{\dimexpr.21\textwidth-2\tabcolsep\relax}>{\raggedright\arraybackslash}p{\dimexpr.24\textwidth-2\tabcolsep\relax}>{\raggedright\arraybackslash}p{\dimexpr.13\textwidth\relax}@{}}
\caption{Application evidence: health, well-being, and care.}\label{tab:applications-health-evidence}\label{tab:vital_signs}\label{tab:mental_health}\label{tab:wellbeing}\label{tab:rehab}\label{tab:l3_health_wellbeing}\label{tab:l3_clinical_rehab}\\
\toprule
\rowcolor{gray!15}
\textbf{Study} & \textbf{Users, task, and context} & \textbf{Platform and capabilities} & \textbf{Contribution and workflow} & \textbf{Evaluation and main finding} & \textbf{Demonstrated reuse} \\
\midrule
\endfirsthead
\caption[]{Application evidence: health, well-being, and care. (continued)}\\
\toprule
\rowcolor{gray!15}
\textbf{Study} & \textbf{Users, task, and context} & \textbf{Platform and capabilities} & \textbf{Contribution and workflow} & \textbf{Evaluation and main finding} & \textbf{Demonstrated reuse} \\
\midrule
\endhead
\midrule
\multicolumn{6}{r}{\footnotesize Continued on the next page}\\
\endfoot
\bottomrule
\endlastfoot
\textbf{AirSpecs comfort study}\newline\cite{ramsamy2025sensors} & Everyday wearers exploring how bodily experience relates to their environment. & Eyewear environmental and physiological sensors; phone/watch interfaces and experience sampling. & Presents local readings and notifications alongside self-reports, supporting interpretation of discomfort and consideration of changes to the surroundings. & Thirty participants across three sites, mostly five days. Users explored spaces and reported actions such as opening windows; many wanted explanations and manual control. The study examines experiences, not a validated autonomous comfort predictor. & Builds on the Captivates open-source platform~\cite{chwalek2021captivates}. \\
\addlinespace[4pt]
\rowcolor{gray!5}
\textbf{Commercial-glasses chewing detection}\newline\cite{chewing_2022} & Dietary monitoring through chew, bout, and eating-event records. & Razer Anzu microphones; phone audio recording and offline computer processing. & Classifies short sound windows and aggregates them into bouts and eating events, making longer behavioral summaries available from ordinary audio glasses. & Five participants, twelve eating sessions and 6 h 14 min of eating, talking, walking, and resting. Leave-one-person-out testing evaluates recognition and event duration; aggregation improves event-level usefulness despite imperfect individual labels. Dietary change and nutrition estimates were not evaluated. & --- \\
\addlinespace[4pt]
\textbf{MunchSonic}\newline\cite{munchsonic2024} & Dietary-action monitoring, intake counts, and chewing summaries during meals. & Glasses-mounted active acoustic sensing emphasizing the face and nearby hand movements. & Recognizes six dietary/non-dietary classes and aggregates actions into summaries. Range selection and reduced gain focus on movements relevant to eating. & Twelve participants in 45-minute meal-containing sessions; leave-one-person-out macro-F1 was 93.5\% for two-second windows. Results establish action logging, not improved diet or clinical outcomes. & Adapts ActSonic hardware~\cite{chen2024actsonic} and trains a separate dietary classifier. \\
\addlinespace[4pt]
\rowcolor{gray!5}
\textbf{EchoGuide}\newline\cite{echoguide2024} & Reviewing and querying recorded eating episodes with temporal context. & Active acoustic glasses and synchronized head-mounted GoPro video; offline processing. & Activity predictions select video intervals, which are captioned and indexed for timestamped descriptions and retrieval-based answers. & Nine participants in 40-minute meal-containing sessions. Offline selection reduced caption-record length by 68\% on average against dense processing while retaining useful summaries; food identity remained difficult. Online capture control and health effects were not evaluated. & Reuses ActSonic\textquotesingle s pretrained model~\cite{chen2024actsonic} and an Ego4D-trained LaViLa captioner~\cite{grauman2022ego4d}. \\
\addlinespace[4pt]
\textbf{EchoBreath}\newline\cite{echobreath_2025} & Recording respiratory behaviors attributable to the wearer. & Glasses microphones and active near-face acoustic sensing; smartphone connection. & Combines sound and facial-motion evidence to recognize respiratory events, distinguish wearer from other sources, and reject unrelated activity. & Twenty-five-person laboratory and ten-person semi-wild studies. Personalized accuracy was 89.8\% in an apartment, 89.5\% in an office, and 82.8\% in a cafeteria. Labels used participant logs and recording checks. The 93.1\% result is laboratory accuracy; clinical diagnosis was not tested. & --- \\
\addlinespace[4pt]
\rowcolor{gray!5}
\textbf{BlinkWise}\newline\cite{blink_mmwave_2025} & Relating blink dynamics to drowsiness, visual engagement, and partial blinking. & Near-eye mmWave radar and on-frame processing estimate continuous eye openness. & Separates blink phases and measures their duration, enabling behavioral associations beyond a binary blink count. & Eight people provided hourly five-minute sessions over twelve hours for drowsiness comparisons; ten completed visual-search tasks. Blink parameters correlated with drowsiness and separated task engagement from passive viewing, but did not distinguish easy from hard search. These probes do not establish direct clinical prediction. & --- \\
\addlinespace[4pt]
\textbf{EchoTouch}\newline\cite{echotouch_2025} & Face-touch awareness and reminders tailored to touch type and activity. & Glasses active acoustic sensing; phone vibration reminders. & Recognizes touch location and intrusive versus non-intrusive actions, enabling reminders for all touches or only intrusive touches. & Twenty-person laboratory and ten-person semi-wild recognition studies. A thirty-person acceptance study included eleven people trying the hardware and reminder modes. Selective reminders were generally preferred at home/work, while outdoor preferences shifted toward more reminders. Durable behavior change or infection reduction was not measured. & --- \\
\addlinespace[4pt]
\rowcolor{gray!5}
\textbf{GymSoles++}\newline\cite{gymsoles_2021} & Posture feedback for people performing squats. & Pressure-sensing insoles measure center of pressure; Google Glass or a monitor displays it; separate tactile alternatives. & Makes body-load information available during movement, allowing a peripheral wearable display to be compared with fixed-screen and vibration feedback. & Ten healthy adults compared four feedback conditions. Usability and preference differences were not significant; qualitative feedback favored seeing guidance without turning toward a monitor. Simultaneous Glass-plus-vibration superiority and rehabilitation outcomes were not tested. & --- \\
\addlinespace[4pt]
\textbf{Mindful Moments}\newline\cite{10.1145/3563657.3596030} & Brief breath-based mindfulness practice during everyday walking breaks. & OHMD visual animation synchronized with audio; wrist physiological sensing for evaluation. & Uses a user-initiated five-minute session to pace inhalation and exhalation, fitting guided practice into short indoor routines. & Eighteen-person comparison with walking meditation and ordinary walking, followed by fifteen-person three-day home/office use. Immediate self-reported mindfulness improved over control; between-condition EDA differences were not significant. Devices were accessed for sessions, not worn all day; long-term treatment effects were not established. & --- \\
\addlinespace[4pt]
\rowcolor{gray!5}
\textbf{EmoGlass}\newline\cite{lim2022emoglass} & Self-monitoring and reflection on everyday emotional experiences. & Glasses camera images part of the face; embedded expression inference and companion phone app. & Connects seven expression labels to reports, contextual diaries, and retrieval of positive activities, giving users ways to review and interpret their records. & Fifteen participants in a three-day field study; mean recognition accuracy was 73.0\% against sparse self-reports. Three low-accuracy participants were excluded from later qualitative analysis. Findings concern recognition and reflection experience rather than diagnosis or sustained emotional-health improvement. & --- \\
\addlinespace[4pt]
\textbf{EMOShip}\newline\cite{smartglasses_dream_2022} & Emotion-aware moment capture and review of associated visual context. & Inward eye imaging, pupil/gaze features, and outward scene imagery. & Selects moments using attention and expression gates, then associates learned emotion categories with descriptions of attended scenes. & Film-elicited, self-labelled responses and an additional unseen-user test. Macro binary accuracy across six non-neutral categories was 80.2\%; scene context sometimes confused negative categories. Outputs are model-derived associations, not identified emotional causes or evidence of mental-health treatment. & Adapts EMO eye features~\cite{guo2020emo} and MemX attention triggering~\cite{liu2021memx}. \\
\addlinespace[4pt]
\rowcolor{gray!5}
\textbf{ActiviSee}\newline\cite{activisee_2023} & Activity-transition recognition intended to support reminders when mobility begins. & Glasses-mounted accelerometer, gyroscope, magnetometer, and barometer; workstation analysis. & Classifies posture/activity windows and examines transitions relevant to changing multifocal glasses before walking. & Twenty-three healthy adults performed five scripted activities in a room. Offline recognition was harder for transitions than for static/mobile categories; merging similar sedentary classes improved classification. Continuous-stream recognition and an implemented reminder system were future work, so fall prevention was not evaluated. & --- \\
\addlinespace[4pt]
\textbf{Captivates}\newline\cite{chwalek2021captivates} & Research platform for physiological observation across everyday contexts. & Open eyeglass hardware with facial thermopiles, blink sensors, inertial tracking, and optical localization. & Integrates sensors and data access to enable later studies of bodily responses, behavior, and context; design work addresses fit and public wearability. & Sensor calibration and small pilot studies include five-person blink validation and a six-person cognitive-task pilot. Across-day home collection and cognitive-state modeling were proposed next steps. The paper establishes a research platform rather than a completed long-term health intervention. & Platform subsequently adapted for AirSpecs comfort research~\cite{ramsamy2025sensors}. \\
\addlinespace[4pt]
\rowcolor{gray!5}
\textbf{Motion-based vital-sign monitoring}\newline\cite{vitalmotion2022} & Feasibility of respiratory and cardiac estimation from a related headset. & Meta Quest accelerometer; ECG and respiration belt supply reference signals. & Filters motion-contaminated windows and separates respiratory/cardiac frequency components from head and face micro-vibrations. & Twenty participants completed one-minute sitting, standing, and seated-controller conditions. Average reported errors were below 3.3\% for respiration and 5.7\% for heartbeat. Evidence concerns brief, mostly static measurements rather than everyday moving-use coverage or clinical monitoring. & --- \\
\addlinespace[4pt]
\textbf{ThermalEye}\newline\cite{thermaleye_2025} & Blink-event measurement with prospective ocular-health applications. & Passive low-resolution thermal array near the eye; RGB camera provides annotations only. & Uses eye/eyelid temperature-pattern changes and temporal filtering to produce eye-state estimates and blink intervals. & Four participants performed prescribed blinking while remaining still and forward-facing under temperature changes. Held-out and leave-one-person-out tests establish blink-detection feasibility. Targeted fatigue, dry-eye, and glaucoma studies were future work rather than demonstrated health applications. & --- \\
\addlinespace[4pt]
\rowcolor{gray!5}
\textbf{Biosensor smart-glasses eye tracker}\newline\cite{biosensor_glasses_2024} & Gaze measurement intended to enable future health and interaction applications. & Huawei Vision Glass-based prototype with inward near-infrared eye cameras, outward camera, and server processing. & Segments pupils and maps eye-image features to gaze points through calibration, supplying a measurement capability for later applications. & Pupil-dataset evaluation and gaze estimation at six screen targets, one metre away; reported gaze RMSE is 1.0 degree. Participant count for that gaze experiment is unspecified. No neurological-biomarker diagnosis or clinical-care outcome was evaluated. & --- \\
\addlinespace[4pt]
\textbf{Privacy-aware HAR for digital therapeutics}\newline\cite{leite2023privacyaware} & Activity records as a prospective input to digital therapeutics. & OCOsense glasses IMU; centralized and federated machine-learning comparisons. & Recognizes basic postures and walking while comparing model size and distributed training choices for privacy-aware activity analysis. & Eighteen adults performed three instructed apartment scenarios, producing thirty-six hours of labelled recordings. Evaluation concerns activity classification and model/training tradeoffs. Clinical populations, treatment delivery, and therapeutic outcomes were not studied. & Uses the OCOsense smart-glasses platform. \\
\addlinespace[4pt]
\rowcolor{gray!5}
\textbf{HoloLens in anatomic pathology}\newline\cite{hanna2018hololens} & Pathologists and trainees inspecting specimens, consulting remotely, and viewing clinical information. & HoloLens cameras, see-through display, gestures/voice, and secure hospital connectivity. & Combines first-person consultation and annotations with 3D specimen manipulation, whole-slide viewing, and virtual clinical windows during pathology workflows. & Clinical-workflow demonstrations establish feasible consultation, viewing, and manipulation. Some slide viewers only displayed content or were incompatible with gestures. No controlled diagnostic-accuracy or patient-outcome study was reported; network and information-system integration remain consequential. & Uses Skype for HoloLens, remote-desktop software, and existing whole-slide viewers, including OpenSeadragon. \\
\addlinespace[4pt]
\textbf{Clinical 3D-model viewing}\newline\cite{zari2023magic} & Surgeons and residents inspecting liver anatomy from radiological images. & Magic Leap 1, HoloLens 2, and laptop running the Verima medical-model viewer. & Holds clinical content and viewing software consistent while comparing visualization, spatial understanding, manipulation, and comfort across devices. & Three liver models; twelve clinicians recruited, with eleven questionnaire responses analyzed after one wearing-compatibility exclusion. Magic Leap was favored for visualization/interaction; HoloLens gesture use could cause fatigue. Results concern model-viewing experience and rendering, not surgical or patient outcomes. & Uses the Verima suite to reconstruct DICOM content and present the same models across platforms. \\
\addlinespace[4pt]
\end{longtable}

\begin{longtable}{@{}>{\raggedright\arraybackslash}p{\dimexpr.12\textwidth-2\tabcolsep\relax}>{\raggedright\arraybackslash}p{\dimexpr.14\textwidth-2\tabcolsep\relax}>{\raggedright\arraybackslash}p{\dimexpr.16\textwidth-2\tabcolsep\relax}>{\raggedright\arraybackslash}p{\dimexpr.21\textwidth-2\tabcolsep\relax}>{\raggedright\arraybackslash}p{\dimexpr.24\textwidth-2\tabcolsep\relax}>{\raggedright\arraybackslash}p{\dimexpr.13\textwidth\relax}@{}}
\caption{Application evidence: professional work and remote collaboration.}\label{tab:applications-work-evidence}\\
\toprule
\rowcolor{gray!15}
\textbf{Study} & \textbf{Users, task, and context} & \textbf{Platform and capabilities} & \textbf{Contribution and workflow} & \textbf{Evaluation and main finding} & \textbf{Demonstrated reuse} \\
\midrule
\endfirsthead
\caption[]{Application evidence: professional work and remote collaboration. (continued)}\\
\toprule
\rowcolor{gray!15}
\textbf{Study} & \textbf{Users, task, and context} & \textbf{Platform and capabilities} & \textbf{Contribution and workflow} & \textbf{Evaluation and main finding} & \textbf{Demonstrated reuse} \\
\midrule
\endhead
\midrule
\multicolumn{6}{r}{\footnotesize Continued on the next page}\\
\endfoot
\bottomrule
\endlastfoot
\textbf{Remote emergency medical teamwork}\newline\cite{seewhatisee_2025} & EMS providers seeking remote physician guidance during emergency care. & Vuzix M400 first-person video; remote web interface, voice/text guidance, and patient-information access. & Shares scene details and clinical context during consultation, letting remote physicians return instructions while the wearer continues hands-on work. & Two-site emergency simulations enrolled 35 EMS providers and 11 physicians; fifteen teams were analyzed. Fewer communication breakdowns than radio, but longer calls than phone. Shared visual attention averaged 55.6\%; missed views and reduced local-team overhearing affected coordination. Patient outcomes were not evaluated. & Uses the authors' previously developed EMS glasses and remote-team web applications. \\
\addlinespace[4pt]
\rowcolor{gray!5}
\textbf{XR virtual displays in public work}\newline\cite{pavanatto2024working} & Mobile workers organizing reference material and completing computer tasks in public settings. & HoloLens 2 plus laptop in the exploratory study; ThinkReality A3, Quest Pro, and laptop in the comparison. & Extends a laptop into arranged virtual screens while retaining physical keyboard/mouse input, supporting access to several documents without frequent window switching. & Twenty participants performed personal tasks; eighteen completed three-condition, thirty-minute work tasks. XR reduced window switches; task accuracy was equivalent across conditions and productivity did not significantly differ. Visual fatigue increased versus laptop. Bystander questionnaires complemented task data; full-day work was not evaluated. & MRTK-based prototype~\cite{mrtk2}; Lenovo Virtual Display Manager and Immersed in the comparison. \\
\addlinespace[4pt]
\end{longtable}

\begin{longtable}{@{}>{\raggedright\arraybackslash}p{\dimexpr.12\textwidth-2\tabcolsep\relax}>{\raggedright\arraybackslash}p{\dimexpr.14\textwidth-2\tabcolsep\relax}>{\raggedright\arraybackslash}p{\dimexpr.16\textwidth-2\tabcolsep\relax}>{\raggedright\arraybackslash}p{\dimexpr.21\textwidth-2\tabcolsep\relax}>{\raggedright\arraybackslash}p{\dimexpr.24\textwidth-2\tabcolsep\relax}>{\raggedright\arraybackslash}p{\dimexpr.13\textwidth\relax}@{}}
\caption{Application evidence: connected devices, human--robot interaction, and robot learning.}\label{tab:applications-robot-evidence}\label{tab:l3_robot_human}\\
\toprule
\rowcolor{gray!15}
\textbf{Study} & \textbf{Users, task, and context} & \textbf{Platform and capabilities} & \textbf{Contribution and workflow} & \textbf{Evaluation and main finding} & \textbf{Demonstrated reuse} \\
\midrule
\endfirsthead
\caption[]{Application evidence: connected devices, human--robot interaction, and robot learning. (continued)}\\
\toprule
\rowcolor{gray!15}
\textbf{Study} & \textbf{Users, task, and context} & \textbf{Platform and capabilities} & \textbf{Contribution and workflow} & \textbf{Evaluation and main finding} & \textbf{Demonstrated reuse} \\
\midrule
\endhead
\midrule
\multicolumn{6}{r}{\footnotesize Continued on the next page}\\
\endfoot
\bottomrule
\endlastfoot
\textbf{BLEselect}\newline\cite{chen2023bleselect} & Select one nearby IoT device through head or hand gestures. & Eyewear Bluetooth antenna array, device transmitters, optional wrist transmitter; laptop processing. & Associates device identity and radio direction with nodding, pointing, or encircling. Selection works without visual object recognition, but requires transmitter-equipped candidates. & Twenty-two participants selected six devices in each of two rooms from fixed positions. Accuracy was 94.8--97.8\%; nodding was fastest, while pointing and nodding were preferred to encircling. Collinear devices remain ambiguous. & --- \\
\addlinespace[4pt]
\rowcolor{gray!5}
\textbf{Gaze robot control}\newline\cite{robot_gaze_iros_2023} & Steer a mobile robot along an obstacle route while following it. & HoloLens 2 eye tracking and registered virtual controls; Summit-XL robot, Unity and ROS. & Gaze displacement drives robot velocity through an admittance controller and a no-motion dead zone. Damping tunes response speed and stability. & Ten participants compared joystick control with three gaze damping settings. The fastest gaze setting differed from the preferred setting. SUS differences were not significant; the joystick scored significantly better on ease of use. Robot slippage can impair alignment. & Modifies ARETT gaze access and uses MRTK eye-tracking interfaces with ROS/rosbridge. \\
\addlinespace[4pt]
\textbf{AR robot assembly}\newline\cite{robot_assembly_2024} & Select parts for a cobot to retrieve during manual assembly. & HoloLens 2 gaze/hand tracking, QR alignment, and robot communication. & Gaze or hand rays identify parts; two pinch actions select and confirm. Visual trajectories and virtual colliders support coordination with the robot. & Six expert formative participants and sixteen evaluation participants. Gaze reduced total part-selection time from 47.5 to 39.0 s; SUS and task-fluency differences were not significant. The significance result concerns selection, rather than overall assembly time. & Uses MRTK, ROS/TCP communication, and QR-based spatial alignment. \\
\addlinespace[4pt]
\rowcolor{gray!5}
\textbf{Legged-manipulator AR interface}\newline\cite{legged_robot_2023} & Navigate a legged robot, inspect surroundings, and choose grasping goals. & HoloLens 2, Spot with Kinova arm, hand/voice input and spatial anchors. & Places robot goals and controls in a shared spatial frame; supports target placement and robot-follow behavior. & Seventeen participants compared AR and a controller during navigation, stairs, and a secondary photo task. The controller was faster and had lower workload; AR increased immersion. Five returning participants tried an improved interface, with practice a possible confound. & Uses MRTK, ROSSharp, and Azure Spatial Anchors for shared coordinates. \\
\addlinespace[4pt]
\textbf{VARIL}\newline\cite{ar_viz_policy_2022} & Coordinate warehouse deliveries with multiple robots. & Unity/WebGL simulated environment; separate tablet AR demonstration with three Turtlebots. & Learns a state-dependent policy from expert feedback to show or hide robot paths, positions, avatars, and destinations, limiting unnecessary visual information. & Twenty-five participants controlled a virtual worker by keyboard and mouse. Adaptive visualizations reduced robot waiting/completion times and improved reported experience relative to fixed policies. The physical example used a tablet; no head-worn field trial was reported. & Uses Unity to implement and evaluate the visualization policy. \\
\addlinespace[4pt]
\rowcolor{gray!5}
\textbf{ARCADE}\newline\cite{arcade_2024} & Provide and validate demonstrations for robot manipulation learning. & HoloLens 2 hand tracking, robot digital twin, and Fetch robot. & Maps hand motion to robot joints, generates variants from a seed demonstration, and combines user review with trajectory-similarity filtering to expand training data. & Ten trials per condition across reaching, pushing, and pick-and-place assessed completion error. Expanded AR data improved reaching and pick-and-place over expanded kinesthetic data; pushing showed no significant difference. A separate pouring demonstration succeeded in 8/10 trials. & Uses Unity Robotics Hub ROS connectors for demonstration transfer. \\
\addlinespace[4pt]
\textbf{EgoMimic}\newline\cite{egomimic_2025} & Learn robot object placement, laundry handling, and grocery manipulation from human and robot demonstrations. & Aria video/hand trajectories for human demonstrations; matched Aria view and wrist cameras on robot. & Aligns action coordinates, distributions, and appearance to jointly learn from human recordings and robot demonstrations. Additional human data can adapt the policy to a new scene. & Robot rollouts on three tasks with policy baselines and a no-human-data ablation. Human data improved results, with task-dependent gains; robot demonstrations remained part of training. & Uses Aria MPS, SAM, and an ACT-based policy; datasets follow robomimic conventions. \\
\addlinespace[4pt]
\rowcolor{gray!5}
\textbf{AINA}\newline\cite{dexterity_2026} & Train multi-fingered robots from human demonstrations of everyday manipulation. & Aria Gen 2 video and hand pose; inferred stereo depth; external robot RGB-D cameras. & Learns object/fingertip trajectories from human recordings, aligned using a calibrated robot scene and one in-scene human demonstration. No robot training demonstrations are required. & Fifteen rollouts on each of nine tasks, with success ranging from 1/15 to 13/15; baseline comparisons covered subsets. Force information, object changes, and camera timing constrain transfer. & Uses FoundationStereo, Grounded-SAM, CoTracker, and a Point-Policy-based learner. \\
\addlinespace[4pt]
\end{longtable}

\endgroup

\clearpage
\subsection[Deployment Evidence]{Deployment Evidence}
\label{app:deployment-evidence}
\begingroup
Tables~\ref{tab:deployment-operation-evidence}--\ref{tab:deployment-social-evidence} summarize the studies discussed in Section~\ref{sec:deployment}. They record the platform and people affected, the approach examined, the evaluation method and findings, and the conditions relevant to use. Technical experiments, user studies, simulations, and conceptual analyses contribute different forms of evidence. Results retain each study's setting and protocol; operational studies already discussed in earlier sections are revisited here for their deployment implications.\par
\small
\setlength{\tabcolsep}{3pt}
\renewcommand{\arraystretch}{1.10}
\setlength{\LTpre}{8pt}
\setlength{\LTpost}{8pt}
\begin{longtable}{@{}>{\raggedright\arraybackslash}p{\dimexpr0.13\textwidth-2\tabcolsep\relax}>{\raggedright\arraybackslash}p{\dimexpr0.18\textwidth-2\tabcolsep\relax}>{\raggedright\arraybackslash}p{\dimexpr0.21\textwidth-2\tabcolsep\relax}>{\raggedright\arraybackslash}p{\dimexpr0.28\textwidth-2\tabcolsep\relax}>{\raggedright\arraybackslash}p{\dimexpr0.2\textwidth\relax}@{}}
\caption{Deployment evidence: operational reliability and wearability.}\label{tab:deployment-operation-evidence}\\
\toprule
\rowcolor{gray!15}
\textbf{Study} & \textbf{Platform, people, and purpose} & \textbf{Approach examined} & \textbf{Evaluation and main finding} & \textbf{Conditions and implications} \\
\midrule
\endfirsthead
\caption[]{Deployment evidence: operational reliability and wearability. (continued)}\\
\toprule
\rowcolor{gray!15}
\textbf{Study} & \textbf{Platform, people, and purpose} & \textbf{Approach examined} & \textbf{Evaluation and main finding} & \textbf{Conditions and implications} \\
\midrule
\endhead
\midrule
\multicolumn{5}{r}{\footnotesize Continued on the next page}\\
\endfoot
\bottomrule
\endlastfoot
\rowcolor{gray!5}
\textbf{MemX}\newline\cite{liu2021memx} & Eye and outward cameras; attention-triggered recording on a prototype with external computation. & Eye behavior triggers richer scene analysis and recording of moments likely to interest the wearer. & Attention evaluation and 11 field pilot scenarios; energy estimated from component and processing models for a target low-power platform. & Savings depend on how often the recording stages run. Estimated runtime does not demonstrate an integrated all-day eyewear deployment. \\
\addlinespace[4pt]
\textbf{SkillSight}\newline\cite{li2026skillsight} & Egocentric skill assessment using existing datasets; gaze and head-related inputs at runtime. & A teacher uses visual and gaze observations; distillation transfers task knowledge to a reduced-input student. & Three datasets covering cooking, music, and sports. Accuracy is evaluated with power estimated from sensing, computation, and memory costs. & Richer training resources reduce runtime input requirements. Some dataset head trajectories use visual-inertial tracking; a reduced rotation-and-gaze variant is also evaluated. \\
\addlinespace[4pt]
\rowcolor{gray!5}
\textbf{WearSE}\newline\cite{sun2025wearse} & Speakers and microphones connected by wire to a smartphone; streaming speech enhancement. & Combines audio and articulation-related acoustics in a causal pipeline that reuses temporal computation. & Phone prototype: 30.1\,ms processing per 50\,ms segment plus 100\,ms window delay, giving 130.1\,ms total delay. Energy measured on the phone. & Runtime depends on the companion device and communication path. Wireless operation was a proposed extension, not the configuration evaluated for these timing results. \\
\addlinespace[4pt]
\textbf{mmET}\newline\cite{lu2025mmet} & Frame-mounted radar eye tracking; repeat wearing and new-user adaptation. & Learns across repeated remounting and jointly estimates gaze and eye geometry; pretraining reduces adaptation data. & Sixteen participants across the experiments. Mean user-dependent errors were 1.49$^\circ$ within sessions and 4.47$^\circ$ across remounting sessions, without per-session recalibration. & New-user adaptation remains necessary. Reported low radar power covers sensor boards, excluding model inference; vigorous activity was not extensively evaluated. \\
\addlinespace[4pt]
\rowcolor{gray!5}
\textbf{Focals user-group report}\newline\cite{pyae2022focus} & Focals by North in daily scheduling, communication, and navigation; university users and their social contacts. & Daily journals and a focus group about incorporating a commercial head-worn display into ordinary routines. & Six participants over three weeks, with approximately three hours of daily use requested. Participants reported distraction, limited added value, social friction, and occasional non-use; COVID restrictions affected the final week. & A glasses-like form and functioning notifications did not ensure sustained usefulness. Findings concern this small group and product rather than smart-glasses adoption generally. \\
\addlinespace[4pt]
\textbf{Going Blank Comfortably}\newline\cite{lyons2023going} & Inactive monocular head-worn displays; optical-combiner placement in the wearer's visual field. & Compare mock combiner shapes and positions, then assess visual comfort while the display is inactive. & Thirty-person preference study followed by a twelve-person stationary comfort study. Moving the nasal combiner boundary toward primary gaze reduced comfort. & The optical element can impose a burden even when no information is displayed. These results concern visual comfort under tested positions, not social acceptance, long-term health effects, or a universal placement threshold. \\
\addlinespace[4pt]
\end{longtable}

\begin{longtable}{@{}>{\raggedright\arraybackslash}p{\dimexpr0.13\textwidth-2\tabcolsep\relax}>{\raggedright\arraybackslash}p{\dimexpr0.18\textwidth-2\tabcolsep\relax}>{\raggedright\arraybackslash}p{\dimexpr0.21\textwidth-2\tabcolsep\relax}>{\raggedright\arraybackslash}p{\dimexpr0.28\textwidth-2\tabcolsep\relax}>{\raggedright\arraybackslash}p{\dimexpr0.2\textwidth\relax}@{}}
\caption{Deployment evidence: privacy and consent.}\label{tab:deployment-privacy-evidence}\label{tab:privacy_capture_recording}\label{tab:privacy_bystander}\label{tab:privacy_consent_permission}\label{tab:l4_capture_privacy_1}\label{tab:l4_capture_privacy_2}\\
\toprule
\rowcolor{gray!15}
\textbf{Study} & \textbf{Platform, people, and purpose} & \textbf{Approach examined} & \textbf{Evaluation and main finding} & \textbf{Conditions and implications} \\
\midrule
\endfirsthead
\caption[]{Deployment evidence: privacy and consent. (continued)}\\
\toprule
\rowcolor{gray!15}
\textbf{Study} & \textbf{Platform, people, and purpose} & \textbf{Approach examined} & \textbf{Evaluation and main finding} & \textbf{Conditions and implications} \\
\midrule
\endhead
\midrule
\multicolumn{5}{r}{\footnotesize Continued on the next page}\\
\endfoot
\bottomrule
\endlastfoot
\rowcolor{gray!5}
\textbf{Privacy-Enhancing Technology and Everyday AR}\newline\cite{guo2022privacy} & Bystanders encountering AR sensing, recording, identification, and inference in everyday scenarios. & Scenario-based survey of awareness, consent preferences, relationships, and prior permissions across uses of captured data. & 102 respondents; awareness and desired consent differed across activities and situations. Many respondents were unaware of physiological or internal-state inference capabilities. & Privacy communication needs to describe intended processing as well as collection. Findings concern reported preferences rather than observed use of a deployed protection mechanism. \\
\addlinespace[4pt]
\textbf{Privacy-aware HAR}\newline\cite{leite2023privacyaware} & OCOsense accelerometer and gyroscope recordings from 18 participants; four posture activities. & Compares model complexity and centralized versus federated training; participant-specific data partitions act as clients for federated averaging. & Table 2 reports macro-F1 of 0.84 for centralized and 0.80 for federated training with the simpler model. Original collection streamed through an iPad to cloud storage. & Addresses training-data organization and model sharing. On-glasses federated deployment and resistance to information leakage were not evaluated. \\
\addlinespace[4pt]
\rowcolor{gray!5}
\textbf{Adopting Smart Glasses Responsibly}\newline\cite{iqbal2023adopting} & Ray-Ban Stories and people affected by everyday camera-glasses use. & Ethical commentary relating wearable capture, inconspicuous appearance, recording indicators, and responsible adoption. & Discusses potential benefits and privacy concerns using consumer-device examples; it does not report a participant study or evaluate a protection system. & Provides context for recording transparency and responsibility. Its arguments guide interpretation of empirical studies without establishing the prevalence of concerns or the effectiveness of indicators. \\
\addlinespace[4pt]
\textbf{BystandAR}\newline\cite{corbett2023bystandar} & HoloLens 2 wearer, interaction subjects, and other people visible in RGB and depth streams. & Combines gaze history and voice to infer interaction subjects, then masks other faces; inference frequency and thresholds control protection and availability. & 16-person evaluation; the default threshold protected 98.14\% of bystanders while retaining access to 96.27\% of interaction subjects. Rapid movement challenges masking between updates. & Inferred relevance does not establish consent. Enforcing interception against other applications requires integration beyond the prototype. \\
\addlinespace[4pt]
\rowcolor{gray!5}
\textbf{MeMic}\newline\cite{zulfikar2024memic} & Speech-recording wearable; snug-neck prototype for technical testing, with eyewear among demonstrated form factors. & IMU-detected speech vibration gates microphone power; a visible indicator signals active recording. & 12-person short technical study and 150-person video-based acceptability study. Selected privacy and social-fear measures favored speech-triggered over continuous recording. & Limits recording between wearer utterances; overlapping bystander speech remains capturable. Technical performance depends on contact with a speech-vibrating body region, and acceptability findings concern demonstrations. \\
\addlinespace[4pt]
\textbf{In Focus, Out of Privacy}\newline\cite{bhardwaj2024focus} & Camera-glasses wearers managing capture around other people in everyday settings. & Two-week diary and interviews investigate recording practices, awareness cues, privacy responsibility, and wearer strategies. & 15 wearers reported that glare, distance, crowds, and unfamiliarity could make indicators difficult for others to notice or interpret, creating a burden of explanation and privacy management. & Device discretion can complicate recording transparency. Evidence captures wearers' experiences and accounts of bystanders, rather than directly measuring bystander detection rates. \\
\addlinespace[4pt]
\rowcolor{gray!5}
\textbf{Securing Bystander Privacy}\newline\cite{corbett2023securing} & Mixed-reality sensing that supports wearer tasks while exposing nearby people. & Research overview distinguishes explicit and implicit protection and relates usability, privacy, legitimate information access, and consent. & Synthesizes protection approaches and their design requirements; no independent experiment establishes an optimal protection level or a universal functionality tradeoff. & Provides vocabulary for evaluating what information remains available after protection. Concrete operating costs and user effects must be established through the associated system and user studies. \\
\addlinespace[4pt]
\textbf{BystandARIA}\newline\cite{al2025bystandaria} & Bystander LED signals and an externally processed Project Aria camera feed. & Decodes modulated LED messages to change face-blurring state; detected faces are tracked over time. & Controlled LED tests report 85\% detection accuracy and 50 ms per processed frame, with sensitivity to lighting, distance, and movement. Current commands affect active face tracks. & Demonstrates preference signaling and blur control. Individual LED-to-face association is described as planned functionality, limiting claims about person-specific control in multi-person scenes. \\
\addlinespace[4pt]
\rowcolor{gray!5}
\textbf{Privacy Equilibrium}\newline\cite{rajaram2025privacy} & Multiple AR users and bystanders with potentially conflicting sensing requirements and privacy preferences. & Constrained optimization reconciles privacy profiles, sensing scopes, and application requirements through a Unity simulation toolkit. & Hospital and office simulations plus eight-expert walkthroughs examine negotiated configurations. Experts highlight task semantics, severity, reversibility, and accessibility needs. & Makes the dependency between permitted sensing and application functionality explicit. Computed satisfaction measures represent policy compatibility; they do not establish resolved disagreements or lived satisfaction in deployment. \\
\addlinespace[4pt]
\textbf{Effective Consent Mechanisms}\newline\cite{windl2025designing} & Device-user privacy choices during brief, repeated, or context-changing AR interactions. & Focus groups and expert co-design develop a scenario taxonomy, a consent continuum, and mechanisms for persistent preferences, contextual prompts, and withdrawal. & 17 focus-group participants and 11 privacy experts contribute design requirements and proposals; the mechanisms were not comparatively evaluated in everyday deployment. & Relates consent effort and persistence to interaction context. Extending these primarily user-side mechanisms to bystander negotiation remains a separate design problem. \\
\addlinespace[4pt]
\rowcolor{gray!5}
\textbf{Mind the Gap}\newline\cite{wang2026mind} & Camera-glasses wearers and bystanders across public, bounded, and sensitive settings in China. & Survey and interviews compare privacy expectations, willingness to accommodate others, and responses to possible privacy-enhancing designs. & 525 survey respondents and 20 interview participants reveal gaps around prior consent and sharing, with preferences varying by setting and stakeholder perspective. & Supports contextual privacy choices and explicit negotiation of responsibilities. Results concern reported expectations and reactions to design concepts, rather than effectiveness of deployed protection. \\
\addlinespace[4pt]
\textbf{Multi-Layered Privacy Permission Framework}\newline\cite{mansour2025multi} & XR applications handling sensor, biometric, behavioral, and contextual information throughout use. & Conceptual framework links governance, data infrastructure, permissions, interaction, and perception; a user journey and checklist organize onboarding, prompts, control, and review. & Illustrative scenarios operationalize the framework; longitudinal studies and real-world deployments are identified as necessary validation. & Provides a design structure for continuing consent and withdrawal across an experience. It does not demonstrate that a particular permission interface improves comprehension, trust, or privacy outcomes. \\
\addlinespace[4pt]
\rowcolor{gray!5}
\textbf{SITARA / Now You See Me}\newline\cite{fan2026nowyousee} & Face privacy across glasses, a companion phone, and a trusted consent/key service. & Default face blur and encrypted originals permit consent-based restoration; optional synthetic replacement preserves a viewable recording. & Raspberry Pi proxy: a 10-second clip required 13.69 s for core protection and 22.88 s for the full pipeline, with energy 1.68 and 2.8 times baseline. & Quantifies protection costs and infrastructure dependencies. User findings overlap with the related See Me If You Can study; full-pipeline real-time glasses operation was not demonstrated. \\
\addlinespace[4pt]
\textbf{See Me If You Can}\newline\cite{li2026seemeifyoucan} & Wearer and bystander perspectives on the related three-tier consent-based face-restoration protocol. & Glasses onboarding, an interactive demonstration with simulated consent requests, and qualitative interviews examine disclosure and control. & 18 participants, evenly split between wearers and bystanders; default protection could reassure bystanders while frustrating wearers, and requester identity and purpose shaped decisions. & Adds user evidence to the linked technical design. It does not establish comparative trust superiority; audio and contextual identification remain outside the face-protection guarantee. \\
\addlinespace[4pt]
\end{longtable}

\begin{longtable}{@{}>{\raggedright\arraybackslash}p{\dimexpr0.13\textwidth-2\tabcolsep\relax}>{\raggedright\arraybackslash}p{\dimexpr0.18\textwidth-2\tabcolsep\relax}>{\raggedright\arraybackslash}p{\dimexpr0.21\textwidth-2\tabcolsep\relax}>{\raggedright\arraybackslash}p{\dimexpr0.28\textwidth-2\tabcolsep\relax}>{\raggedright\arraybackslash}p{\dimexpr0.2\textwidth\relax}@{}}
\caption{Deployment evidence: security and access control.}\label{tab:deployment-security-evidence}\label{tab:privacy_authentication}\label{tab:l4_authentication}\\
\toprule
\rowcolor{gray!15}
\textbf{Study} & \textbf{Platform, people, and purpose} & \textbf{Approach examined} & \textbf{Evaluation and main finding} & \textbf{Conditions and implications} \\
\midrule
\endfirsthead
\caption[]{Deployment evidence: security and access control. (continued)}\\
\toprule
\rowcolor{gray!15}
\textbf{Study} & \textbf{Platform, people, and purpose} & \textbf{Approach examined} & \textbf{Evaluation and main finding} & \textbf{Conditions and implications} \\
\midrule
\endhead
\midrule
\multicolumn{5}{r}{\footnotesize Continued on the next page}\\
\endfoot
\bottomrule
\endlastfoot
\rowcolor{gray!5}
\textbf{C-Auth}\newline\cite{lim2023c} & Glasses-mounted downward-facing RGB camera; access by an enrolled wearer. & Segments nose and cheek contours, aligns them, and matches them to enrolled references. & Twenty participants across three days. Adding second-session reference contours raised average true-positive rate to 98\%, with false-positive rate about 5\%. & Wearing position affects observations; reference updating improved access, while similar facial contours remained difficult to separate. \\
\addlinespace[4pt]
\textbf{SonicID}\newline\cite{li2024sonicid} & Speaker--microphone pairs on the frame; acoustic verification of the enrolled wearer. & Selects stable facial echo samples and adapts a pretrained model using per-user enrollment data. & Forty participants over three days. With one-minute enrollment, true-positive rate declined from 97.4\% on the same day to 79.6\% across days; false-positive rate was 3.7\% in both conditions. Processing used a local server and selected static samples. & Reported sample duration is not end-to-end unlock latency. Daily placement variation and selecting usable samples remain deployment considerations. \\
\addlinespace[4pt]
\rowcolor{gray!5}
\textbf{SkullID}\newline\cite{shin2024skullid} & Bone-conduction playback and contact microphones on glasses; wearer verification. & Combines skull-transformed acoustic responses from separated contact locations to tolerate placement variation. & Three studies cover placement, remounting, noise, and repeated visits. The one-week recall study (17 genuine users and 10 impostors) reports 2.94\% equal-error rate. Attack-specific training reduced tested replay successes. & Spatially complementary contacts improve repeatability. Reported spoofing resistance concerns the tested replay/imitation conditions; recall sessions do not establish continuous daily deployment. \\
\addlinespace[4pt]
\textbf{Symmetric gaze stimuli}\newline\cite{dondi2026gaze} & Desktop display and remote eye tracker; related behavioral biometric method. & Classifies eye-movement features collected while participants view symmetric animated stimuli, without eye-tracker calibration. & Thirty-six retained participants across three sessions. Horizontal Multiple stimuli with random forests achieved equal-error rates of 5.0\% (closed set) and 6.09\% (open set) on later-session testing; each animation lasted ten seconds. & Supports cross-session biometric feasibility in the tested setup. Eyewear integration and authentication during ordinary activity were not evaluated; viewing the stimulus adds an explicit task. \\
\addlinespace[4pt]
\rowcolor{gray!5}
\textbf{PXRA}\newline\cite{cha2025lightweight} & XR client, gateway, and cloud authentication server; device/session identity and message protection. & Combines hardware-derived identity, cloud-assisted cryptography, and authenticated context binding. & Formal protocol analysis and a Raspberry Pi client testbed with an emulated physically unclonable function, desktop gateway, and cloud backend. & Offloading reduces client computation while depending on backend trust and availability. Proxy evaluation and symbolic verification do not establish resistance to all implementation or side-channel failures. \\
\addlinespace[4pt]
\textbf{Virtual Touch to Tesla Command}\newline\cite{yu2024virtualtouch} & Commodity glasses paired with phones and configured vehicle-automation services; authorization of connected actions. & Studies whether accepted touch/voice inputs can propagate through an automation chain without sufficient user verification. & Demonstrations on multiple glasses and two vehicles show unauthorized operations under the tested pairing, connectivity, and automation configurations. & A genuine connected device can relay an unauthorized command. Security checks must cover consequential actions as well as the identity of the device sending them. \\
\addlinespace[4pt]
\end{longtable}

\begin{longtable}{@{}>{\raggedright\arraybackslash}p{\dimexpr0.13\textwidth-2\tabcolsep\relax}>{\raggedright\arraybackslash}p{\dimexpr0.18\textwidth-2\tabcolsep\relax}>{\raggedright\arraybackslash}p{\dimexpr0.21\textwidth-2\tabcolsep\relax}>{\raggedright\arraybackslash}p{\dimexpr0.28\textwidth-2\tabcolsep\relax}>{\raggedright\arraybackslash}p{\dimexpr0.2\textwidth\relax}@{}}
\caption{Deployment evidence: social acceptability, agency, and trust.}\label{tab:deployment-social-evidence}\label{tab:privacy_social_acceptance}\label{tab:l4_social_acceptance_1}\label{tab:l4_social_acceptance_2}\\
\toprule
\rowcolor{gray!15}
\textbf{Study} & \textbf{Platform, people, and purpose} & \textbf{Approach examined} & \textbf{Evaluation and main finding} & \textbf{Conditions and implications} \\
\midrule
\endfirsthead
\caption[]{Deployment evidence: social acceptability, agency, and trust. (continued)}\\
\toprule
\rowcolor{gray!15}
\textbf{Study} & \textbf{Platform, people, and purpose} & \textbf{Approach examined} & \textbf{Evaluation and main finding} & \textbf{Conditions and implications} \\
\midrule
\endhead
\midrule
\multicolumn{5}{r}{\footnotesize Continued on the next page}\\
\endfoot
\bottomrule
\endlastfoot
\rowcolor{gray!5}
\textbf{Cross-cultural AR-glasses acceptance}\newline\cite{kim2025cultural} & AR smart glasses; prospective wearers in the United States and South Korea. & Technology-acceptance questionnaires relating usefulness, ease of use, privacy concern, attitudes, and intention. & Online surveys: 402 U.S. and 898 South Korean respondents, sampled in different years. Usefulness related to attitudes and intention in both; privacy relationships differed by sample. & Reported attitudes and intentions provide context-specific acceptance evidence. Privacy concern had no significant direct relationship with intention in either sample; cultural explanations remain interpretive. \\
\addlinespace[4pt]
\textbf{Everyday smart-glasses perceptions in Australia}\newline\cite{mcgregor2024exploring} & Australian owners and non-owners; survey framed around Ray-Ban Stories camera glasses. & WEAR-scale and personal-innovativeness survey covering social fears, aspirational desires, familiarity, and reported use. & Cross-sectional online convenience sample of 1,037 respondents. Owners reported higher acceptability, greater fulfillment of aspirational desires, and lower social fears than non-owners. & Ownership and favorable perceptions were associated; the design does not establish that ownership causes acceptance or that an inconspicuous form factor guarantees adoption. \\
\addlinespace[4pt]
\rowcolor{gray!5}
\textbf{Wearable AR video ethnography}\newline\cite{fleck2025wearable} & Published accounts of wearable AR use; 74\% of videos concern Apple Vision Pro or Quest. & Digital ethnography of unsponsored YouTube videos describing use beyond initial unboxing. & 112 videos totaling 26 h 53 min. Analysis identified physical and technical burdens, privacy concerns, social uncertainty, and differences between device types. & The selected videos reveal reported everyday experiences, not representative usage rates. Headset-heavy entertainment findings should not be generalized to glasses; Ray-Ban accounts emphasized capture and communication. \\
\addlinespace[4pt]
\textbf{Subtle heads-up interaction}\newline\cite{orlosky2021ubiquitous} & Wearers performing everyday activities; field prototype combined ring input with Vuzix Blade. & Compare subtle input techniques, then examine the preferred finger technique during ongoing activities. & Sixteen-person Wizard-of-Oz comparison and eight-person field study, with 59 valid prompted instances. Preferences depended on occupied body parts and activity; ring use could interrupt conversation mentally. & Small physical gestures still compete for attention and can interfere with gripping objects. Reported preferences do not establish one universally unobtrusive input method. \\
\addlinespace[4pt]
\rowcolor{gray!5}
\textbf{Wearable-display design space}\newline\cite{lyons2021design} & Wearable displays across body locations, including smart glasses; wearer and other intended audiences. & Literature-derived design space and interactive tool organizing display properties, placement, content, audience, and temporal characteristics. & Analysis and classification of existing projects supply a vocabulary for comparing designs; the contribution is not a user study demonstrating social acceptance. & Provides a vocabulary for analyzing information audience and visibility; social comfort and trust require empirical evaluation. \\
\addlinespace[4pt]
\textbf{AR-kward encounters}\newline\cite{zhang2025arkward} & Quest 3 in outdoor presentation and farmers' market tasks; wearers, vendors, and passersby. & In-the-wild observation and interviews examining social encounters and technical barriers during public AR use. & Sixteen wearers and 22 non-users. Reactions included curiosity, discomfort, indifference, and habituation; eye visibility and unfamiliar behavior affected interaction. & Public acceptance varied within the studied situations. The bulky pass-through headset and recruitment of interested bystanders limit transfer to ordinary-looking glasses and broader populations. \\
\addlinespace[4pt]
\rowcolor{gray!5}
\textbf{Paracentral progress visualization}\newline\cite{tu2022paracentral} & HoloLens 2 displaying secondary progress information while wearers maintain a conversation. & Compare circular and linear progress displays with text labels near the conversation partner's face. & Twelve wearers in a simulated-conversation study and twelve in a subsequent conversation study with trained partners. Circular progress displays were generally preferred and supported monitoring with less reported distraction. & Visual encoding and timing influence attention during conversation. Controlled tasks, fixed viewing distance, and a university sample constrain generalization; this design informed ParaGlassMenu. \\
\addlinespace[4pt]
\textbf{ParaGlassMenu}\newline\cite{tu2023paraglassmenu} & Head-worn menu interaction for IoT control during conversation; both wearer and partner experiences. & Extend paracentral information presentation into subtle bidirectional menu control and compare alternative interfaces. & Twenty-person interface comparison and twelve conversation pairs. The menu supported conversational control, while more complex tasks increased demands and partners could misinterpret the amount of digital interaction. & Builds on paracentral visualization. Discreet control can preserve participation in tested tasks, but low movement does not remove cognitive competition or guarantee legibility to partners. \\
\addlinespace[4pt]
\rowcolor{gray!5}
\textbf{AI-eyewear user acceptance}\newline\cite{niu2025exploring} & U.S. respondents reporting experience with AI-powered eyewear. & Questionnaire analysis using structural equation modeling and fuzzy-set qualitative comparative analysis. & 629 MTurk respondents. Acceptance was associated with perceived conversational and task intelligence, naturalness, aesthetics, usefulness, and privacy concerns. & The measured constructs describe perceived value and anticipated use. Cross-sectional self-reports do not demonstrate causal effects, sustained everyday adoption, or benefits attributable specifically to embodiment or wearing comfort. \\
\addlinespace[4pt]
\textbf{Smart glasses in science classrooms}\newline\cite{arantes2026too} & Classroom use and institutional responsibility, considered alongside road-safety and other policy contexts. & Conceptual comparative policy analysis using classroom scenarios to examine governance, risk, and responsibility. & Analyzes mismatches between policy assumptions and potential educational use; the scenarios are analytical devices rather than observations or an acceptance experiment. & Individual willingness to use glasses does not resolve institutional responsibilities. Supplies governance questions for deployment; it does not quantify adoption or establish that schools are safer or more accepting than roads. \\
\addlinespace[4pt]
\rowcolor{gray!5}
\textbf{Impressions of healthcare professionals}\newline\cite{sobieraj2023can} & Observers viewing nurses and physicians wearing different smart-glasses designs in standardized photographs. & Photo-based impression experiment with qualitative focus groups addressing appearance and professional perception. & 415 experimental respondents. Glasses designs affected warmth, competence, and trustworthiness ratings; controlling for technology willingness attenuated some differences. & Device appearance can affect first impressions within professional roles. The study does not measure actual clinical relationships, diagnostic performance, or patient outcomes during glasses use. \\
\addlinespace[4pt]
\textbf{Negotiating dyadic AR interactions}\newline\cite{chung2023negotiating} & Five wearer--partner pairs using AR face filters in an exploratory laboratory interaction. & Observe conversations and interview both participants about the augmented view, self-presentation, privacy, and control. & Ten participants. Wearers could feel more relaxed while partners lacked knowledge of their augmented representation or control over it; recording capability was not always understood. & Wearer benefit and partner agency can diverge. Transparency and control suggestions are design implications; the study did not establish their effectiveness in longitudinal public use. \\
\addlinespace[4pt]
\rowcolor{gray!5}
\textbf{XR work in public settings}\newline\cite{pavanatto2024working} & Virtual-display work with AR/VR headsets compared with laptops; workers and nearby bystanders. & Study work in public settings alongside observed use, interviews, and bystander responses. & The second study involved 18 workers and 209 distinct bystanders. Observers inferred XR work less accurately than laptop work; display limits, comfort, and surroundings shaped workers' experiences. & Application support must coexist with legible activity and suitable physical surroundings. Headset results and predominantly university samples do not establish general acceptance of lightweight glasses. \\
\addlinespace[4pt]
\textbf{Healthcare adoption and social interaction}\newline\cite{zuidhof2025social} & Healthcare professionals anticipating smart-glasses use and its social implications. & Questionnaire combining selected technology-acceptance variables with social-interaction constructs. & 450 healthcare respondents, predominantly from the United States. Adding selected social variables improved explained variance in anticipated adoption, with mixed individual effects and some unsupported hypotheses. & Professional acceptance involves social expectations as well as utility. The findings are anticipated and self-reported; they do not directly measure patient experience or show that social discomfort is unimportant in practice. \\
\addlinespace[4pt]
\end{longtable}

\endgroup

\end{document}